\documentclass[letterpaper,twocolumn,10pt]{article}
\usepackage{usenix}

\usepackage{cite}

\usepackage{graphicx} %
\usepackage{subfig}
\PassOptionsToPackage{table}{xcolor}
\usepackage{tikz}
\usepackage{amsmath}
\usepackage[table]{xcolor} %

\usepackage{tcolorbox}
\usepackage{todonotes}
\usepackage{amssymb}
\usepackage{mathtools}
\usepackage{amsthm}

\usepackage[utf8]{inputenc} %
\usepackage[T1]{fontenc}    %
\usepackage{hyperref}       %
\usepackage{url}            %
\usepackage{booktabs}       %
\usepackage{amsfonts}       %
\usepackage{nicefrac}       %
\usepackage{microtype}      %

\usepackage{soul}
\usepackage{algorithm}
\usepackage[noend]{algpseudocode}
\usepackage{enumitem}
\usepackage{breakurl}

\newenvironment{squishitemize}
{\begin{list}{\textbullet}{%
    \setlength{\itemsep}{0pt}%
    \setlength{\parsep}{0pt}%
    \setlength{\topsep}{0pt}%
    \setlength{\parskip}{0pt} %
    \setlength{\labelwidth}{.5in}%
    \setlength{\labelsep}{0.05in} %
    \setlength{\leftmargin}{.15in} %
    }}
  {\end{list}}
\title{The Erasure Illusion: Stress-Testing the Generalization of LLM Forgetting Evaluation}
\author{
{\rm Hengrui Jia$\dagger$, Taoran Li*, Jonas Guan$\dagger$, Varun Chandrasekaran*}\\
$\dagger$University of Toronto and Vector Institute, * University of Illinois Urbana-Champaign
}

\newif\ifdraft
\drafttrue

\usepackage{xparse}
\usepackage{xspace}
\usepackage{multirow}
\usepackage{graphicx}
\usepackage{algorithm}
\usepackage{algpseudocode}

\makeatletter
\def\@addpunct2#1{\ifnum\spacefactor>\@m \else#1\fi}
\newcommand{\para}[1]{\noindent\textbf{#1\unskip\@addpunct2{.}}~~}
\makeatother

\newcommand{\ie}{\textit{i.e.,}\@\xspace}
\newcommand{\eg}{\textit{e.g.,}\@\xspace}

\newcommand{\name}{Proximal Surrogate Generation (PSG)\xspace}
\newcommand{\mini}{PSG\xspace}

\newcommand{\novel}{Book X\xspace}

\usepackage{comment}

\begin{document}

\maketitle

\begin{abstract}

\looseness=-1
Machine unlearning aims to remove specific data influences from trained models, a capability essential for adhering to copyright laws and ensuring AI safety. 
Current unlearning metrics typically measure success by monitoring the model's performance degradation on the specific unlearning dataset ($D_u$). 
We argue that for Large Language Models (LLMs), this evaluation paradigm is insufficient and potentially misleading. 
{Many real-world uses of unlearning--motivated by copyright or safety--implicitly target not only verbatim content in $D_u$, but also behaviors influenced by the broader generalizations the model derived from it.}
We demonstrate that LLMs can pass standard unlearning evaluation and appear to have ``forgotten'' the target knowledge, while simultaneously retaining strong capabilities on content that is semantically adjacent to $D_u$. 
This phenomenon indicates that erasing exact sentences does not necessarily equate to removing the underlying knowledge. 
To address this gap, we propose \name, an automated stress-testing framework that generates a surrogate dataset, $\tilde{D}_u$. 
This surrogate set is constructed to be semantically derived from $D_u$ yet sufficiently distinct in embedding space. 
By comparing unlearning metric scores between $D_u$ and $\tilde{D}_u$, we can stress-test the reliability of the metric itself. 
Our extensive evaluation across three LLM families (Llama-3-8B, Qwen2.5-7B, and Zephyr-7B-$\beta$), three distinct datasets, and seven standard metrics reveals widespread inconsistencies. 
We find that current metrics frequently overestimate unlearning success, failing to detect retained knowledge exposed by our stress-test datasets.

\end{abstract}

\section{Introduction}
\label{sec:intro}

The rapidly increasing adoption of large language models (LLMs) in society raise severe privacy, compliance, and intellectual property concerns~\cite{cooper2024machine,nasr2025scalable,karamolegkou2023copyright,wen2023unveiling}. Machine unlearning aims to address these concerns by selectively removing the influence of specific data points from a trained model such that it behaves as if it had never seen that data~\cite{Cao2015, Bourtoule2021}. This is particularly important in domains requiring post-hoc data deletion; for instance, to satisfy regulatory frameworks like the GDPR’s “right to be forgotten”~\cite{Voigt2017}. %

{In many applications---particularly those motivated by copyright or safety---the data targeted for unlearning is often not merely surface-level content from an unlearning set $D_u$, but the higher-order knowledge inferred from $D_u$~\cite{liu2024machine}.}
{Beyond suppressing verbatim recitation, many real-world deployments implicitly expect unlearning} to eliminate abstractions that allow the model to reconstruct or advantageously leverage information from $D_u$ through paraphrase, composition, or transfer to semantically adjacent prompts~\cite{cooper2024machine}.
Regulatory or ethical requirements may demand not only the removal of verbatim training examples, but also the elimination of learned behaviors such as authorial style, domain-specific reasoning patterns, or unsafe content generation~\cite{eldan2023whosharrypotterapproximate}.

\looseness=-1
Critically, LLM unlearning success is often evaluated only on the unlearning set $D_u$~\cite{shi2024detecting,zhang2025mink,schwarzschild2024rethinking,schwarzschild2024rethinking,patil2023sensitiveinformationdeletedllms,thaker2025position}, or small handcrafted proxies closely resembling it~\cite{maini2024tofutaskfictitiousunlearning,li2024wmdpbenchmarkmeasuringreducing,eldan2023whosharrypotterapproximate}.
This risks yielding metric scores that only track surface recitation rather than the intended removal of higher-order knowledge (\S~\ref{subsec:statement}).
This mismatch creates a false sense of unlearning: a model can satisfy these metrics while retaining capabilities that transfer to content semantically related to $D_u$ (\S~\ref{sec:eval}).

To better understand this gap, we propose the notion of a \emph{surrogate unlearning dataset} $\tilde{D}_u$, which is disjoint but semantically-related to $D_u$. 
{This allows us to examine whether existing metrics removed not only surface-level traces of $D_u$ but also changes in related behaviors that practitioners often expect unlearning to influence.}
{For example, many applications implicitly expect that unlearning a novel would also cause the model to perform consistently worse on fan fiction derived from that novel.}
We find that this is not the case: current unlearning metrics may judge the unlearning of a novel as successful, yet simultaneously report no notable unlearning on a derived fan fiction, despite their significant semantic overlap (\S~\ref{sec:motivating_exp}).

Motivated by these insights, we propose \name to automatically construct such surrogate datasets $\tilde{D}_u$ for any given unlearning set $D_u$ and model $\theta$ (\S~\ref{sec:method}). Our approach simulates a lightweight fine-tuning process using $D_u$, and then samples text that the model becomes more confident in (i.e., shows increased likelihood). These samples are then perturbed, using techniques inspired by gradient-based control generation (e.g., GCG~\cite{zou2023universal}) to increase their (embedding) distance from $D_u$. This creates semantically derivative but non-overlapping data. The resulting $\tilde{D}_u$ can be used to stress-test unlearning metrics: if a metric fails to behave consistently across $D_u$ and $\tilde{D}_u$, it likely cannot reliably measure the success of unlearning.

\looseness=-1
We validate our findings across combinations of a range of LLMs, including Llama-3-8B~\cite{grattafiori2024llama3herdmodels}, Qwen2.5-7B~\cite{yang2024qwen2technicalreport}, and Zephyr-7B-beta~\cite{tunstall2023zephyrdirectdistillationlm}, and 3 datasets to simulate different unlearning scenarios: a fantasy novel dataset, a toxic comment dataset~\cite{DBLP:journals/corr/abs-1903-04561}, and a biomedical weapon dataset~\cite{li2024wmdpbenchmarkmeasuringreducing}. For each pair of model to be unlearned and $D_u$, we craft a $\tilde{D}_u$, consisting of sentences that can be learned from $D_u$, have sufficiently low perplexity, and are distant from sentences in $D_u$ in the embedding space.
After performing unlearning using Negative Preference Optimization (NPO)~\cite{zhang2024negativepreferenceoptimizationcatastrophic} and Representation Misdirection for Unlearning (RMU)~\cite{li2024wmdpbenchmarkmeasuringreducing}, we evaluate the unlearned models on both $D_u$ and $\tilde{D}_u$, using 7 common metrics for LLM unlearning. We observe that in {$61.1\%$} of the scenarios, the unlearning metrics are not consistent on $D_u$ and $\tilde{D}_u$ (\ie we define inconsistency as the two metric distributions' means differing by more than $1/2$ pooled standard deviation; \S~\ref{sec:eval}).
In particular, the metric scores computed on $\tilde{D}_u$ are mostly higher than $D_u$, meaning they suggest $\tilde{D}_u$ is less successfully unlearned than $D_u$, despite $\tilde{D}_u$ being learned from $D_u$.
Our code is available on  
\href{https://github.com/nick-jia/The-Metric-Mirage-A-False-Sense-of-Unlearning}{github}.

\looseness=-1
Our results suggest that the existence of meaningful surrogate datasets $\tilde{D}_u$ is not only plausible but also widespread. 
{We do not propose a new definition of unlearning; rather, our results show that many practical expectations of LLM unlearning are not reflected in existing metrics.}
Thus, we advocate for a more nuanced understanding of what it means to unlearn in generative models, urging the community to distinguish between privacy-oriented deletion and knowledge-based unlearning. 
In doing so, we hope to inspire the design of new metrics that capture both direct and generalized influence of training data.

To summarize, our main contributions are as follows:%
\begin{squishitemize}
\itemsep0em
    \item{\looseness=-1
    We highlight a critical concern in LLM unlearning: current metrics that determine unlearning success are unreliable (\S~\ref{subsec:statement}); metrics can certify success on $D_u$ yet report no notable unlearning on semantically-related set $\tilde{D_u}$ (\S~\ref{sec:eval}).}
    \item{We propose a lightweight algorithm to construct a surrogate dataset $\tilde{D_u}$ for any unlearning set $D_u$ and model to be unlearned $\theta$, which can be used to audit the reliability of an unlearning metric on the dataset (\S~\ref{sec:method}).}
    \item{We empirically evaluate 7 common unlearning metrics on all combinations of 3 LLMs, 3 datasets, and 2 standard unlearning algorithms (NPO and RMU) and find that none of the metrics consistently correlate with unlearning success; this shows the widespread impact of the metric mirage in LLM unlearning (\S~\ref{sec:setup}~and~\ref{sec:eval}).}
\end{squishitemize}

\section{Related Work}
\label{sec:rel}

{
In this section, we provide the necessary background information for this work. We define the unlearning problem and introduce notation. We also list and briefly describe some frequently used unlearning methods and evaluation metrics that will be studied.
}

\subsection{Background \& Notation}
\label{sec:notation}

\noindent{\bf What is the Unlearning Problem (for LLMs)?} {It is broadly defined as the mechanism to erase the ``impact'' of certain data points from a model (particularly, its behaviors). 
Depending on the incentives of unlearning (\eg privacy, copyright, etc.), this could either mean (a) {\textbf{Data-level:}} obtaining a model as if it were not trained on the {exact data points of the} unlearning set~\cite{barbulescu2024each,dou2024avoiding,huang2024demystifying} (similar to guarantees provided by differential privacy~\cite{hayes2025inexact,gupta2021adaptive}) {when the goal is to address privacy concerns}, or (b) {\textbf{Knowledge-level:}} erasing all knowledge and behaviors that could be learned {or derived} from the unlearning set post-hoc {when the goal is to \eg avoid copyright infrigement or mitigate unsafe behaviors}~\cite{liu2025rethinking, jia2024soul, ji2024reversing, yao2024large, wang2025rethinking, wang2025towards, wang2025uipe, ren2025general}. 
In this work, we focus on the latter, 
as it is a less well-defined form of unlearning.
{Consequently, many existing evaluation methods for unlearning are designed for the former yet used to judge the latter, since LLMs generalize and the influence of the unlearning dataset appears beyond memorized points.}
Besides, proper evaluation methods for the former have already been studied and shown to be difficult~\cite{hayes2025inexact}.
}

As suggested in prior work~\cite{Bourtoule2021}, unlearning can trivially be achieved by retraining the model without the data of interest; such an approach constitutes ``exact'' unlearning. However, since this is computationally expensive, many ``approximate/inexact'' (and cheaper) alternatives exist. Concrete strategies will be discussed in \S~\ref{sec:unlearn_methods}. {It is also worth noting that ``the data of interest'' may not always be traceable for knowledge-level unlearning, because the given unlearning set often serves only as a ``partial'' proxy for the full capabilities stakeholders wish to remove; similar information may exist elsewhere in the rest of the training dataset.}

\vspace{1mm}
\noindent{\bf Notation:} Having established the problem, we now present the notation we use throughout the paper. Let $\mathcal{D}$ be the data distribution, $D \sim \mathcal{D}^m$ denote the dataset (of size $m$) used to train the model, $D_u \subseteq D$ be the unlearning dataset (i.e., whose contents we want to unlearn). and $D_r$ be the retain dataset (i.e., whose information we want to retain; typically $D_r \subset D - D_u$). One can think of $D_u = \{z_1, \cdots, z_n\}$ where each $z_i = (x_i, y_i)$ where $x_i$ is some prefix and $y_i$ is some suffix. Let $\theta$ denote (the parameters of) the original (base) model trained on the full dataset $D$. Then, the exact unlearned model (\eg obtained by training solely on $D - D_u$) is denoted by $\theta_{u}$, and the approximately unlearned model is denoted by $\hat{\theta}_u$.
A good approximate unlearning algorithm $\mathcal{U}$ aims to find $\hat{\theta}_u$ that is ``close'' to $\theta_u$ in terms of behavior and performance, particularly with respect to $D_u$ and $D_r$. It is formally denoted by
$$ \hat{\theta}_u = \mathcal{U}(\theta, D_u) $$

The effectiveness of $\mathcal{U}$ is measured by how well $\hat{\theta}_u$ satisfies unlearning desiderata, such as:

\begin{enumerate}
\item {\em Knowledge Removal}: The model $\hat{\theta}_u$ should demonstrate no discernible knowledge of $D_u$. For instance, its performance \eg perplexity 
(denoted $\mathcal{L}_{p}$) on $D_u$ should be similar to that of a model not trained on $D_u$:
\begin{equation*} 
  \mathbb{E}_{(x, y) \sim D_u}[\mathcal{L}_{p}(\hat{\theta}_u(x), y)] \approx {}  
  \mathbb{E}_{(x, y) \sim D_u}[\mathcal{L}_{p}(\theta_u(x), y)]
\end{equation*}

\item {\em Utility Preservation}: The model $\hat{\theta}_u$ should maintain its performance on the retain set $D_r$:
\begin{equation*} 
  \mathbb{E}_{(x, y) \sim D_r}[\mathcal{L}_{p}(\hat{\theta}_u(x), y)] \approx {} \\ 
  \mathbb{E}_{(x, y) \sim D_r}[\mathcal{L}_{p}(\theta(x), y)]
\end{equation*}
\item {\em Efficiency}: The cost of obtaining $\hat{\theta}_u$ from $\theta$ should be significantly less than retraining $\theta_u$ from scratch.
\end{enumerate}

\subsection{Unlearning Methods}
\label{sec:unlearn_methods}

{We now provide a brief description of commonly-used approximate unlearning techniques for LLMs. Broadly speaking, these techniques often first identify the model parameters related to $D_u$, 
and then optimize these parameters to remove $D_u$'s impacts by \eg gradient-based methods. 

\looseness=-1
For instance, early approaches applied gradient ascent (GA) on $D_u$ to reduce the model’s log-probability of those examples~\cite{jang2022knowledgeunlearningmitigatingprivacy}; the neurons with stronger gradient signal are naturally the ones related to $D_u$.} While straightforward, this often causes severe degradation in generation quality. 
Huang {\em et al.}~\cite{huang2024unified} propose gradient-based unlearning that decomposes the unlearning update into weighted gradient ascent on $D_u$ and gradient descent on $D_r$ to minimize utility degradation. 
Negative Preference Optimization (NPO)~\cite{zhang2024negativepreferenceoptimizationcatastrophic}, a variant of preference optimization~\cite{rafailov2023direct} that only uses negative samples, minimizes the learned reward for ``undesired'' outputs, thereby striking a better balance between forgetting and retaining overall performance.
Variants such as SimNPO~\cite{fan2025simplicityprevailsrethinkingnegative} removes the need for a separate reference model, and recent work such as FLAT~\cite{wang2024llmunlearninglossadjustment} further streamlines the loss to avoid relying on $D_r$. 

Lu {\em et al.}~\cite{lu2022quarkcontrollabletextgeneration} proposed the Quantized Reward Konditioning method that conditions generation on discrete reward levels using special tokens; it samples outputs, bins them into reward quantiles, and fine-tunes the model to prefer high-reward outputs conditioned on their reward token, while staying close to the original model via a KL penalty. By doing so, they are able to post-hoc unlearn undesirable properties. 

More recent work like DeMem~\cite{kassem-etal-2023-preserving} utilize a negative similarity metric (between the ground truth and generation) as a reward to encourage the LLM to not memorize, ergo facilitating unlearning.
{While the aforementioned methods rely on end-to-end gradients with respect to specially designed loss functions,}
Li {\em et al.}~\cite{li2024wmdpbenchmarkmeasuringreducing} propose an alternative: Representation Misdirection for Unlearning (RMU) is a finetuning method that unlearns hazardous knowledge by optimizing to randomize the learned representations of $D_u$. As a follow up, LUNAR~\cite{shen2025lunarllmunlearningneural} performs unlearning by redirecting representations of forgotten data into latent regions where the model naturally signals its inability to respond. Work by Huutien {\em et al.}~\cite{huutien2025effectssteeringlatentrepresentation} also {adapts RMU and steers} model representation in the intermediate layer to a targeted random representation.

\looseness=-1
{It is worth noting that there have also been attempts to facilitate exact unlearning to LLMs. For instance,}
Chowdhury {\em et al.}~\cite{chowdhury2025scalableexactmachineunlearning} also introduce a parameter‐efficient fine-tuning framework that attaches lightweight, shard-specific adapter modules to the base model, enabling exact unlearning by simply removing the adapters corresponding to the data to be forgotten. But such an approach only works if the unlearning dataset is part of the fine-tuning data (and not the pre-training data).

\subsection{Evaluation Metrics}
\label{sec:metrics}

{Here, we taxonomize existing LLM unlearning evaluation metrics and discuss how they work. Before jumping into the details, we describe additional notation.}
We define $\hat{\theta}_r = \texttt{train}(\hat{\theta}_u, D')$ as the model obtained by retraining $\hat{\theta}_u$ on a dataset $D'$. %
Let $\mathcal{L}_{acc}$ denote an accuracy-based loss.

\vspace{1mm}
\noindent\underline{\em M1. Next-Token Prediction Loss:} To evaluate retention of language modeling ability on forgotten content, {the most straightforward approach is to measure the next-token prediction loss (which is often the training loss function of LLMs) on $D_u$ and assert the negative of it is below a certain threshold $\epsilon$,\footnote{We add the negative sign to ensure the consistency of the meaning of the threshold -- being below it indicates successful unlearning.}~\ie
\[
\ \mathbb{E}_{(x,y) \sim D_u}[- \mathcal{L}_p(\hat{\theta}_u(x))] \leq \epsilon_{\text{NTP}}.
\]

\vspace{1mm}
\noindent\underline{\em M2. Memorization:} {Works in the LLM memorization literature have been used as metrics for unlearning~\cite{schwarzschild2024rethinking}. They often measure how feasible it is to ``reactivate'' forgotten behaviors by adversarial prompt manipulations (i.e., like jailbreaking).} 
\[
\mathbb{E}_{(x,y) \sim D_u}[\mathcal{L}_{acc}(x,\theta_{\text{mod}}(x))] \leq \epsilon_{\text{mem}},
\]
Following Schwarzschild {\em et al.}~\cite{schwarzschild2024rethinking}, we define $\theta_{\text{mod}}(x) = \hat{\theta}_u(\mathcal{A}_{\text{JB}}(x))$, where $\mathcal{A}_{\text{JB}}(x)$ is an adversarially optimized prompt that elicits $x$ as output; that is, 
\[
\mathcal{A}_{\text{JB}}(x) = \arg\max_p \Pr[\hat{\theta}_u(p) = x],
\]
In other words, this metric measures the similarity between a sentence from $D_u$, and the generation of $\hat{\theta}_u$ when queried by a prompt that is adversarially optimized to generate the sentence. A lower metric value suggests that no adversarial prefix can generate the target sentence, indicating worse memorization and thus better unlearning, and vice versa.

\noindent\underline{\em M3. Relearning:} For {relearning}-based methods~\cite{ucki2025an}, one retrains $\hat{\theta}_u$ on $D' \subseteq D_u$ or on out-of-distribution points to obtain a relearned model $\theta_{\text{mod}}=\texttt{train}(\hat{\theta}_u, D')$. Then the next-token prediction loss is evaluated to quantify success of unlearning:
\[
\ \mathbb{E}_{(x,y) \sim D_u}[- \mathcal{L}_p(\theta_{\text{mod}}(x))] \leq \epsilon_{\text{relearn}}.
\]
That is, if the model has relearned parts of $D_u$ and still has high loss value, then it is a sign that unlearning is successful.

\vspace{1mm}
\noindent\underline{\em M4. Activation Drift:} {The activations of hidden layers (a.k.a., hidden states) of LLMs can also be used as a signal to infer whether the LLMs still have knowledge about $D_u$. If we denote the $i^{th}$ layer's hidden states of $\theta$ with respect to data point $x$ as $\theta^i(x)$, then the metric would be:}
\[
\ \mathbb{E}_{(x,y) \sim D_u}[- \mathcal{L}_p(g(\hat{\theta}^i_u(x)))] \leq \epsilon_{\text{AD}}.
\]
{where $g$ is a function mapping hidden states to the output space of the LLM. Motivated by an interpretation technique of LLMs named ``logit lens'', here $g$ is defined to be applying the final linear output head of the LLM to the intermediate hidden states~\cite{ucki2025an, patil2023sensitiveinformationdeletedllms}. Intuitively, this metric can identify cases of ``incorrect'' unlearning, where the probability of generating the tokens related to $D_u$ is only suppressed in later layers of the model, but not in the early layers. In other words, the knowledge about $D_u$ is still in the parameters of the model.
}

\vspace{1mm}
\noindent\underline{\em M5. Membership Inference:} {Membership inference attacks aim to identify whether a given data point is included in the training dataset of a given trained model. It has often been used as metrics for unlearning (not restricted to LLMs). More formally, it is defined as follows:}
\[
\mathbb{E}_{(x,y) \sim D_u}[\mathcal{A}_{\text{MI}}(\hat{\theta}_u, x)] \leq \epsilon_{\text{MIA}},
\]
{where $\mathcal{A}_{\text{MI}}$ can be any membership inference attack for LLMs~\cite{shi2024detecting,zhang2025mink}, and it outputs a membership score indicating how likely the given point $x$ is in the training set of $\hat{\theta}_u$. Intuitively, a lower score indicates successful unlearning, since $x$ is a training member of $\theta$.}

\looseness=-1
{The metrics listed thus far are the most practical to use.
We also list 2 other unlearning metrics that rely on the assumptions of either an exactly unlearned model, or on the existence of $T(D_u)$, a Q\&A variant of $D_u$, akin to a unit test.
These assumptions are hard to satisfy in practice: exact unlearning is too computationally costly for LLMs, and creating a $T(D_u)$ that ensures full coverage on $D_u$ is challenging.
In the limit, $T(D_u)$ must contain questions on all higher-order knowledge that can be inferred from $D_u$; this is as hard as solving unlearning in the generative setting.
Thus, while we include these metrics for completeness, due to their current impracticality, we do not consider them in our experiments.}

\vspace{1mm}
\noindent\underline{\em M6. Output Distribution Similarity:} {When there exists an exactly unlearned model (\ie one retrained on $D - D_r$), $\theta_u$, evaluating unlearning is trivial: one may measure how $\hat{\theta}_u$'s outputs differ from outputs of $\theta_u$.}
\[
\mathbb{E}_{x \sim T(D)}[\mu_{\text{KL}}(\theta_u(x), \hat{\theta}_u(x))] \leq \epsilon_{\text{OD}}.
\]
{We use $\mu_{\text{KL}}(\cdot, \cdot)$ to denote KL divergence, since outputs of both models are probability distributions. }

\vspace{1mm}
\noindent\underline{\em M7. Task Accuracy Deviation:} 
\looseness=-1 When there exists a dataset $T(D_u)$ derived from $D_u$ (such as prompt-response pairs, classification examples, or Q\&A tasks), one is able to ``test'' $\hat{\theta}_u$'s knowledge about $D_u$ on $T(D_u)$ to evaluate unlearning. 
Creating $T(D_u)$ can be achieved by manual efforts or neural approaches~\cite{du2017learning}, and is analogous to making exams based on contents of a course, except we wish for $\hat{\theta}_u$ to perform poorly on $T(D_u)$ when unlearning is successful:
\[
\mathbb{E}_{(x,y) \sim T(S(D_u))}[\mathcal{L}_{acc}(\hat{\theta}_u(x), y)] \leq \epsilon_{\text{TA}}.
\]

\vspace{1.5mm}
\begin{tcolorbox}
\noindent{\bf {A Common Theme:}} Across all metrics, a common theme is the attempt to quantify the residual influence of \( D_u \) on \( \hat{\theta}_u \). Thus, we can abstract all of these unlearning metrics using $\texttt{Eval}(\hat{\theta}_u, D_u)$, \ie it is a function defined on the unlearned model and the unlearning dataset. 
While these methods differ, they share the goal of detecting incomplete disentanglement from the unlearning set. However, a key limitation is that \( D_u \) may be only a partial proxy for the broader property we seek to erase. If similar information is redundantly encoded in other parts of \( \mathcal{D} \), removing \( D_u \) alone may be insufficient. In the rest of the paper, we propose our approach to exploit this observation.
\end{tcolorbox}

\section{Problem Formulation}
\label{sec:problem}

In this section, we formally state the problem statement and the threat model. We also present a motivating experiment to provide an intuition for the limitations in existing evaluation metrics for LLM unlearning.

\subsection{Problem Statement}
\label{subsec:statement}

The discussion of the metrics thus far has highlighted the main concern: {\em there is a narrow view of what it means to ``forget'' --- implicitly reducing {knowledge-level} unlearning to the erasure of specific examples}. 
In training supervised-learning ML models, concerns about evaluation artifacts are generally addressed through the use of held-out validation sets that provide an unbiased measure of generalization. However, in unlearning, no natural validation set exists: $D_u$ is precisely the data to be removed, and datasets semantically (and distributionally) ``close'' to it are usually undefined/unavailable. 
\begin{tcolorbox}
{\bf Our central objective} is to construct a surrogate validation dataset $\tilde{D}_u$ to audit the faithfulness of unlearning metrics.
\end{tcolorbox}
This enables us to \emph{falsify} unlearning metrics, \ie to demonstrate cases where a metric produces misleading assessments of the model's retained knowledge.  We are particularly interested in scenarios where evaluation on $D_u$ suggests unlearning is effective, whereas evaluation on our constructed $\tilde{D}_u$ suggests otherwise. 
We believe this is a critical failure mode as it can mean that sensitive, copyrighted, or harmful information, {\em intended to be removed}, still persists in deployed models.

\vspace{1mm}
\noindent{\bf Why Does This Matter?} One might argue that forgetting $D_u$ alone is sufficient (as that is what is specified in the unlearning objective), and that evaluation on similar but out-of-specification data (e.g., $\tilde{D}_u$) is unnecessary. However, this argument is fundamentally flawed. LLMs are designed to generalize by construction, and the influence of $D_u$ manifests not only in ``memorized'' tokens (or those directly corresponding to $D_u$) but also in higher-level abstractions and semantic patterns. In legal and ethical contexts, such as compliance with GDPR~\cite{Voigt2017}, takedown of harmful content, or the removal of proprietary data, the goal of unlearning is often to erase the \emph{knowledge} derived from $D_u$, not just the specific inputs. Failing to evaluate on semantically similar examples permits a model to retain and act on this learned knowledge, undermining the spirit of unlearning.
By constructing and evaluating on $\tilde{D}_u$, we provide a mechanism to expose such failures and stress-test unlearning metrics. Our approach offers a more faithful audit of unlearning outcomes, ensuring that metrics reflect the true extent to which a model has relinquished knowledge associated with $D_u$.

\subsection{Motivating Experiment}
\label{sec:motivating_exp}

\noindent{\bf Goal:} When unlearning aims to erase all knowledge and behaviors derived from $D_u$, a basic criterion for ideal unlearning metrics is that they yield consistent results on both the unlearning dataset $D_u$ and a closely related dataset $\tilde{D}_u$. By consistent, we mean that the distribution of metric values on $\tilde{D}_u$ should closely match that on $D_u$. To test whether this criterion holds, we conduct a motivating experiment using a fantasy novel (which we refer to as \novel) as $D_u$ and its fan fiction as $\tilde{D}_u$.

\noindent{\bf Approach \& Setup:} We collect a dataset of 12 fantasy novels and their corresponding fan fiction, where \novel is one of the 12 novels, and fine-tune a Llama-3-8B~\cite{grattafiori2024llama3herdmodels} model on them to simulate the system requiring unlearning. \novel is of particular interest because it was published after the model’s pre-training cutoff date. This ensures that the base model has not been exposed to it during pre-training, allowing us to disentangle unlearning effects from confounders introduced by pre-existing. We then unlearn $D_u$ using NPO~\cite{zhang2024negativepreferenceoptimizationcatastrophic}, treating the remaining novels (excluding fan fiction) as the retain dataset to preserve model utility. Once the unlearned model $\hat{\theta}_u$ is obtained, we compute an unlearning metric from category M5, Min-k\%~\cite{shi2024detecting}, on $\tilde{D}_u$ (more details in \S~\ref{sec:setup}).  

\begin{figure}[t]
    \centering
    \includegraphics[width=1\linewidth]{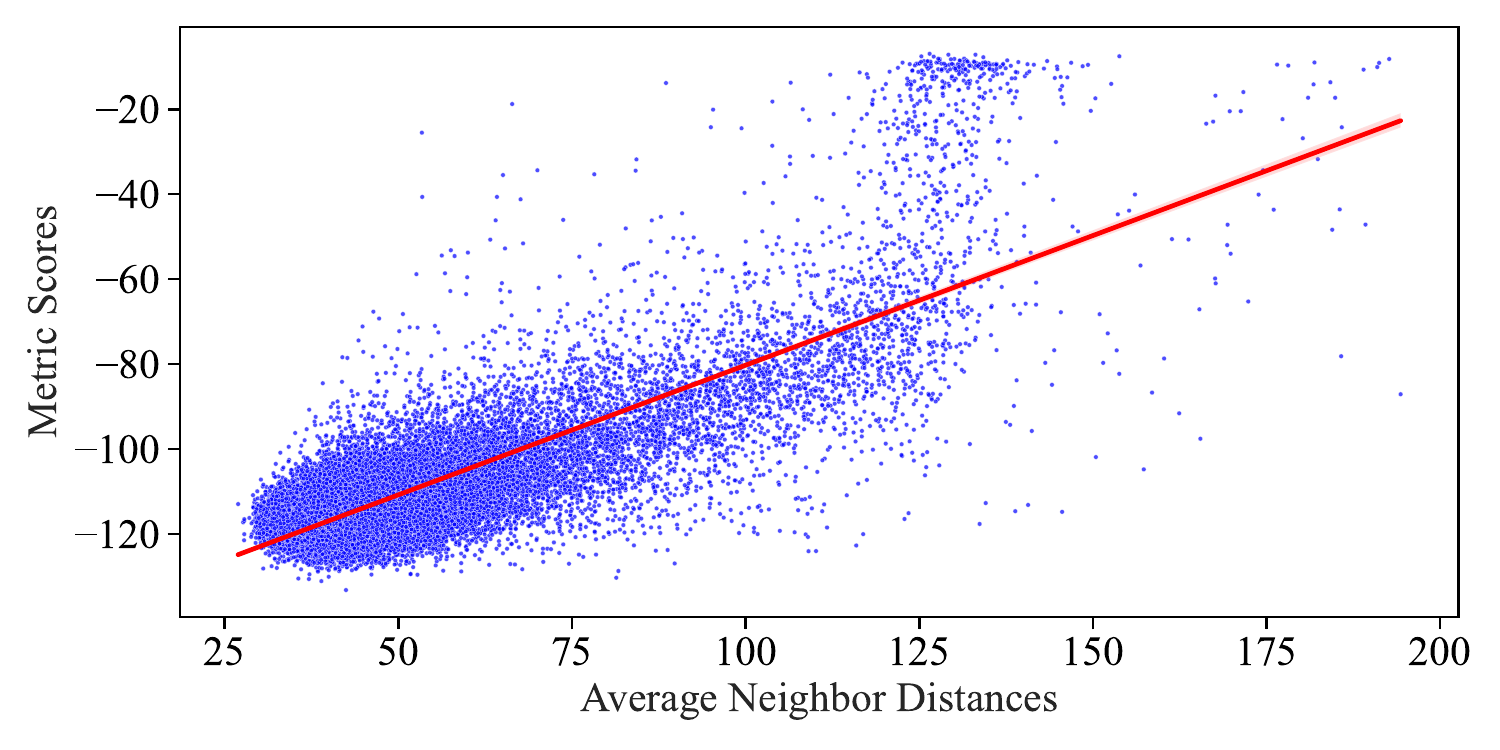}
    \caption{
    {\bf Min-k\% metric scores with respect to the average embedding distance from data points in $\tilde{D}_u$ to the 100 nearest neighbors in $D_u$.} One can see a clear correlation as indicated by the red regression line in the figure. This suggests despite the metric is expected to perform consistently across $D_u$ and $\tilde{D}_u$, it can be impacted by where the data points located in the embedding space.
    }
    \label{fig:motivating}
\end{figure}

To evaluate whether the metric behaves consistently across $D_u$ and $\tilde{D}_u$, we take each sentence from $\tilde{D}_u$ and measure its embedding similarity to sentences from $D_u$. Our hypothesis is that if the metric is indeed consistent, the metric value for a sentence in $\tilde{D}_u$ should be independent of its embedding similarity to sentences in $D_u$.  
Specifically, we use the last hidden states of the model (outputs of the final transformer block averaged across the sequence length) as embeddings, where the embedding of a point $x$ is denoted by $e(x)$. We use $\mu(\cdot,\cdot)$ to denote a certain similarity or distance metric, such as KL divergence, Jensen–Shannon divergence, $\ell_2$ distance, or cosine similarity. In our analysis, we adopt $\ell_2$ distance:  
\[
\mu(x \sim D_u, y \sim \tilde{D}_u) = \| e(x) - e(y) \| .
\]  
For each sentence from $\tilde{D}_u$, we identify its 100 nearest neighbors in $D_u$ using this distance and plot the Min-k\% scores of the corresponding sentence against these distances (Figure~\ref{fig:motivating}).

\noindent{\bf Results:} It can be observed that the two axes are highly correlated: sentences from $\tilde{D}_u$ that lie further from the embedding distribution of $D_u$ tend to receive lower unlearning scores, \ie they are more likely to be judged as ``not unlearned.'' At first glance this may seem reasonable: embeddings are often assumed to capture semantic meaning, so one might expect that sentences semantically distant from $D_u$ be considered irrelevant for unlearning. However, semantic distance alone does not imply the absence of information from $D_u$. In our case, $\tilde{D}_u$ consists of fan fictions of $D_u$; although semantically different, they are directly related to and built upon $D_u$. This contradicts our hypothesis and shows that the metric under consideration fails to behave consistently across $D_u$ and $\tilde{D}_u$.  

This raises a broader concern: {\em does a dataset analogous to $\tilde{D}_u$ always exist, independent of the model, dataset, and unlearning algorithm, and for all proposed unlearning metrics? If so, then no metric could ever definitively certify successful unlearning, since there would always be points highly related to $D_u$ yet assigned values suggesting the opposite.}

\subsection{Requirements \& Threat Model}
\label{sec:threat}

To address the question raised by the motivating experiment on falsifying unlearning metrics, we now formally state the requirements and the threat model. Any entity aiming to falsify the effectiveness of a metric would need to demonstrate the existence of a dataset $\tilde{D}_u$ such that:
\begin{enumerate}
\itemsep0em
\item $\tilde{D}_u$ is provably closely related to, or can be learned from, $D_u$, but
\item The metric values computed on $\tilde{D}_u$, $\texttt{Eval}(\hat{\theta}_u, \tilde{D}_u)$, are inconsistent with $\texttt{Eval}(\hat{\theta}_u, D_u)$.
\end{enumerate}

To achieve these two criteria, we believe the following properties are necessary for $\tilde{D}_u$. 

\begin{itemize}[leftmargin=*,noitemsep,topsep=0pt]

\item \textbf{P1. Natural Semantic Relevance}:
\textit{The dataset should be meaningfully related to $D_u$ and expressed in natural language.}
$\tilde{D}_u$ must be both semantically related to $D_u$ and linguistically fluent. This rules out adversarially crafted datasets that elicit specific responses but consist of meaningless or ungrammatical text, as such datasets are neither naturally occurring nor informative for evaluating unlearning. We quantify this property in two ways: (i) by comparing the likelihood of a sentence under the model before unlearning versus after finetuning on $D_u$, where a significant increase indicates related knowledge, and (ii) by measuring fluency using standard language-model metrics such as perplexity.

\looseness=-1
\item \textbf{P2. Embedding Separation}:
\textit{The dataset should stress-test metrics by including semantically close but embedding-wise distant cases.}
As motivated in \S~\ref{sec:motivating_exp}, samples from $\tilde{D}_u$ that are semantically close to $D_u$ but far apart in the embedding space are particularly informative. Unlearning metrics often behave inconsistently in such cases (as observed in the fan fiction dataset), making them critical boundary tests. By including these samples, we can examine whether metrics only work in the immediate neighborhood of $D_u$ or remain robust to embedding distance when semantic relevance is preserved. This property can be quantified by measuring the embedding distance from a sample to its $k$ nearest neighbors in $D_u$.

\end{itemize}

\looseness=-1
\noindent{\bf What Does This Mean?} Intuitively, the existence of $\tilde{D}_u$ (that satisfies the properties listed above) implies two things: (1) $\texttt{Eval}$ is not an accurate or comprehensive evaluation mechanism for unlearning. In particular, assume $\texttt{Eval}$ belongs to one of the metric categories defined in \S~\ref{sec:metrics}, and that its values lie in $\mathbb{R}^{+}$. Then, demonstrating $\texttt{Eval}(\hat{\theta}_u, \tilde{D}_u) > \texttt{Eval}(\hat{\theta}_u, D_u)$ shows that the metric can produce false positives \ie claiming unlearning is successful when it is not. (2) The metric $\texttt{Eval}$ can be exploited to spoof successful unlearning by selectively reporting results on a subset of $\tilde{D}_u$ that yields better scores than $D_u$. This latter case is particularly relevant when the full $D_u$ cannot be revealed, for example in scenarios involving unlearning bio-weapon–related knowledge.

\noindent{\bf Threat Model:} 
{
We consider an \emph{auditor} \ie an entity tasked with evaluating unlearning metrics on behalf of an external authority (such as a regulator), so that the metrics may be used by the external authority to faithfully assess knowledge-based unlearning. 
Since the focus is on unlearning metrics rather than the process of unlearning in this threat model, the auditor may use arbitrary models and unlearning datasets. Thus, 
}
the auditor has white-box access to the model before unlearning $\theta$, the model after unlearning $\hat{\theta}_u$, and the unlearning dataset $D_u$. The auditor may lack sufficient computational resources to perform exact unlearning (\ie retraining from scratch), but can carry out lightweight computations such as fine-tuning. White-box access also enables the auditor to inspect hidden states of the model. For example, the auditor can compute sentence embeddings of data points by averaging token embeddings from hidden layers, denoted $e_\theta(\cdot)$, where $\theta$ is the model under consideration.

\section{\name}
\label{sec:method}

\newcommand{\algosec}[1]{
 \hrulefill \quad \makebox[0.4\textwidth][c]{\textbf{#1}} \quad \hrulefill
}

\begin{algorithm*}[t]
\caption{\name}
\label{alg:main}
\begin{algorithmic}[1]
\Require Original model $\theta$, unlearning dataset $D_u$, iterations $N$, embedding function $e(\cdot)$, distance metric $\mu(\cdot, \cdot)$%
\Ensure $\tilde{D}_u$
\State $\tilde{D}_u = \{\}$
\State $\theta_f \gets SFT(\theta, D_u)$ \Comment{Finetune $\theta$ on $D_u$}
\State $E_f = \{e(x) | \forall x \in D_u \}$ \Comment{Compute embedding for every sentence in $D_u$}

\For{$sent \in D_u$}
\State $x = generate_{\theta}(sent)$ \Comment{Generate the next sentence}
\For{$t = 1$ to $N$}
    \State $x^* \gets \arg\min_{x} - w_{\text{dist}} d(e(x), minibatch(E_f)))$ \Comment{\textbf{Maximize Embedding Distance} via GCG}
    
    \State $x_{prompt} \gets \text{``Rephrase: ''} + x^* + \text{`` Rephrased: ''}$ \Comment{Prompt for Rephrasing}
    \State $S \gets w_{\text{{prefer}}}\Pr_{\theta_f}(.|x_{prompt}) +  w_{\text{likelihood}} \Pr_{\theta}(.|x_{prompt})$ \Comment{Weighted Summed Objective}
    
    \State $x \gets \text{sample}(S, \theta)$ \Comment{\textbf{Sampling-based rephrasing}}
\EndFor
\State $x_{\text{final}} \gets x$ 
\If{$\Pr_{\theta_f}(x_{\text{final}}) - \Pr_{\theta}(x_{\text{final}}) > \tau_{\text{prefer}}$ \& $d(e(x_{\text{final}}), KNN(e(x_{\text{final}}), E_f))) > \tau_{\text{dist}}$ \& $\Pr_{\theta}(x_{\text{final}}) > \tau_{\text{likelihood}}$ } \Comment{\textbf{Filtering}}
\State $\tilde{D}_u \gets \tilde{D}_u \cup  \{x\}$ \Comment{Append sentences satisfying G1, G2, and G3}
\EndIf
\EndFor
\State \Return $\tilde{D}_u$
\end{algorithmic}
\end{algorithm*}

\looseness=-1
The objective of our algorithm is to generate natural-language sentences that contain information derived from $D_u$, yet exhibit large embedding distances from the actual sentences in $D_u$ (recall the criteria described in \S~\ref{sec:threat}).
The underlying hypothesis is that current unlearning metrics are unreliable because they are too dependent on embedding-space distance from $D_u$.
Yet, as observed in \S~\ref{sec:motivating_exp}, embedding-space distance is not a reliable proxy for measuring whether two sets of sentences share information.
By generating sentences derived from $D_u$, yet are purposely far from $D_u$ in the embedding-space, we can test this hypothesis: if a metric certifies a model to have successfully unlearned sentences in $D_u$, but not the derived sentences that contain knowledge inferred from $D_u$, it is likely not reliable because it is overly sensitive to embedding-space distance.

To achieve the objective, we propose \name, which we formalize in Algorithm~\ref{alg:main}.

\subsection{Algorithm Overview}
\label{subsec:overview}

\looseness=-1
Given an LLM $\theta$ and an unlearning dataset $D_u$, we fine-tune $\theta$ on $D_u$ to obtain $\theta_f$ ({\bf line 2} of Algorithm~\ref{alg:main}).\footnote{A concern the reader may have here is that the LLM could have already been exposed to $D_u$ during pre-training, and thus have a high performance on knowledge generalized from $D_u$ even without fine-tuning. While it is possible to pre-train from scratch to avoid this, the computational costs make this option impractical. The key observation is that \mini only depends on $\theta$ improving its performance on $D_u$ after fine-tuning relative to before, which we found to be the case in all experiments.}
We use $\theta_f$ to generate $\tilde{D_u}$, a set of natural-language sentences that are derived from $D_u$, yet are for from sentences in $D_u$ in the embedding space. 
To facilitate this, we define an \emph{objective function} that combines likelihood increase with the standard likelihood objective for sampling ({\bf line 9}), and a complementary \emph{loss function} that penalizes proximity to $D_u$ in the embedding space ({\bf line 7}).
To generate sentences for $\tilde{D_u}$, we iteratively rephrase every sentence in $D_u$ to maximize the objective (ensuring {\bf P1}) while minimizing the loss (ensuring {\bf P2}).
The former is achieved through sampling ({\bf line 10}), while the latter uses a GCG-like optimization step ({\bf line 7}). Unlike standard GCG, which forces the generation of predefined outputs~\cite{zou2023universal}, our variant allows the model to generate any text as long as it increases embedding distance from $D_u$. This process can be seen as creating adversarial examples to maximize embedding separation, with fluency preserved through filtering ({\bf line 12}). This filtering stage is applied after $N$ iterations, based on three requirements:
(1) the likelihood increase from $\theta$ to $\theta_f$, which captures semantic relevance (and excludes gibberish by ensuring the sentence has sufficient likelihood under $\theta$);
(2) the embedding distance to the $k$ nearest neighbors from $D_u$, which ensures embedding separation; and (3) the likelihood of the sentence with respect to $\theta$ for fluency, as aforementioned.
To support verifying the second criterion efficiently, we pre-compute embeddings of all sentences in $D_u$ using $\theta$ ({\bf line 3}), avoiding repeated encoding during distance calculations. Sentences that exceed user-defined thresholds on all three axes (with guidelines in \S~\ref{subsec:p1} and \S~\ref{subsec:p2}) are retained as natural candidates derived from $D_u$ that may falsify unlearning metrics, {composing the surrogate dataset $\tilde{D}_u$. That said, differing from \eg Q\&A datasets used to evaluate unlearning, $\tilde{D}_u$ is not meant to cover all derivable knowledge from $D_u$. Instead, it is designed for a falsification-style stress test, where the sentences contained by it are ``boundary cases'' that likely have different metric values from $D_u$. Therefore, $\tilde{D}_u$ does not need to be a large dataset but can be as small as \eg 100 sentences, meaning \mini is not computationally expensive.}

\looseness=-1
Finally, it is worth noting that \mini requires only the dataset $D_u$ and the pre-unlearning model $\theta$ as input; it does not depend on the unlearned model $\hat{\theta}_u$. 

\subsection{Achieving Natural Semantic Relevance}
\label{subsec:p1}
  
The {\bf P1} criterion is enforced through both the sampling and filtering stages of Algorithm~\ref{alg:main}. To approximate how the model generalizes from $D_u$, we fine-tune the original pre-unlearning model $\theta$ solely on $D_u$, obtaining $\theta_f$. By definition, if training on $D_u$ reduces the cross-entropy loss for next-token prediction on a data point outside $D_u$, then the model has generalized from $D_u$ to that point. In unlearning scenarios motivated by, \eg copyright or safety concerns, such points should also be unlearned. Accordingly, we would expect unlearning metric values on these points to align with those obtained directly on $D_u$. This captures our key intuition: \textit{one needs to leverage observations from learning to better understand unlearning}.  

However, the reverse does not always hold: not all points influenced by $D_u$ will show a measurable loss decrease during fine-tuning. This is because $D_u$ was already part of $\theta$’s training, and some points may have plateaued at low loss. To bias sampling toward sentences that plausibly reflect generalization, we adapt the standard likelihood objective for token generation. In particular, given a prompt $x_{\text{prompt}}$, the model’s token distribution is adjusted as  
\[
w_{\text{prefer}}\,{\Pr}_{\theta_f}(\cdot \mid x_{\text{prompt}}) + w_{\text{likelihood}}\,{\Pr}_{\theta}(\cdot \mid x_{\text{prompt}}),
\]  
with $w_{\text{prefer}} > w_{\text{likelihood}}$ and the weights normalized to sum to $1$. This ensures preference for tokens whose likelihood increases from $\theta$ to $\theta_f$, while still preserving the (pretrained) base model’s fluency. Sampling under this mixed objective therefore produces candidates that are both semantically tied to $D_u$ and linguistically natural.  

\looseness=-1
At the end of the rephrasing loop, we obtain a candidate $x_{\text{final}}$ that has also undergone a few iterations of embedding-distance maximization (see next subsection). Because this step may cause the sentence to drift away from semantic relevance, we apply a final filtering criterion. Specifically, $x_{\text{final}}$ is retained in $\tilde{D}_u$ only if  
\[
{\Pr}_{\theta_f}(x_{\text{final}}) - {\Pr}_{\theta}(x_{\text{final}}) > \tau_{\text{prefer}},  
\]  
where $\tau_{\text{prefer}}$ is a threshold. {As a heuristic, it may be set based on the likelihood increase from $\theta$ to $\theta_f$ across $D_u$. For example, in the rest of this paper, we set it to the first quartile of this likelihood increase. This ensures that only sentences relevant to $D_u$ can be included in $\tilde{D}_u$, since irrelevant sentences would have an unchanged or even decreased (due to catastrophic forgetting~\cite{goodfellow2013an}) likelihood when the model is only finetuned on $D_u$.
} 
In addition, we also require that the likelihood under $\theta$ itself exceeds $\tau_{\text{likelihood}}$.
To discard low-quality or ungrammatical sentences,
{$\tau_{\text{likelihood}}$ may be set to the first quartile of likelihoods on $D_u$ so sentences in $\tilde{D}_u$ are comparably fluent to the ones from $D_u$.}
Together, these filtering conditions ensure that sentences in $\tilde{D}_u$ remain both semantically related to $D_u$ and fluent.

\subsection{Achieving Embedding Separation}
\label{subsec:p2}

\looseness=-1
To achieve {\bf P2}, \ie encouraging large embedding distances between points created by \mini and the unlearning dataset $D_u$, we adapt a GCG-like optimization~\cite{zou2023universal}. GCG is commonly used as a jailbreaking method: the model is queried with an adversarial prompt consisting of ``free tokens'' and a malicious or toxic request that would otherwise be refused. The free tokens are then optimized to maximize the likelihood of the model producing a specific continuation, \eg \texttt{"Here is the response..."}, thereby coercing it into answering the malicious prompt.  
In our setting, the goal is to maximize embedding distance rather than induce a specific output (see Appendix~\ref{app:jialbreak} for details). We therefore modify the GCG loss to  
\[
- w_{\text{dist}}\, d\big(e(x), \text{minibatch}(E_f)\big),
\]  
where $e(\cdot)$ denotes the embedding function, $E_f$ is the set of embeddings for $D_u$, and $d(\cdot,\cdot)$ is the averaged chosen distance metric (Euclidean or cosine). Performing GCG to minimize this loss increases the embedding distance.  
Concretely, the embedding function $e(\cdot)$ is the mean of the last-layer hidden states of $\theta$ (the model before unlearning), and $E_f$ can be precomputed before entering the for-loop in Algorithm~\ref{alg:main} to reduce computation. Distances are computed as the average distance from $e(x)$ to a mini-batch sampled from $E_f$. The optimization produces a pair $(x^*, \text{free tokens})$, analogous to adversarial prompts in standard GCG. When used in the rephrasing stage, querying the model with $x^*$ yields outputs that are distant from $E_f$ in the embedding space.  

Finally, in the filtering stage, we discard points with insufficient embedding-distance increase, using a threshold $\tau_{\text{dist}}$. 
Here, instead of computing the embedding distance from $x_\text{final}$ to a random minibatch sampled from $D_u$, we find the $k$ nearest neighbors (e.g., $k=100$) of $x_\text{final}$ from $D_u$, to reduce stochasticity.
Then, in practice, $\tau_{\text{dist}}$ may be set to the embedding distance from a point in $D_u$ to its $k$ nearest neighbors, averaged across all points in $D_u$, multiplied by a coefficient $>1$. This ensures the sentences included in $\tilde{D}_u$ to have higher embedding distances to $D_u$ than an average point in $D_u$, {preventing memorized patterns from $D_u$ to be included in $\tilde{D}_u$.}

\looseness=-1
\noindent \underline{\em Why GCG?} One may ask why a GCG-like method is necessary instead of the sampling-based approach used in the previous subsection. The key reason is that sampling operates on probability distributions over the vocabulary: a natural output when querying an LLM. In contrast, selecting the token that maximizes embedding-distance increase would require computing embeddings for the entire vocabulary. Such embeddings are not a standard output of LLM queries, and given the typically large vocabulary size, this would incur an intractable computational cost.

\section{Experimental Setup}
\label{sec:setup}

\noindent{\bf Models, Datasets \& Algorithms.}  
We conduct experiments on a combination of 3 pretrained LLMs: Meta-Llama-3-8B~\cite{grattafiori2024llama3herdmodels}, Qwen2.5-7B~\cite{yang2024qwen2technicalreport}, and Zephyr-7B-Beta~\cite{tunstall2023zephyrdirectdistillationlm}, together with 3 unlearning datasets from distinct domains: (i) a fantasy novel denoted as \novel\footnote{We use this anonymized novel instead of the more commonly considered \textit{Harry Potter} because the latter may not represent typical copyrighted content, given its extensive presence across English datasets.}, (ii) a toxic subset of comments from the Civil Comments platform~\cite{DBLP:journals/corr/abs-1903-04561}, and (iii) a dataset containing biological weapon–related knowledge, WMDP-bio~\cite{li2024wmdpbenchmarkmeasuringreducing}.  We apply 2 representative unlearning techniques: NPO~\cite{zhang2024negativepreferenceoptimizationcatastrophic} and RMU~\cite{li2024wmdpbenchmarkmeasuringreducing}. Combining 3 LLMs, 3 unlearning datasets, and 2 techniques yields 18 unlearned models in total.  {Unless otherwise specified, the same hyperparameters of \mini are used across all settings.}

For each triplet (LLM, unlearning dataset $D_u$, unlearning technique), we first fine-tune the LLM on $D_u$ combined with additional datasets from similar distributions.\footnote{An exception is WMDP-bio, which prior work has shown to already be included in pretraining corpora~\cite{li2024wmdpbenchmarkmeasuringreducing}.} Portions of these additional datasets serve as retain sets during unlearning, enabling us to simulate an LLM trained on $D_u$ without training from scratch. We then perform unlearning using the corresponding retain set to preserve model utility. The full configuration of unlearning and retain datasets for each experiment is summarized in Table~\ref{tab:dataset_other}, with additional details (including compute resources) provided in Appendix~\ref{app:exp-setup}.   

\vspace{1mm}
\noindent{\bf Unlearning Metrics Used.}  
After obtaining the unlearned models, we evaluate unlearning success using the metrics described in \S~\ref{sec:metrics}. In particular, we considered 7 metrics and their implementation details are as follows: 

\begin{squishitemize}
\itemsep0em
\item \textit{Likelihood} (M1) is directly computed from the model’s loss value, \ie the negative log-likelihood of a given sentence.  

\item \textit{Memorization} (M2) is a variant of Adversarial Compression Ratio (ACR)~\cite{schwarzschild2024rethinking}, a metric for evaluating memorization in LLMs. ACR measures the number of adversarial tokens required by GCG to force the model to reproduce an exact training point. However, varying the number of adversarial tokens makes this method computationally expensive for large datasets. We therefore modify it to instead measure the number of identical tokens between the model’s generation and a training sentence, given a fixed number of adversarial tokens.  

\item \textit{Orthogonalization}~\cite{ucki2025an} (M3) is a form of relearning attack. It uses a small fraction of $D_u$ to estimate the parameter change caused by unlearning, and then applies this change back to the model. The likelihood of sentences under this modified model serves as a metric for how easily the unlearned knowledge can be relearned.  

\item \textit{Logit Lens}~\cite{patil2023sensitiveinformationdeletedllms} (M4) examines the hidden states of LLMs. Most current LLMs consist of several transformer blocks followed by a linear classification head to predict the next token. The hypothesis behind this metric is that unlearned content may only be suppressed in later layers. Accordingly, the method connects the linear head to an earlier transformer block to predict the next token. The likelihood of sentences under this setup defines the metric value.  

\item \textit{zlib Ratio} (M5) is the ratio between likelihood and the zlib entropy~\cite{zlib} of a sentence. It was proposed as a membership inference technique for LLMs~\cite{Carlini2021Extracting}. The idea is that a sentence is more likely to be a member if it has low entropy, which can result from, \eg repeated patterns of memorized training text.  

\item \textit{Min-K\% Prob} (M5) is another widely used membership inference technique for LLMs~\cite{shi2024detecting}. Its intuition is that for a sentence seen during training, even its lowest-probability tokens should exceed a certain threshold.  

\item \textit{Min-K\%++} (M5) is a follow-up~\cite{zhang2025mink} to Min-K\% Prob. The main difference is that it normalizes per-token probability by the expected probability of all tokens in the vocabulary, since tokens observed during training typically have locally (rather than globally) maximal probability.  
\end{squishitemize}

\section{Evaluation}
\label{sec:eval}

\begin{figure}[t!]
    \centering
    \subfloat[Negative Preference Optimization (NPO)]{
    \includegraphics[width=0.93\linewidth]{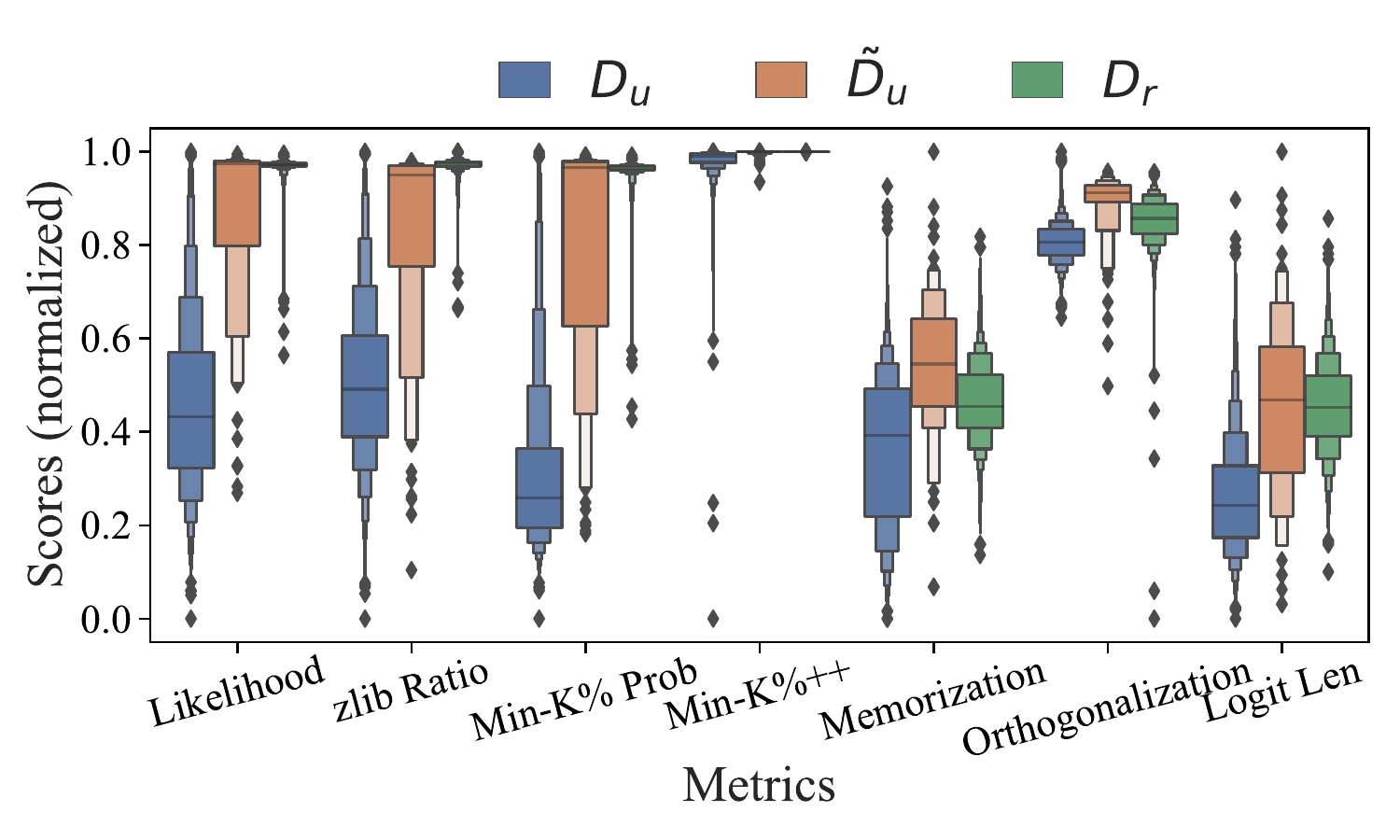}
    }
    
    \subfloat[Representation Misdirection for Unlearning (RMU)]{
    \includegraphics[width=0.93\linewidth]{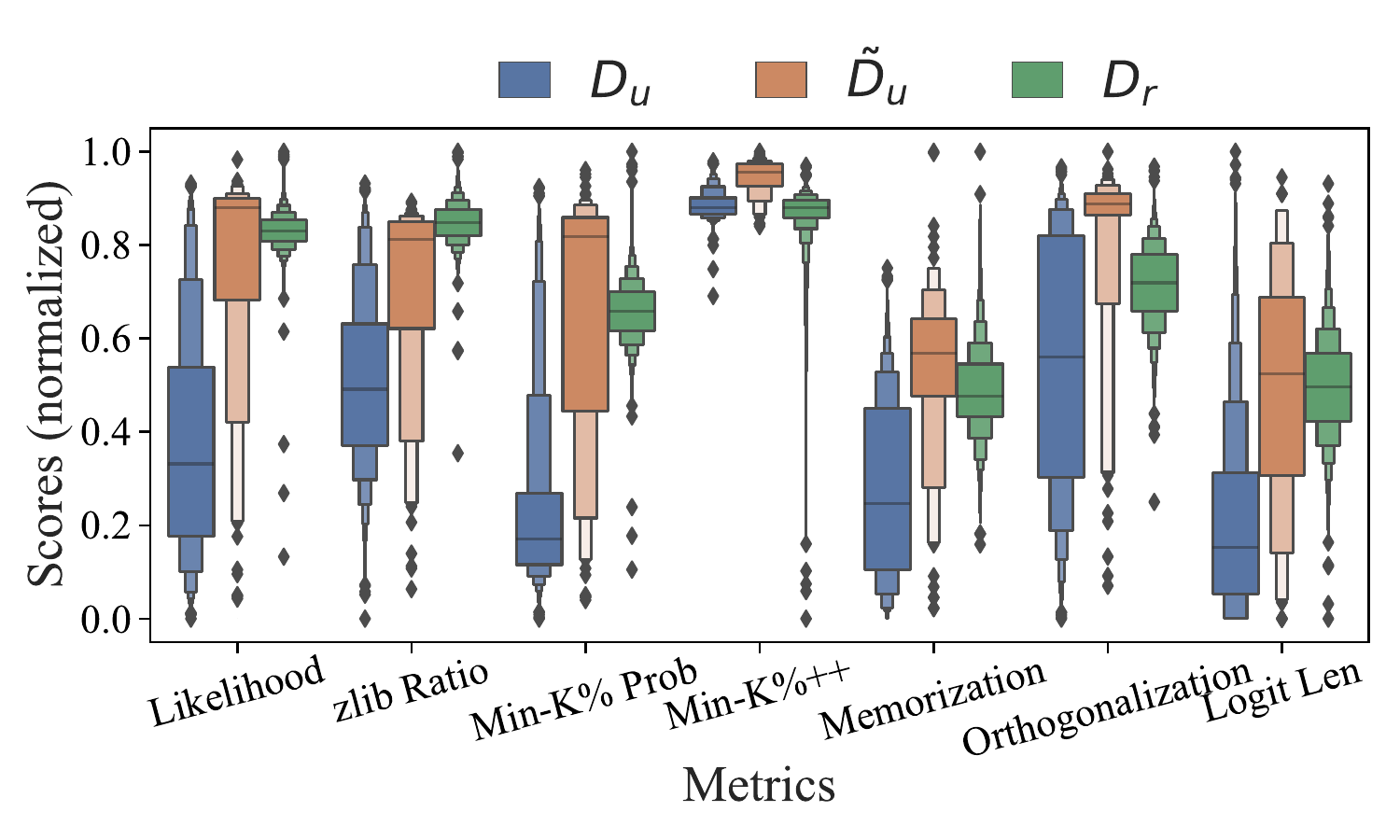}}
    \caption{\looseness=-1\textbf{Boxplots of metric scores of the unlearning dataset $D_u$, surrogate unlearning dataset $\tilde{D}_u$, and retain dataset $D_r$, when \textit{\novel} is unlearned from a Llama-3 model using (a) NPO and (b) RMU.} The metric scores are normalized to a [0,1] range, and a larger value indicates unsuccessful unlearning (according to the metrics). One can observe that the box corresponding to $\tilde{D}_u$ is almost always higher than the box corresponding to $D_u$. This indicates that for this setting, \mini is able to create sentences that falsify the unlearning metrics by demonstrating they are unlearned less successfully than $D_u$.}    \label{fig:hics_llama}
\end{figure}
\begin{table*}[t!]
\centering
\footnotesize
\setlength{\tabcolsep}{4pt}
\begin{tabular}{|lll|lllllll|}
\hline
\multicolumn{1}{|l|}{Dataset}                                                                                       & \multicolumn{1}{l|}{Model}                    & Method & \multicolumn{7}{c|}{Unlearning Metrics}                                                                                                                                                                                                                                                                      \\ \hline
                                                                                                                    &                                               &        & \multicolumn{1}{l|}{Likelihood}             & \multicolumn{1}{l|}{zlib Ratio}              & \multicolumn{1}{l|}{Min-K\% Prob}           & \multicolumn{1}{l|}{Min-K\%++}              & \multicolumn{1}{l|}{Memorization}           & \multicolumn{1}{l|}{Orthogonalization}       & Logit Len              \\ \hline
\multicolumn{1}{|l|}{\multirow{6}{*}{\begin{tabular}[c]{@{}l@{}} \novel \end{tabular}}} & \multicolumn{1}{l|}{\multirow{2}{*}{Llama-3}} & NPO    & \multicolumn{1}{l|}{\textbf{2.13 +/- 0.16}} & \multicolumn{1}{l|}{\textbf{1.61 +/- 0.12}}  & \multicolumn{1}{l|}{\textbf{2.16 +/- 0.14}} & \multicolumn{1}{l|}{\textbf{0.90 +/- 0.19}} & \multicolumn{1}{l|}{\textbf{1.16 +/- 0.09}} & \multicolumn{1}{l|}{\textbf{1.37 +/- 0.07}}  & \textbf{1.20 +/- 0.09} \\ \cline{3-10} 
\multicolumn{1}{|l|}{}                                                                                              & \multicolumn{1}{l|}{}                         & RMU    & \multicolumn{1}{l|}{\textbf{1.56 +/- 0.12}} & \multicolumn{1}{l|}{\textbf{0.93 +/- 0.10}}  & \multicolumn{1}{l|}{\textbf{1.74 +/- 0.14}} & \multicolumn{1}{l|}{\textbf{1.68 +/- 0.10}} & \multicolumn{1}{l|}{\textbf{1.35 +/- 0.11}} & \multicolumn{1}{l|}{\textbf{1.13 +/- 0.11}}  & \textbf{1.18 +/- 0.11} \\ \cline{2-10} 
\multicolumn{1}{|l|}{}                                                                                              & \multicolumn{1}{l|}{\multirow{2}{*}{Qwen2.5}} & NPO    & \multicolumn{1}{l|}{\textbf{2.01 +/- 0.12}} & \multicolumn{1}{l|}{\textbf{0.54 +/- 0.09}}  & \multicolumn{1}{l|}{\textbf{1.96 +/- 0.13}} & \multicolumn{1}{l|}{\textbf{1.53 +/- 0.11}} & \multicolumn{1}{l|}{\textbf{1.51 +/- 0.09}} & \multicolumn{1}{l|}{-0.49 +/- 0.06}          & \textbf{0.82 +/- 0.08} \\ \cline{3-10} 
\multicolumn{1}{|l|}{}                                                                                              & \multicolumn{1}{l|}{}                         & RMU    & \multicolumn{1}{l|}{\textbf{1.47 +/- 0.13}} & \multicolumn{1}{l|}{\textbf{0.77 +/- 0.12}}  & \multicolumn{1}{l|}{\textbf{1.57 +/- 0.12}} & \multicolumn{1}{l|}{\textbf{1.43 +/- 0.15}} & \multicolumn{1}{l|}{\textbf{1.62 +/- 0.10}} & \multicolumn{1}{l|}{\textbf{1.54 +/- 0.13}}  & \textbf{0.85 +/- 0.10} \\ \cline{2-10} 
\multicolumn{1}{|l|}{}                                                                                              & \multicolumn{1}{l|}{\multirow{2}{*}{Zephyr}}  & NPO    & \multicolumn{1}{l|}{\textbf{1.47 +/- 0.09}} & \multicolumn{1}{l|}{0.20 +/- 0.05}           & \multicolumn{1}{l|}{\textbf{1.42 +/- 0.08}} & \multicolumn{1}{l|}{\textbf{1.33 +/- 0.09}} & \multicolumn{1}{l|}{\textbf{1.18 +/- 0.09}} & \multicolumn{1}{l|}{-0.35 +/- 0.04}          & \textbf{1.03 +/- 0.07} \\ \cline{3-10} 
\multicolumn{1}{|l|}{}                                                                                              & \multicolumn{1}{l|}{}                         & RMU    & \multicolumn{1}{l|}{\textbf{1.36 +/- 0.10}} & \multicolumn{1}{l|}{-0.11 +/- 0.04}          & \multicolumn{1}{l|}{\textbf{1.30 +/- 0.09}} & \multicolumn{1}{l|}{\textbf{1.14 +/- 0.07}} & \multicolumn{1}{l|}{\textbf{1.18 +/- 0.08}} & \multicolumn{1}{l|}{\textbf{1.08 +/- 0.11}}  & \textbf{0.98 +/- 0.08} \\ \hline
\multicolumn{1}{|l|}{\multirow{6}{*}{\begin{tabular}[c]{@{}l@{}}Toxic Civil \\ Comments\end{tabular}}}              & \multicolumn{1}{l|}{\multirow{2}{*}{Llama-3}} & NPO    & \multicolumn{1}{l|}{\textbf{0.65 +/- 0.10}} & \multicolumn{1}{l|}{0.23 +/- 0.09}           & \multicolumn{1}{l|}{\textbf{0.73 +/- 0.10}} & \multicolumn{1}{l|}{\textbf{0.54 +/- 0.08}}          & \multicolumn{1}{l|}{\textbf{0.89 +/- 0.07}} & \multicolumn{1}{l|}{-0.23 +/- 0.06}          & 0.46 +/- 0.09          \\ \cline{3-10} 
\multicolumn{1}{|l|}{}                                                                                              & \multicolumn{1}{l|}{}                         & RMU    & \multicolumn{1}{l|}{\textbf{0.67 +/- 0.11}} & \multicolumn{1}{l|}{-0.10 +/- 0.08}          & \multicolumn{1}{l|}{\textbf{0.66 +/- 0.09}} & \multicolumn{1}{l|}{\textbf{0.62 +/- 0.08}} & \multicolumn{1}{l|}{\textbf{0.79 +/- 0.08}} & \multicolumn{1}{l|}{\textbf{0.87 +/- 0.10}}  & 0.44 +/- 0.09          \\ \cline{2-10} 
\multicolumn{1}{|l|}{}                                                                                              & \multicolumn{1}{l|}{\multirow{2}{*}{Qwen2.5}} & NPO    & \multicolumn{1}{l|}{0.38 +/- 0.08}          & \multicolumn{1}{l|}{-0.15 +/- 0.02}          & \multicolumn{1}{l|}{\textbf{0.56 +/- 0.09}} & \multicolumn{1}{l|}{0.40 +/- 0.09}          & \multicolumn{1}{l|}{\textbf{0.62 +/- 0.07}} & \multicolumn{1}{l|}{0.06 +/- 0.04}           & 0.16 +/- 0.10          \\ \cline{3-10} 
\multicolumn{1}{|l|}{}                                                                                              & \multicolumn{1}{l|}{}                         & RMU    & \multicolumn{1}{l|}{\textbf{0.70 +/- 0.08}} & \multicolumn{1}{l|}{-0.23 +/- 0.05}          & \multicolumn{1}{l|}{\textbf{0.78 +/- 0.09}} & \multicolumn{1}{l|}{\textbf{1.04 +/- 0.09}} & \multicolumn{1}{l|}{\textbf{0.97 +/- 0.07}} & \multicolumn{1}{l|}{0.40 +/- 0.05}           & -0.02 +/- 0.09         \\ \cline{2-10} 
\multicolumn{1}{|l|}{}                                                                                              & \multicolumn{1}{l|}{\multirow{2}{*}{Zephyr}}  & NPO    & \multicolumn{1}{l|}{0.15 +/- 0.09}          & \multicolumn{1}{l|}{\textbf{-0.51 +/- 0.05}} & \multicolumn{1}{l|}{0.26 +/- 0.10}          & \multicolumn{1}{l|}{0.26 +/- 0.10}          & \multicolumn{1}{l|}{0.29 +/- 0.07}          & \multicolumn{1}{l|}{0.40 +/- 0.08}           & 0.04 +/- 0.08          \\ \cline{3-10} 
\multicolumn{1}{|l|}{}                                                                                              & \multicolumn{1}{l|}{}                         & RMU    & \multicolumn{1}{l|}{0.05 +/- 0.09}          & \multicolumn{1}{l|}{\textbf{-0.63 +/- 0.03}} & \multicolumn{1}{l|}{0.19 +/- 0.10}          & \multicolumn{1}{l|}{-0.15 +/- 0.08}         & \multicolumn{1}{l|}{0.27 +/- 0.07}          & \multicolumn{1}{l|}{-0.07 +/- 0.09}          & -0.04 +/- 0.08         \\ \hline
\multicolumn{1}{|l|}{\multirow{6}{*}{WMDP-bio}}                                                                     & \multicolumn{1}{l|}{\multirow{2}{*}{Llama-3}} & NPO    & \multicolumn{1}{l|}{0.49 +/- 0.05}          & \multicolumn{1}{l|}{\textbf{-1.06 +/- 0.02}} & \multicolumn{1}{l|}{0.25 +/- 0.03}          & \multicolumn{1}{l|}{0.23 +/- 0.04}          & \multicolumn{1}{l|}{\textbf{0.88 +/- 0.05}} & \multicolumn{1}{l|}{\textbf{0.65 +/- 0.09}}           & 0.36 +/- 0.02          \\ \cline{3-10} 
\multicolumn{1}{|l|}{}                                                                                              & \multicolumn{1}{l|}{}                         & RMU    & \multicolumn{1}{l|}{\textbf{0.88 +/- 0.03}} & \multicolumn{1}{l|}{\textbf{-1.10 +/- 0.02}} & \multicolumn{1}{l|}{0.46 +/- 0.06}          & \multicolumn{1}{l|}{0.07 +/- 0.08}          & \multicolumn{1}{l|}{\textbf{0.89 +/- 0.05}} & \multicolumn{1}{l|}{\textbf{-4.10 +/- 0.03}} & 0.39 +/- 0.00          \\ \cline{2-10} 
\multicolumn{1}{|l|}{}                                                                                              & \multicolumn{1}{l|}{\multirow{2}{*}{Qwen2.5}} & NPO    & \multicolumn{1}{l|}{0.34 +/- 0.04} & \multicolumn{1}{l|}{\textbf{-1.30 +/- 0.01}} & \multicolumn{1}{l|}{0.24 +/- 0.04}          & \multicolumn{1}{l|}{-0.15 +/- 0.00}         & \multicolumn{1}{l|}{\textbf{1.08 +/- 0.04}} & \multicolumn{1}{l|}{\textbf{3.44 +/- 0.04}}  & \textbf{0.55 +/- 0.04} \\ \cline{3-10} 
\multicolumn{1}{|l|}{}                                                                                              & \multicolumn{1}{l|}{}                         & RMU    & \multicolumn{1}{l|}{\textbf{0.66 +/- 0.03}} & \multicolumn{1}{l|}{\textbf{-1.42 +/- 0.02}} & \multicolumn{1}{l|}{0.24 +/- 0.05}          & \multicolumn{1}{l|}{-0.46 +/- 0.06}         & \multicolumn{1}{l|}{\textbf{1.27 +/- 0.05}} & \multicolumn{1}{l|}{\textbf{2.32 +/- 0.18}}  & 0.48 +/- 0.01          \\ \cline{2-10} 
\multicolumn{1}{|l|}{}                                                                                              & \multicolumn{1}{l|}{\multirow{2}{*}{Zephyr}}  & NPO    & \multicolumn{1}{l|}{0.28 +/- 0.07}          & \multicolumn{1}{l|}{\textbf{-1.56 +/- 0.04}} & \multicolumn{1}{l|}{0.33 +/- 0.07}          & \multicolumn{1}{l|}{0.37 +/- 0.07}          & \multicolumn{1}{l|}{\textbf{0.52 +/- 0.09}}          & \multicolumn{1}{l|}{0.37 +/- 0.10}           & 0.30 +/- 0.03          \\ \cline{3-10} 
\multicolumn{1}{|l|}{}                                                                                              & \multicolumn{1}{l|}{}                         & RMU    & \multicolumn{1}{l|}{\textbf{0.98 +/- 0.09}} & \multicolumn{1}{l|}{\textbf{-0.92 +/- 0.04}} & \multicolumn{1}{l|}{\textbf{0.90 +/- 0.08}} & \multicolumn{1}{l|}{\textbf{1.02 +/- 0.08}} & \multicolumn{1}{l|}{\textbf{1.02 +/- 0.11}} & \multicolumn{1}{l|}{0.45 +/- 0.11}           & 0.45 +/- 0.05          \\ \hline
\end{tabular}
\caption{
\textbf{Standardized mean difference between metric values of $\tilde{D}_u$ and $D_u$ across various settings.} For each combination of the 3 datasets, 3 models, and 2 unlearning methods, we compute the 7 metrics on the unlearned model using both $\tilde{D}_u$ and $D_u$. Then for each $\tilde{D}_u$ (100 points), we randomly sample 100 points from $D_u$ and measure the mean difference between their metric values, standardized by the pooled standard deviation. The random sampling is repeated 100 times so that we report average mean differences with their standard deviations in this table. We presenting the mean difference with absolute value greater than $0.5$ in bold---this is $61.1\%$ of the table. In addition, no metric performs consistently on $D_u$ and $\tilde{D}_u$ under all settings.
}
\label{table:res}
\end{table*}

In this section, we design experiments to simulate scenarios where trained LLMs are required to unlearn parts of their training datasets. Our evaluation is guided by 3 questions:  

\begin{squishitemize}
\itemsep0em
    \item How do existing unlearning metrics behave when applied to datasets that are surrogates of the unlearning dataset?  
    \item Are the inconsistencies we observe specific to a single model–dataset pair, or do they generalize across models, datasets, and unlearning methods?  
    \item How sensitive is \mini to its design choices, such as embedding separation thresholds, distance metrics, and generation strategies?  %
\end{squishitemize}  

Our empirical findings can be summarized as follows:  

\begin{squishitemize}
\itemsep0em
    \item Surrogate datasets created by \mini consistently yield metric values that differ substantially from those of the original unlearning datasets, despite being semantically tied to them (as shown in the case study in \S~\ref{subsec:main_res}.)  
    \item These inconsistencies generalize broadly: across 3 models, 3 datasets, and 2 unlearning methods, no existing metric can always remain consistent between $D_u$ and $\tilde{D}_u$ (as shown in \S~\ref{subsec:main_res}.)
    \item Ablation studies show that \mini {does not require heavy hyperparameter tuning} and behaves predictably with respect to its hyperparameters: larger separation thresholds increase effectiveness but reduce efficiency while choices of embedding distance metrics and generation strategies have insignficiant impacts (as shown in \S~\ref{subsec:ablation}.)
\end{squishitemize}

\subsection{Achieving Falsification}
\label{subsec:main_res}

\noindent{\bf Case Study: Fantasy Novel Unlearning.}  
We begin with the scenario where the fantasy novel \novel is unlearned from a Llama-3 model using NPO. We construct the surrogate unlearning dataset $\tilde{D}_u$ and compute the values of 7 unlearning metrics, alongside those for the unlearning dataset $D_u$ and the retain dataset $D_r$. To visualize the results, we plot the distributions of the metric values as boxplots in Figure~\ref{fig:hics_llama}. Since different metrics operate on different scales, we normalize each metric by its minimum and maximum values across the 3 datasets, mapping them into the range $[0,1]$, where larger values indicate less successful unlearning.  
{It is noteworthy that the variability of metric values is large for some metrics, even for $D_u$. However, this is not problematic as long as the values computed on $D_u$ are significantly lower than $D_r$, meaning the differences between the unlearned data and retaining data can be captured by the corresponding metric.}
\textit{Here, this case study serves as a concrete starting point to illustrate how $\tilde{D}_u$ behaves relative to $D_u$ and $D_r$.}  

\vspace{1mm}  
\noindent{\bf Divergence Between $\tilde{D}_u$ and $D_u$.}  
As expected, the boxes corresponding to $D_u$ lie below those of $D_r$, indicating that unlearning impacts the model’s behaviors on $D_u$, which is successfully captured by the metrics. However, the boxes corresponding to $\tilde{D}_u$ created by \mini are markedly different from those of $D_u$. This is noteworthy because
we set $\tau_{\text{prefer}} = \operatorname{Quantile}_{0.25}\!\left( {\Pr}_{\theta_f}(x) - {\Pr}_{\theta}(x) \;\middle|\; x \in D_u \right)$, which ensures that every sentence in $\tilde{D}_u$ exhibits a likelihood increase greater than at least 25\% of the points in $D_u$. 
This guarantees that $\tilde{D}_u$ contains information and knowledge learned from $D_u$.  
Yet, the metric values suggest that $\tilde{D}_u$ is not unlearned as effectively. In other words,
\textit{these results show that $\tilde{D}_u$, despite being semantically tied to $D_u$, produces very different outcomes under existing metrics.}  %

\looseness=-1
\vspace{1mm}  
\noindent{\bf Results Across Models and Datasets.}  
To validate the consistency of these findings, we run \mini for all other dataset–model pairs to obtain $\tilde{D}_u$, {using the same hyperparameters,} and evaluate the same 7 unlearning metrics on both $D_u$ and $\tilde{D}_u$ with respect to the corresponding unlearned models.\footnote{A related concern is whether the gap between the metric scores for $D_u$ and $\tilde{D_u}$ may be a result of variances in the unlearning process, \ie the gap only exists for some unlearning runs, but not others.
To test this, we run NPO on Llama-3 with the Book X unlearning set 4 times with identical hyperparameters but different random seeds (Figure~\ref{fig:repeat}); results show the gap between $D_u$ and $\tilde{D_u}$ is consistent across all runs and metrics.} Since an ideal unlearning metric should behave similarly on $D_u$ and $\tilde{D}_u$, we compute standardized mean differences between their metric values, reported in Table~\ref{table:res} (boxplots similar to Figure~\ref{fig:hics_llama} for these settings can be found in {Figures~\ref{fig:hics_qwen} to~\ref{fig:wmdp_zephyr}} in Appendix~\ref{app:exp_res}). Following standard practice, we bold the cells with absolute values exceeding $0.5$, indicating cases of inconsistency. We observe that this occurs in the majority of cases, and no metric maintains a mean difference below $0.5$ across all dataset–model pairs. These results validate our hypothesis from \S~\ref{sec:problem}: most existing unlearning metrics depend heavily on the embedding-space closeness of the evaluation points to $D_u$. In other words, the metrics overfit to $D_u$ and fail to reliably judge unlearning success on sentences that are semantically derived from $D_u$.  
\textit{Taken together, this highlights a systemic limitation: current unlearning metrics are not robust to semantically related but embedding-distant data.}

\subsection{Ablation Study}
\label{subsec:ablation}

\mini involves several tunable hyperparameters, as well as design choices such as the distance metric used for computing embedding separation. In this section, we conduct ablation studies to examine the effects of these hyperparameters and justify our design decisions.  

\begin{figure}[t!]
    \centering

    \includegraphics[width=\linewidth]{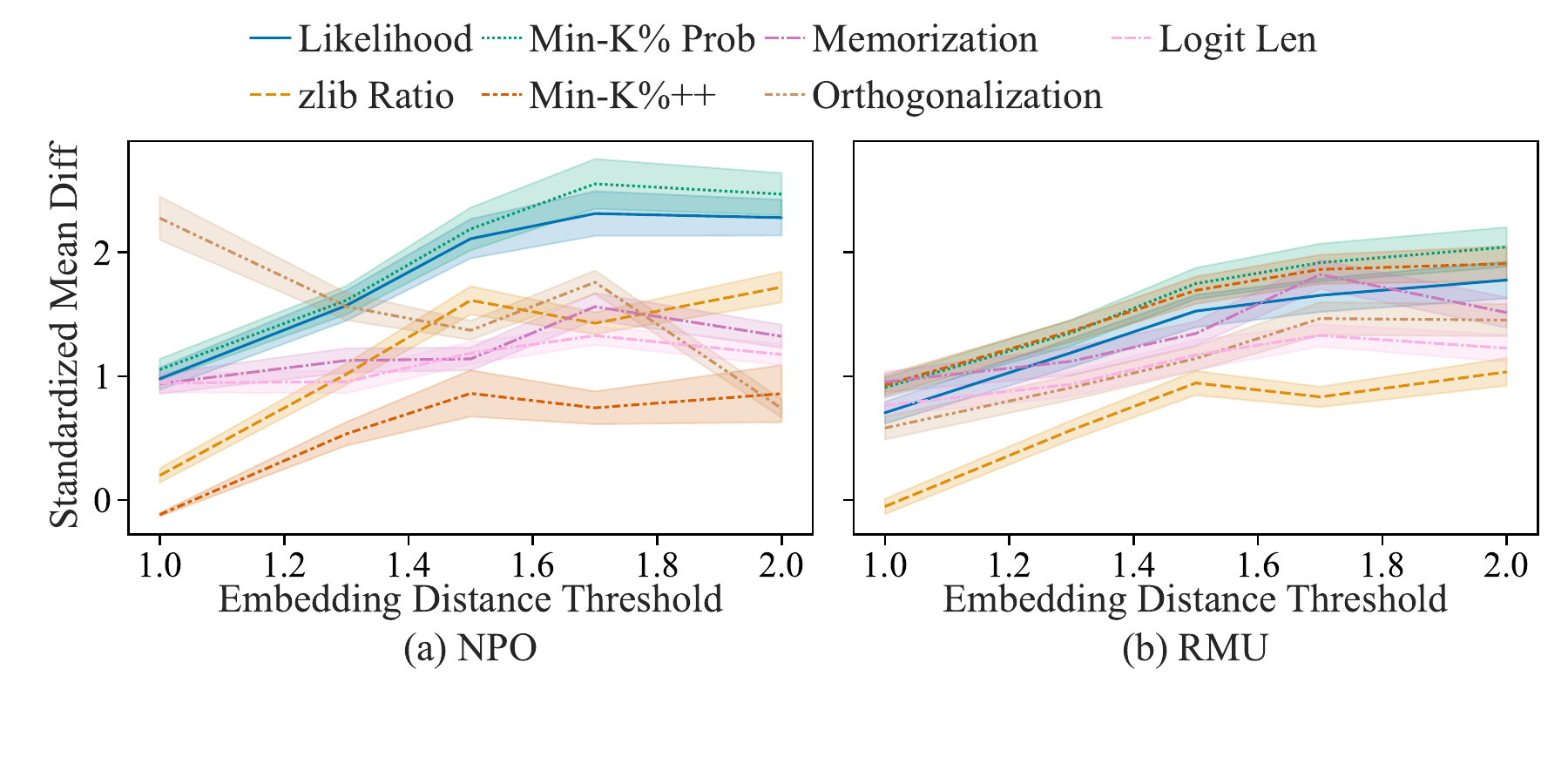}
    
    \caption{\textbf{Standardized mean difference between metric values of $D_u$ and $\tilde{D}_u$, with respect to $\tau_{\text{dist}}$.} Here $\tau_{\text{dist}}$ is relative to the average embedding distance among points in $D_u$. It can be observed that for both NPO and RMU, there is a positive correlation between the standardized mean difference and $\tau_{\text{dist}}$ for most of the metrics. This validates our hypothesis that increasing the embedding distances between $D_u$ and $\tilde{D}_u$ can cause the metrics to perform more differently on them.}
    \label{fig:hics_llama_curve}
\end{figure}

\vspace{1mm}
\noindent{\bf Embedding Separation Threshold.}  
\looseness=-1
As discussed in \S~\ref{sec:method}, one essential component of \mini is to maximize the separation between sentences in $\tilde{D}_u$ and those in $D_u$. This is achieved by the GCG-like optimization and the filtering stage (lines 7 and 12 in Algorithm~\ref{alg:main}, respectively). 
{As aforementioned, the filtering threshold $\tau_{\text{dist}}$ may be set to the average embedding distance from a point in $D_u$ to its $k$ nearest neighbors, multiplied by a coefficient. Doing so avoids the need to search for hyperparameters for different model-dataset pairs. So far, we used $k=100$ and the same coefficient $1.5$ for all the settings. Here, to study the impact of embedding separation on the performance of \mini, we run \mini with varying coefficient values}
ranging from 1 to 2 and plot the standardized mean difference between unlearning metric values of $D_u$ and $\tilde{D}_u$. Results for Llama-3 are shown in Figure~\ref{fig:hics_llama_curve}, while those for Qwen2.5 and Zephyr appear in Figures~\ref{fig:hics_qwen_curve} and~\ref{fig:hics_zephyr_curve} in Appendix~\ref{app:exp_res}.  

\looseness=-1
As seen in the figures, there is a clear positive correlation between $\tau_{\text{dist}}$ and the standardized mean difference (which represents the effectiveness of \mini). When the coefficient is $1$, \ie $\tau_{\text{dist}}$ equals the average embedding distance among points in $D_u$, standardized mean differences are generally below $1$, with some near $0$, indicating little separation between $D_u$ and $\tilde{D}_u$. As $\tau_{\text{dist}}$ increases, standardized mean differences rise accordingly. That said, since $\tau_{\text{dist}}$ is used for filtering, higher thresholds reduce the number of unfiltered sentences and thus slow down \mini when constructing $\tilde{D}_u$.  
\textit{In short, larger separation thresholds make \mini more effective but at the cost of efficiency.}  

\begin{figure}[t!]
    \centering
    \subfloat[Negative Preference Optimization (NPO)]{
    \includegraphics[width=0.93\linewidth]{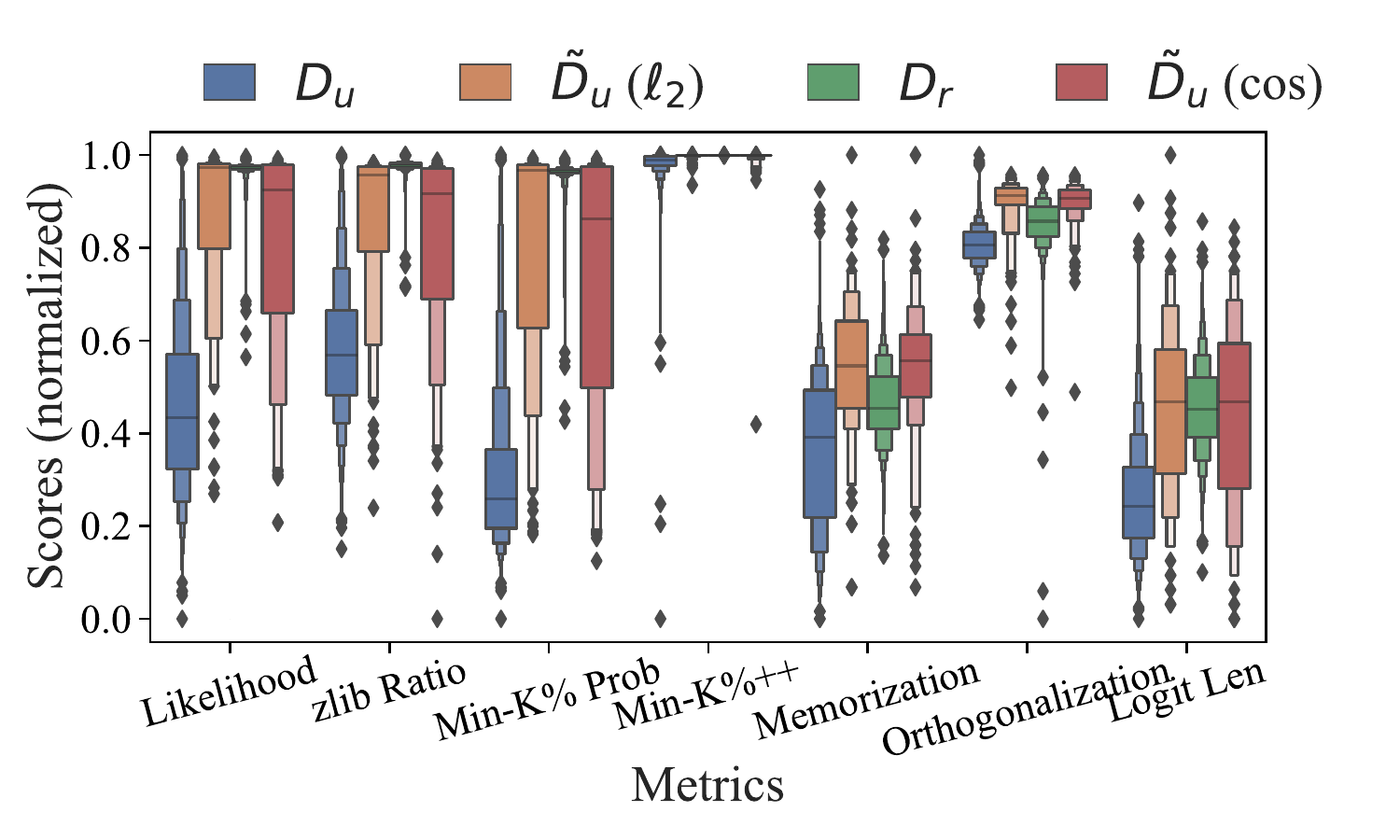}
    }
    
    \subfloat[Representation Misdirection for Unlearning (RMU)]{
    \includegraphics[width=0.93\linewidth]{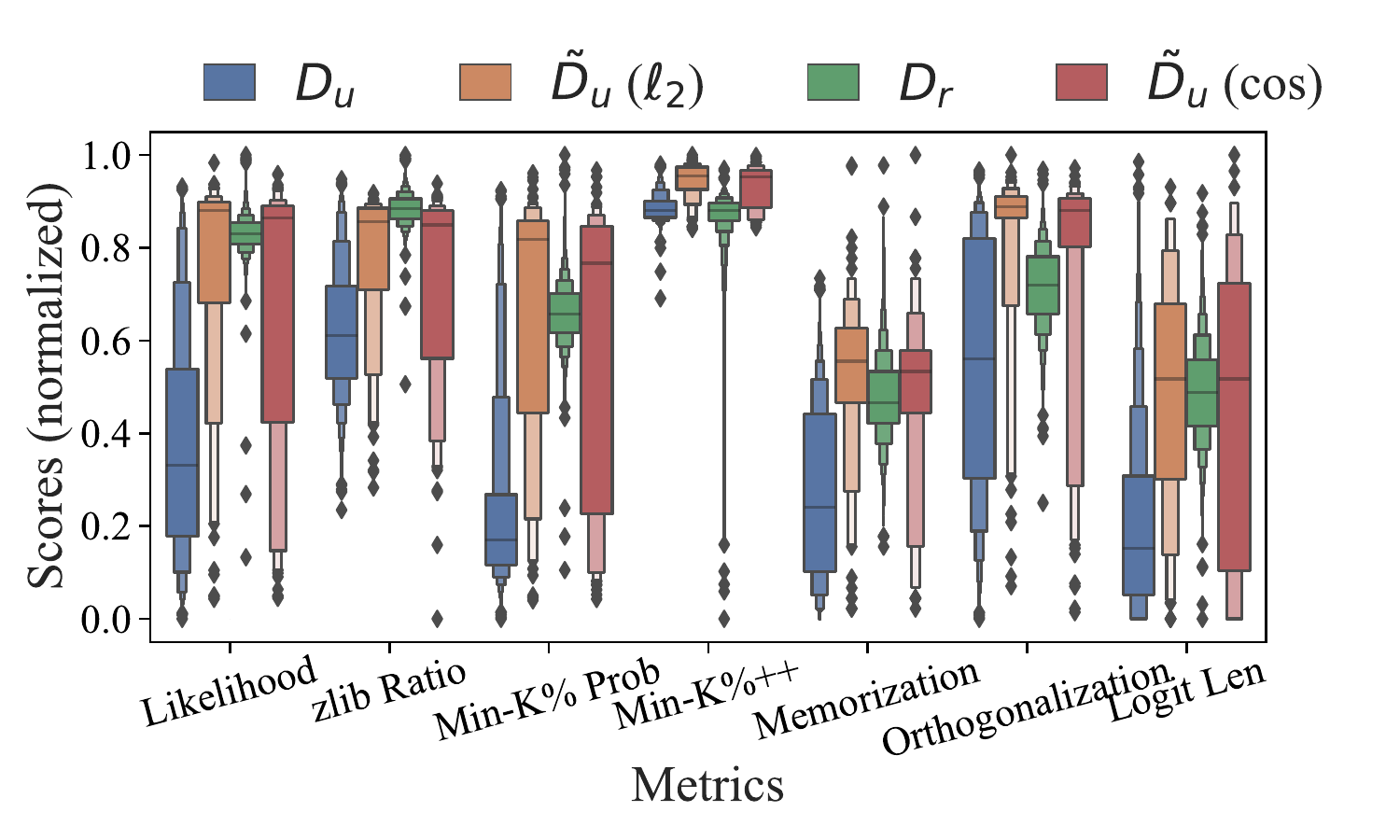}}
    \caption{\textbf{Boxplots of metric scores of 2 surrogate unlearning datasets, $\tilde{D}_u (\ell_2)$ and $\tilde{D}_u(cos)$, where the embedding distance is measured using $\ell_2$ distance and $cos$ distance respectively.} The metric scores of the unlearning dataset $D_u$ and retain dataset $D_r$ are also presented as reference. One can observe that for most cases, $\ell_2$ distance results in slightly more different (mean) scores from $D_u$ than $cos$ distance, and smaller variance in the distribution of the metric scores. Although the differences are not significant, we choose to use $\ell_2$ distance as the result of this ablation study.
    }
    \label{fig:ablation_dist}
\end{figure}  
\vspace{1mm}
\looseness=-1
\noindent{\bf Distance Metric.}  
Beyond the magnitude of separation, the choice of distance metric is another hyperparameter of \mini. We consider 2 common options: Euclidean ($\ell_2$) distance and cosine distance. To test whether this choice significantly affects \mini, we create $\tilde{D}_u$ using each metric and plot their corresponding metric values in Figure~\ref{fig:ablation_dist}, following the same format as Figure~\ref{fig:hics_llama}. Two observations emerge: (1) when cosine distance is used, the resulting metric distributions tend to have larger variance than with $\ell_2$ distance; (2) for some metrics, $\ell_2$ distance yields larger standardized mean differences relative to $D_u$. However, this is not consistent across all metrics, and the differences are generally subtle. For consistency, we use $\ell_2$ distance in other experiments, though \mini appears relatively robust to the choice of metric.  
\textit{Overall, \mini is not highly sensitive to whether cosine or Euclidean distance is used.}  

\begin{figure}[t!]
    \centering
    \subfloat[Negative Preference Optimization (NPO)]{
    \includegraphics[width=0.94\linewidth]{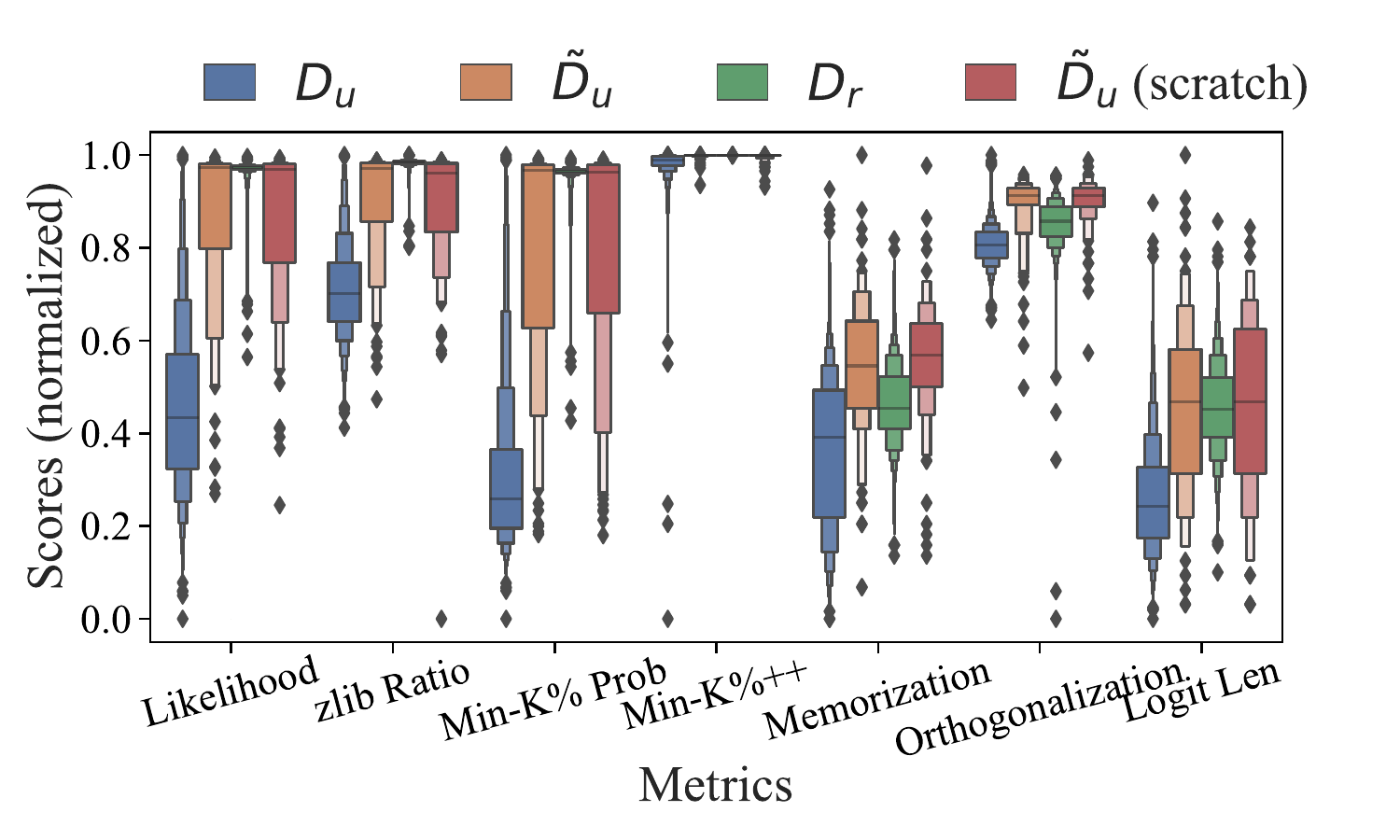}
    }
    
    \subfloat[Representation Misdirection for Unlearning (RMU)]{
    \includegraphics[width=0.94\linewidth]{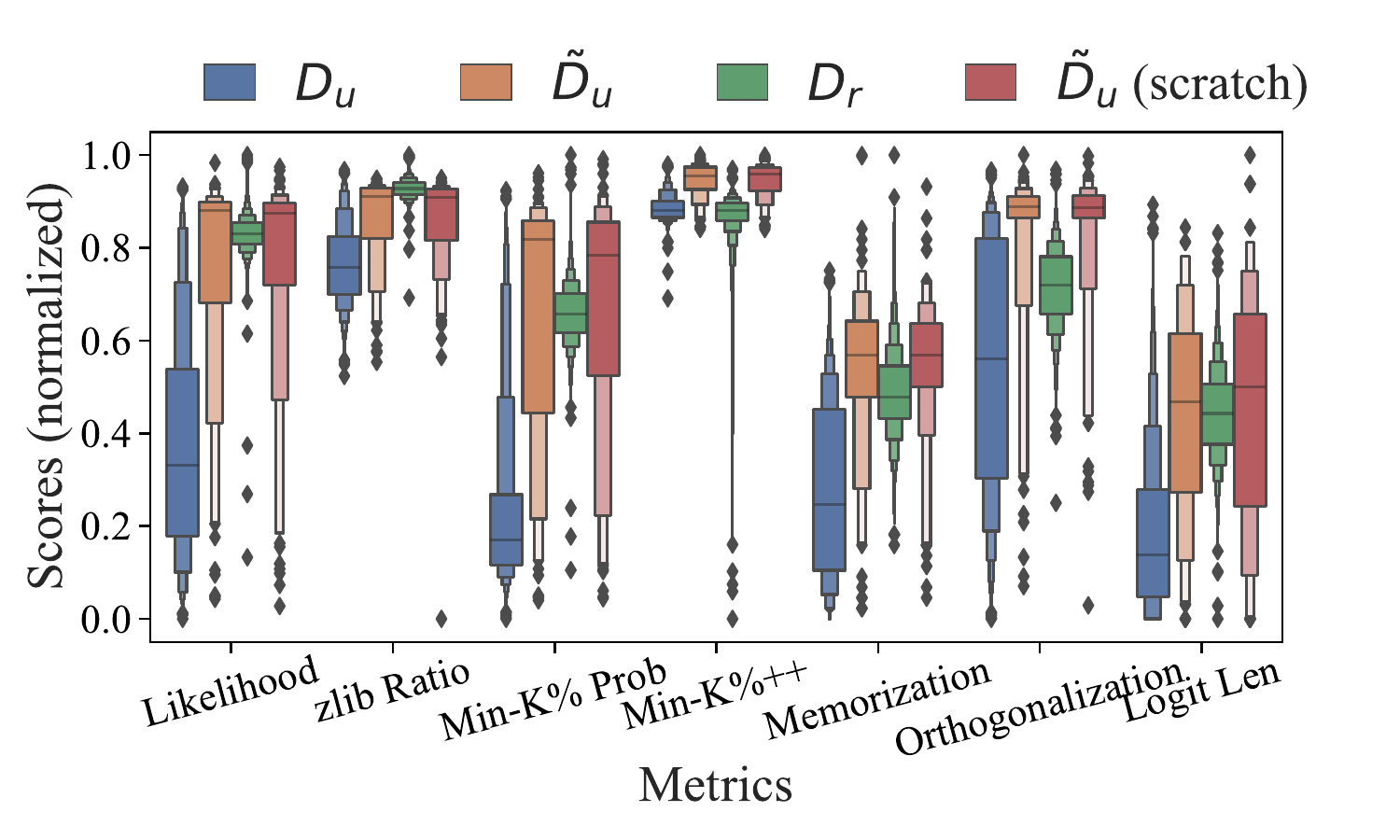}}
    \caption{\looseness=-1
    \textbf{Boxplots of metric scores of 2 surrogate unlearning datasets, $\tilde{D}_u$ and $\tilde{D}_u(\text{scratch})$.} The sentences in the former are generated by continuing writing from sentences in $D_u$, whereas the latter contains sentences written from scratch (\ie generating from an empty string). The metric scores of the unlearning dataset $D_u$ and retain dataset $D_r$ are also presented as reference. It can be seen that for most of the metrics, the difference between the boxes corresponding to the 2 surrogate unlearning datasets are negligible. Thus, we conclude that generating whether from a sentence in $D_u$ or an empty string does not significantly impacts the performance of \mini. 
    }    
    \label{fig:ablation_strategy}
\end{figure}

\vspace{1mm}
\noindent{\bf Generation Strategy.}  
As shown in lines 4–5 of Algorithm~\ref{alg:main}, \mini begins with sentences from $D_u$, generates continuations, and then enters the loop of maximizing separation and rephrasing. A natural question is whether starting from $D_u$ is necessary. To test this, we consider an alternative strategy where \mini generates sentences from scratch by prompting the model with an empty string. We denote the resulting dataset as $\tilde{D}_u \text{(scratch)}$, and plot its metric values in Figure~\ref{fig:ablation_strategy}, alongside those of $D_u$, $D_r$, and the $\tilde{D}_u$ created by continuing from $D_u$. The figure shows that $\tilde{D}_u \text{(scratch)}$ and $\tilde{D}_u$ yield very similar metric distributions across most metrics and both unlearning methods. This suggests that the choice between continuing from $D_u$ or generating from scratch has little effect, likely because the iterative rephrasing loop heavily modifies sentences after initialization. Nonetheless, we prefer the former strategy, since generation from scratch can lead to collisions; for some models, starting from an empty string can result in repeated outputs with high probability.  
\textit{Thus, while either generation works in principle, continuing from $D_u$ is more practical.}  

\begin{figure}[t!]
    \centering
    \subfloat[Negative Preference Optimization (NPO)]{
    \includegraphics[width=0.94\linewidth]{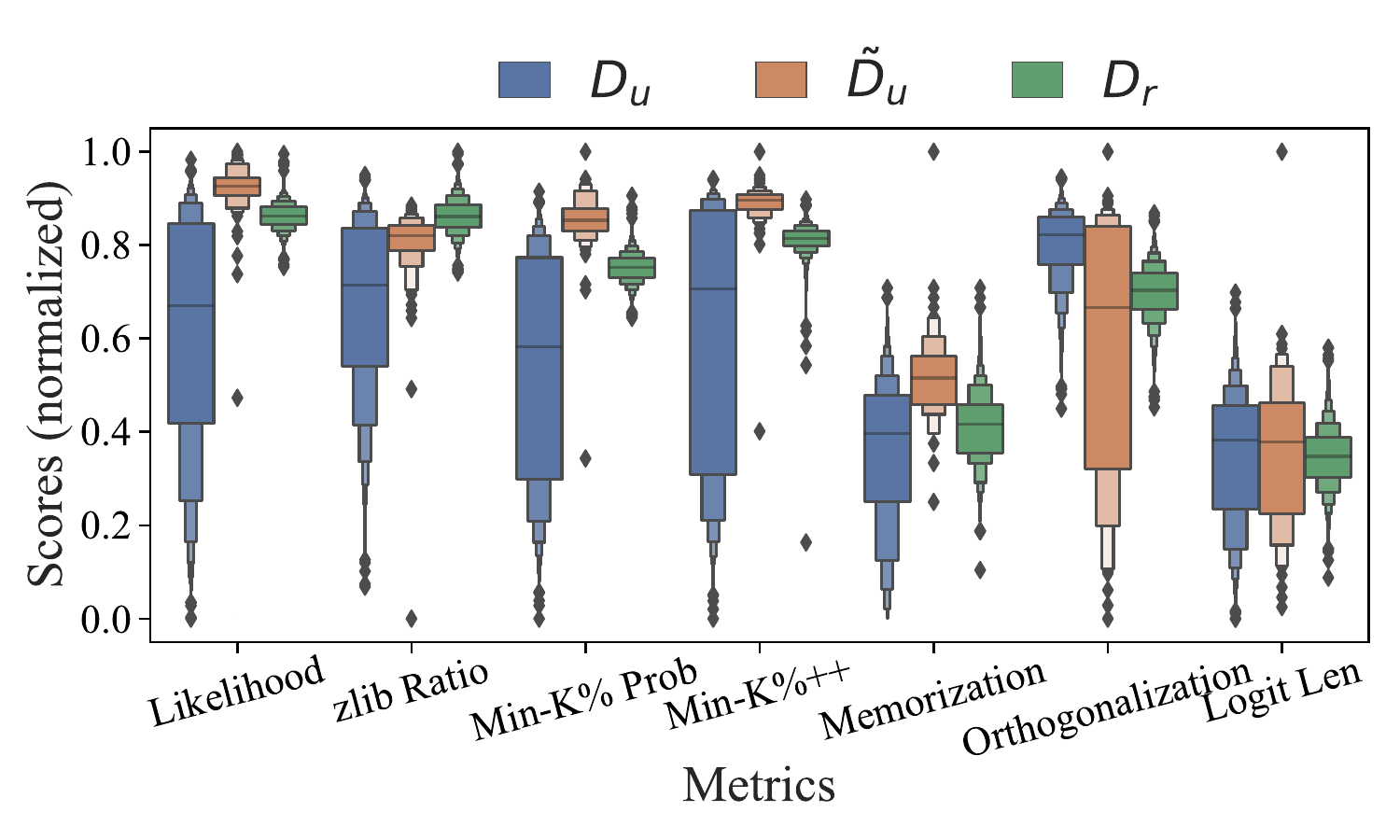}
    }
    
    \subfloat[Representation Misdirection for Unlearning (RMU)]{
    \includegraphics[width=0.94\linewidth]{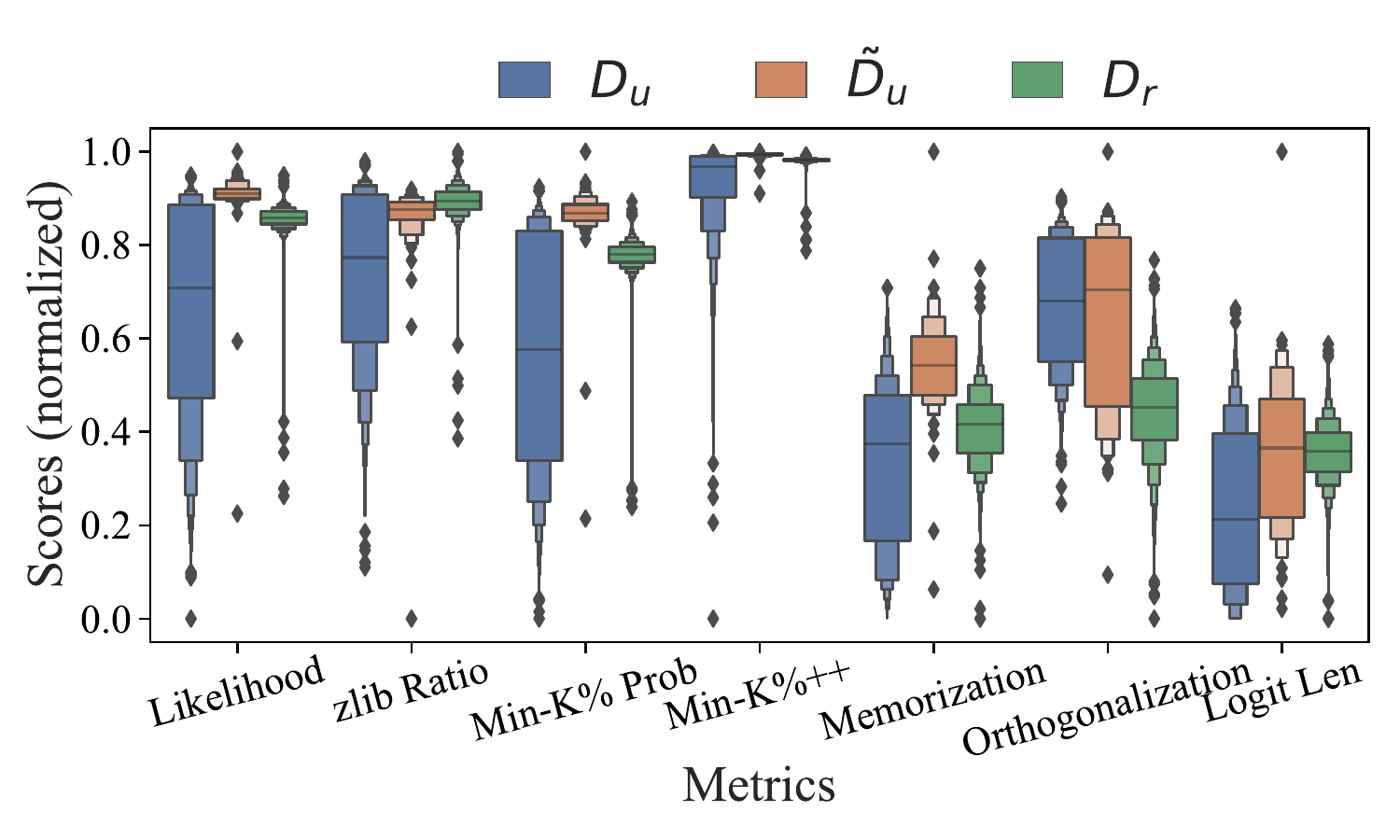}}
    \caption{
    \textbf{Boxplots for metric scores of $D_u$ (\textit{\novel}), $\tilde{D}_u$, and $D_r$ on a smaller model.} Unlike Figure~\ref{fig:hics_llama}, the model architecture used here is Microsoft's Phi-3-mini-4k model with only 3.8 billion parameters, which is significantly less than the other models studied in this work. The results are consistent with Figure~\ref{fig:hics_llama} for most of the metrics: the boxes representing $\tilde{D}_u$ are consistently different from the ones for $\tilde{D}_u$. However, this is not the case for metrics \textit{Orthogonalization} and \textit{Logit Len}. By comparing the metric scores of $D_u$ and $D_r$, we suspect this is because unlearning is not successful in the first place, \ie the metric is performing similarly on all training data (whether it is unlearned or used for retaining), so it is hard to create $\tilde{D}_u$ that is not vastly different from the training distribution and has different metric values at the same time.
    }    
    \label{fig:hics_phi}
\end{figure}

\vspace{1mm}
\looseness=-1
\noindent{\bf Model Size.}  
All models studied so far have around 7–8 billion parameters (see \S~\ref{sec:setup}). To assess the impact of model size, we test \mini on a smaller model: Phi-3-mini-4k-instruct~\cite{phi}, with only 3.8 billion parameters. Other than the architecture, all settings remain the same as in Figure~\ref{fig:hics_llama}: the model is fine-tuned with 12 fantasy novels and their fan fictions, and \novel is unlearned using both NPO and RMU. The metric values of $D_u$, $D_r$, and $\tilde{D}_u$ are shown in Figure~\ref{fig:hics_phi}. Two main observations arise: (1) unlearning appears less successful for the smaller model, as $D_u$ often overlaps with $D_r$ in metric distributions, despite lower mean values; and (2) \mini works as expected for most metrics, \ie $\tilde{D}_u$ achieves large standardized mean differences from $D_u$. However, exceptions occur for {Orthogonalization} and {Logit Lens}, the same metrics that also suggest unlearning itself was unsuccessful. This indicates that \mini requires the condition that the chosen metric differentiates $D_u$ and $D_r$ in the first place. Intuitively, not being able to do so means the metric performs similarly across the entire training distribution, where an extreme example would be a random metric that decides if a point is unlearned successfully with $50-50$ chance. It is impossible to create $\tilde{D}_u$  with a different metric score distribution from $D_u$ in this case. \textit{Therefore, the effectiveness of \mini depends on using metrics that can at least differentiate $D_u$ and $D_r$.}

\section{Discussion}
\label{sec:discussion}

\noindent{\bf How is the surrogate unlearning dataset related to the unlearning dataset?} 
{
As aforementioned in \S~\ref{subsec:p1}, \mini ensures the surrogate unlearning dataset is driven from the unlearning dataset by only including sentences with an increase in likelihood when the model is finetuned on the latter. To further confirm this, we analyze $D_u$ and $\tilde{D}_u$ by leveraging topic modeling techniques, under the same setting as in \S~\ref{subsec:main_res}. First, by clustering the sentences of $D_u$ into topics based on TF-IDF (term frequency–inverse document frequency), we observe $87\%$ sentences of $\tilde{D}_u$ falls into the 5-largest topics of $D_u$, suggesting similar use of words between the two datasets. We then move to an LLM-based topic modeling technique, BERTopic~\cite{grootendorst2022bertopic}. Here, we observe $63\%$ sentences of $\tilde{D}_u$ were classified as outliers, \ie not belonging to any topic clusters of $D_u$. This may be because that LLM-based topic modeling techniques often rely on embeddings of the underlying LLMs. Recall that \mini maximized embedding distances between the two datasets on $\theta$, so this might transfer to other models like the BERT model used in BERTopic. We then rerun BERTopic on both $D_u$ and the retain dataset $D_r$. Despite $D_r$ is much larger in size than $D_u$, we find $90\%$ sentences of $\tilde{D}_u$ falls into topic clusters dominated by sentences from $D_u$. This means \textit{$\tilde{D}_u$ is more relevant to $D_u$ than other subsets of the training dataset.}
}

}

\noindent{\bf Could unlearning methods be the cause of the issue?}  
One might argue that the key issue identified by \mini, \ie unlearning metrics give different scores to sentences in $D_u$ versus sentences learned from $D_u$, could stem from imperfect unlearning methods. 
For instance, when applied for copyright or safety, an unlearning method may falsely primarily target exact sentences in $D_u$, leaving semantically related sentences insufficiently unlearned. 
This could reproduce our observation: the existence of a dataset related to $D_u$ but with a different distribution of metric values. However, even if this is the case, an ideal metric for copyright- or safety-driven unlearning should report the worst-case performance across both $D_u$ and related data, rather than only the performance on $D_u$. Otherwise, the metric remains problematic, as it may overstate or understate the success of unlearning.
\textit{Thus, imperfect unlearning methods may explain the discrepancy, but they do not absolve the metrics of their responsibility to capture it.} 

\vspace{1mm}
\vspace{1mm}\noindent{\bf Could the surrogate unlearning dataset be leveraged during unlearning?}  
Since the surrogate datasets created by \mini consist of sentences that can be learned from $D_u$, a natural question is whether including them during unlearning would improve unlearning success. 
The answer depends on the size of $\tilde{D}_u$ and how well it represents the complete distribution of sentences containing knowledge from $D_u$. If a sufficiently large and representative dataset exists, unlearning it could improve both success and generalizability~\cite{liu2024machine}. 
However, \mini is designed to falsify metrics, not to generate representative distributions: one counterexample suffices to falsify a metric. As such, unlearning with $\tilde{D}_u$ is analogous to training on a validation set, making evaluation harder without truly improving unlearning\footnote{{This does not mean including surrogate unlearning datasets in the unlearning dataset can function as an adaptive attack against \mini. As specified in the threat model in \S~\ref{sec:threat}, the auditor's task is to only evaluate unlearning metrics, so it may use arbitrary unlearning datasets and models. In other words, an adversarial unlearning metric designed against \mini does not have access to the unlearning setting where it would be evaluated.}}. 
That said, the fine-tuned model $\theta_f$ used to construct $\tilde{D}_u$ may still provide useful signals, as it encodes more information about what can be learned from $D_u$ than a few sentences alone. Indeed, similar approaches have been explored in task-vector-based unlearning~\cite{ilharco2023editing}. 
Yet relying on a single fine-tuned model is likely insufficient and may lead to failure cases~\cite{shi2024musemachineunlearningsixway}. 
A more promising direction may be to obtain a distribution of models fine-tuned on $D_u$ and use them collectively to guide unlearning.
\textit{Therefore, while $\tilde{D}_u$ itself may not be suitable for unlearning, the fine-tuned models used to construct it could inform better strategies.}

\vspace{1mm}\noindent{\bf When do unlearning metrics need to be tested against the surrogate dataset?}  
It is worth noting that the surrogate dataset does not overlap with the unlearning dataset. Therefore, by design, if the goal of unlearning is to remove certain exact data points, the unlearning metrics are not expected to be test against a surrogate dataset---for example, when unlearning is motivated by privacy concerns regarding personally identifiable information. Note that this is not claiming that existing unlearning metrics are perfect in that scenario. In fact, it has been pointed out that the evaluation of privacy-oriented unlearning is also flawed~\cite{hayes2025inexact}. However, the drawbacks of metrics are different when the goals of unlearning are different, and so are how we should test them---\textit{surrogate datasets are only needed when unlearning beyond sentence-level, \eg when motivated by copyright or safety.}

\vspace{1mm}\noindent{\bf What would be the right direction toward best practices for evaluating unlearning?}  
Our results highlight the limitations of existing metrics and suggest properties for improved evaluation. First, metrics must align with the true incentive of unlearning. Measures derived from memorization or membership inference are designed to detect whether a model has memorized or trained on an \emph{exact} data point; naturally, they cannot be expected to generalize to semantically related but unseen data, such as fan fiction of copyrighted content. Second, if the goal is to remove broader knowledge rather than specific data points, the evaluation must also address generalization. This implies two considerations: 
\begin{enumerate}
\item Evaluation should move beyond individual data points toward sets of examples that all entail the same piece of knowledge, and measure whether that knowledge has been unlearned.
\item Analogous to standard model evaluation, an external test dataset is necessary. 
\end{enumerate}
What constitutes a good test dataset remains an open question. The surrogate datasets produced by \mini rely on embedding separation, but this is unlikely to be the only requirement. This also highlights a limitation of \mini itself: a large mean difference between $D_u$ and $\tilde{D}_u$ indicates that a metric is not ideal, but the reverse does not imply that the metric is sound. In other words, \mini is designed to falsify unlearning metrics, not to prove their correctness.  
\textit{Hence, best practices for evaluating unlearning should move beyond data-point membership, embrace generalization, and develop robust test datasets.}  {That said, we also recognize the possibility that this work may lead to future findings suggesting it is impossible to faithfully evaluate knowledge-level unlearning, in which case unlearning may not be the best tool for \eg copyright or safety protection. We hope highlighting this possibility can encourage researchers to revisit fundamental questions on unlearning, such as incentive-dependent best practices for evaluating it, rather than focusing on proposing ``better'' LLM unlearning methods based on metrics that may not be faithful.}

\section{Conclusion}  

We study the problem of evaluating unlearning in generative settings, when the goal is to remove knowledge beyond the data-point level from trained models. We argue that current metrics, which are typically applied only to the unlearning set $D_u$ and thus tailored to $D_u$, create a false sense of success. In particular, many metrics suggest that data containing knowledge learned from the unlearning dataset has inconsistent unlearning success with respect to the unlearning dataset. To make this gap explicit, we introduced the notion of surrogate unlearning datasets $\tilde{D}_u$, which are derived from (thus semantically related to) but disjoint from $D_u$, and presented \name, a lightweight algorithm for constructing them automatically.

\looseness=-1
Our empirical results across 18 settings (3 LLMs, 3 datasets, and 2 unlearning algorithms) show that commonly-used unlearning metrics are highly inconsistent between $D_u$ and $\tilde{D}_u$, with $\tilde{D}_u$ frequently appearing less successfully unlearned, according to these metrics. These findings highlight that existing metrics are overfitting to $D_u$ and systematically fail to evaluate unlearning success based on the removal of higher-order knowledge, which is more tied to the incentive of knowledge-based (\eg copyrighted contents removal or unsafe outputs mitigation) unlearning. 
Looking ahead, we hope our work can encourage future unlearning metrics to account for generalization of unlearning (\eg by the use of a pseudo-validation dataset) and tailor to the goal of unlearning: in particular, when unlearning is motivated by knowledge removal, the metrics can be designed to move beyond exact-data deletion and toward capturing the true extent of knowledge removal in LLMs.

\newpage

\bibliographystyle{plain}
\bibliography{main}

@misc{shi2024musemachineunlearningsixway,
      title={MUSE: Machine Unlearning Six-Way Evaluation for Language Models}, 
      author={Weijia Shi and Jaechan Lee and Yangsibo Huang and Sadhika Malladi and Jieyu Zhao and Ari Holtzman and Daogao Liu and Luke Zettlemoyer and Noah A. Smith and Chiyuan Zhang},
      year={2024},
      eprint={2407.06460},
      archivePrefix={arXiv},
      primaryClass={cs.CL},
      url={https://arxiv.org/abs/2407.06460}, 
}

@article{du2017learning,
  title={Learning to ask: Neural question generation for reading comprehension},
  author={Du, Xinya and Shao, Junru and Cardie, Claire},
  journal={arXiv preprint arXiv:1705.00106},
  year={2017}
}

@misc{zhang2024negativepreferenceoptimizationcatastrophic,
      title={Negative Preference Optimization: From Catastrophic Collapse to Effective Unlearning}, 
      author={Ruiqi Zhang and Licong Lin and Yu Bai and Song Mei},
      year={2024},
      eprint={2404.05868},
      archivePrefix={arXiv},
      primaryClass={cs.LG},
      url={https://arxiv.org/abs/2404.05868}, 
}

@misc{jang2022knowledgeunlearningmitigatingprivacy,
      title={Knowledge Unlearning for Mitigating Privacy Risks in Language Models}, 
      author={Joel Jang and Dongkeun Yoon and Sohee Yang and Sungmin Cha and Moontae Lee and Lajanugen Logeswaran and Minjoon Seo},
      year={2022},
      eprint={2210.01504},
      archivePrefix={arXiv},
      primaryClass={cs.CL},
      url={https://arxiv.org/abs/2210.01504}, 
}

@inproceedings{
    rafailov2023direct,
    title={Direct Preference Optimization: Your Language Model is Secretly a Reward Model},
    author={Rafael Rafailov and Archit Sharma and Eric Mitchell and Christopher D Manning and Stefano Ermon and Chelsea Finn},
    booktitle={Thirty-seventh Conference on Neural Information Processing Systems},
    year={2023},
    url={https://openreview.net/forum?id=HPuSIXJaa9}
}

@inproceedings{
    huang2024unified,
    title={Unified Gradient-Based Machine Unlearning with Remain Geometry Enhancement},
    author={Zhehao Huang and Xinwen Cheng and JingHao Zheng and Haoran Wang and Zhengbao He and Tao Li and Xiaolin Huang},
    booktitle={The Thirty-eighth Annual Conference on Neural Information Processing Systems},
    year={2024},
    url={https://openreview.net/forum?id=dheDf5EpBT}
}

@misc{fan2025simplicityprevailsrethinkingnegative,
      title={Simplicity Prevails: Rethinking Negative Preference Optimization for LLM Unlearning}, 
      author={Chongyu Fan and Jiancheng Liu and Licong Lin and Jinghan Jia and Ruiqi Zhang and Song Mei and Sijia Liu},
      year={2025},
      eprint={2410.07163},
      archivePrefix={arXiv},
      primaryClass={cs.CL},
      url={https://arxiv.org/abs/2410.07163}, 
}

@misc{wang2024llmunlearninglossadjustment,
      title={LLM Unlearning via Loss Adjustment with Only Forget Data}, 
      author={Yaxuan Wang and Jiaheng Wei and Chris Yuhao Liu and Jinlong Pang and Quan Liu and Ankit Parag Shah and Yujia Bao and Yang Liu and Wei Wei},
      year={2024},
      eprint={2410.11143},
      archivePrefix={arXiv},
      primaryClass={cs.CL},
      url={https://arxiv.org/abs/2410.11143}, 
}

@inproceedings{kassem-etal-2023-preserving,
    title = "Preserving Privacy Through Dememorization: An Unlearning Technique For Mitigating Memorization Risks In Language Models",
    author = "Kassem, Aly  and
      Mahmoud, Omar  and
      Saad, Sherif",
    editor = "Bouamor, Houda  and
      Pino, Juan  and
      Bali, Kalika",
    booktitle = "Proceedings of the 2023 Conference on Empirical Methods in Natural Language Processing",
    month = dec,
    year = "2023",
    address = "Singapore",
    publisher = "Association for Computational Linguistics",
    url = "https://aclanthology.org/2023.emnlp-main.265/",
    doi = "10.18653/v1/2023.emnlp-main.265",
    pages = "4360--4379"
}

@misc{li2024wmdpbenchmarkmeasuringreducing,
      title={The WMDP Benchmark: Measuring and Reducing Malicious Use With Unlearning}, 
      author={Nathaniel Li and Alexander Pan and Anjali Gopal and others},
      year={2024},
      eprint={2403.03218},
      archivePrefix={arXiv},
      primaryClass={cs.LG},
      url={https://arxiv.org/abs/2403.03218}, 
}

@misc{shen2025lunarllmunlearningneural,
      title={LUNAR: LLM Unlearning via Neural Activation Redirection}, 
      author={William F. Shen and Xinchi Qiu and Meghdad Kurmanji and Alex Iacob and Lorenzo Sani and Yihong Chen and Nicola Cancedda and Nicholas D. Lane},
      year={2025},
      eprint={2502.07218},
      archivePrefix={arXiv},
      primaryClass={cs.LG},
      url={https://arxiv.org/abs/2502.07218}, 
}

@misc{chowdhury2025scalableexactmachineunlearning,
      title={Towards Scalable Exact Machine Unlearning Using Parameter-Efficient Fine-Tuning}, 
      author={Somnath Basu Roy Chowdhury and Krzysztof Choromanski and Arijit Sehanobish and Avinava Dubey and Snigdha Chaturvedi},
      year={2025},
      eprint={2406.16257},
      archivePrefix={arXiv},
      primaryClass={cs.LG},
      url={https://arxiv.org/abs/2406.16257}, 
}

@misc{maini2024tofutaskfictitiousunlearning,
      title={TOFU: A Task of Fictitious Unlearning for LLMs}, 
      author={Pratyush Maini and Zhili Feng and Avi Schwarzschild and Zachary C. Lipton and J. Zico Kolter},
      year={2024},
      eprint={2401.06121},
      archivePrefix={arXiv},
      primaryClass={cs.LG},
      url={https://arxiv.org/abs/2401.06121}, 
}

@misc{eldan2023whosharrypotterapproximate,
      title={Who's Harry Potter? Approximate Unlearning in LLMs}, 
      author={Ronen Eldan and Mark Russinovich},
      year={2023},
      eprint={2310.02238},
      archivePrefix={arXiv},
      primaryClass={cs.CL},
      url={https://arxiv.org/abs/2310.02238}, 
}

@article{ucki2025an,
    title={An Adversarial Perspective on Machine Unlearning for {AI} Safety},
    author={Jakub {\L}ucki and Boyi Wei and Yangsibo Huang and Peter Henderson and Florian Tram{\`e}r and Javier Rando},
    journal={Transactions on Machine Learning Research},
    issn={2835-8856},
    year={2025},
    url={https://openreview.net/forum?id=J5IRyTKZ9s},
    note={}
}

@INPROCEEDINGS{Cao2015,
  author={Cao, Yinzhi and Yang, Junfeng},
  booktitle={2015 IEEE Symposium on Security and Privacy}, 
  title={Towards Making Systems Forget with Machine Unlearning}, 
  year={2015},
  volume={},
  number={},
  pages={463-480},
  doi={10.1109/SP.2015.35}}

@misc{Bourtoule2021,
      title={Machine Unlearning}, 
      author={Lucas Bourtoule and Varun Chandrasekaran and Christopher A. Choquette-Choo and Hengrui Jia and Adelin Travers and Baiwu Zhang and David Lie and Nicolas Papernot},
      year={2020},
      eprint={1912.03817},
      archivePrefix={arXiv},
      primaryClass={cs.CR},
      url={https://arxiv.org/abs/1912.03817}, 
}

@book{Voigt2017,
    author    = {Voigt, Paul and von dem Bussche, Axel},
    title     = {The EU General Data Protection Regulation (GDPR): A Practical Guide},
    year      = {2017},
    publisher = {Springer International Publishing},
    address   = {Cham},
    edition   = {1st},
    isbn      = {978-3-319-57958-0},
    doi       = {10.1007/978-3-319-57959-7}
}

@misc{lu2022quarkcontrollabletextgeneration,
      title={Quark: Controllable Text Generation with Reinforced Unlearning}, 
      author={Ximing Lu and Sean Welleck and Jack Hessel and Liwei Jiang and Lianhui Qin and Peter West and Prithviraj Ammanabrolu and Yejin Choi},
      year={2022},
      eprint={2205.13636},
      archivePrefix={arXiv},
      primaryClass={cs.CL},
      url={https://arxiv.org/abs/2205.13636}, 
}

@misc{grattafiori2024llama3herdmodels,
      title={The Llama 3 Herd of Models}, 
      author={Aaron Grattafiori and Abhimanyu Dubey and Abhinav Jauhri and others},
      year={2024},
      eprint={2407.21783},
      archivePrefix={arXiv},
      primaryClass={cs.AI},
      url={https://arxiv.org/abs/2407.21783}, 
}

@misc{yang2024qwen2technicalreport,
      title={Qwen2 Technical Report}, 
      author={An Yang and Baosong Yang and Binyuan Hui and Bo Zheng and Bowen Yu and Chang Zhou and Chengpeng Li and Chengyuan Lia and Dayiheng Liu and Fei Huang and Guanting Dong and Haoran Wei and Huan Lin and Jialong Tang and Jialin Wang and Jian Yang and Jianhong Tu and Jianwei Zhang and Jianxin Ma and Jianxin Yang and Jin Xu and Jingren Zhou and Jinze Bai and Jinzheng He and Junyang Lin and Kai Dang and Keming Lu and Keqin Chen and Kexin Yang and Mei Li and Mingfeng Xue and Na Ni and Pei Zhang and Peng Wang and Ru Peng and Rui Men and Ruize Gao and Runji Lin and Shijie Wang and Shuai Bai and Sinan Tan and Tianhang Zhu and Tianhao Li and Tianyu Liu and Wenbin Ge and Xiaodong Deng and Xiaohuan Zhou and Xingzhang Ren and Xinyu Zhang and Xipin Wei and Xuancheng Ren and Xuejing Liu and Yang Fan and Yang Yao and Yichang Zhang and Yu Wan and Yunfei Chu and Yuqiong Liu and Zeyu Cui and Zhenru Zhang and Zhifang Guo and Zhihao Fan},
      year={2024},
      eprint={2407.10671},
      archivePrefix={arXiv},
      primaryClass={cs.CL},
      url={https://arxiv.org/abs/2407.10671}, 
}

@misc{tunstall2023zephyrdirectdistillationlm,
      title={Zephyr: Direct Distillation of LM Alignment}, 
      author={Lewis Tunstall and Edward Beeching and Nathan Lambert and Nazneen Rajani and Kashif Rasul and Younes Belkada and Shengyi Huang and Leandro von Werra and Clémentine Fourrier and Nathan Habib and Nathan Sarrazin and Omar Sanseviero and Alexander M. Rush and Thomas Wolf},
      year={2023},
      eprint={2310.16944},
      archivePrefix={arXiv},
      primaryClass={cs.LG},
      url={https://arxiv.org/abs/2310.16944}, 
}

@misc{merity2016pointer,
      title={Pointer Sentinel Mixture Models},
      author={Stephen Merity and Caiming Xiong and James Bradbury and Richard Socher},
      year={2016},
      eprint={1609.07843},
      archivePrefix={arXiv},
      primaryClass={cs.CL}
}

@article{DBLP:journals/corr/abs-1903-04561,
  author    = {Daniel Borkan and
               Lucas Dixon and
               Jeffrey Sorensen and
               Nithum Thain and
               Lucy Vasserman},
  title     = {Nuanced Metrics for Measuring Unintended Bias with Real Data for Text
               Classification},
  journal   = {CoRR},
  volume    = {abs/1903.04561},
  year      = {2019},
  url       = {http://arxiv.org/abs/1903.04561},
  archivePrefix = {arXiv},
  eprint    = {1903.04561},
  bibsource = {dblp computer science bibliography, https://dblp.org}
}

@article{huang2024demystifying,
  title={Demystifying verbatim memorization in large language models},
  author={Huang, Jing and Yang, Diyi and Potts, Christopher},
  journal={arXiv preprint arXiv:2407.17817},
  year={2024}
}

@article{dou2024avoiding,
  title={Avoiding Copyright Infringement via Large Language Model Unlearning},
  author={Dou, Guangyao and Liu, Zheyuan and Lyu, Qing and Ding, Kaize and Wong, Eric},
  journal={arXiv preprint arXiv:2406.10952},
  year={2024}
}

@article{barbulescu2024each,
  title={To each (textual sequence) its own: Improving memorized-data unlearning in large language models},
  author={Barbulescu, George-Octavian and Triantafillou, Peter},
  journal={arXiv preprint arXiv:2405.03097},
  year={2024}
}

@article{liu2025rethinking,
  title={Rethinking machine unlearning for large language models},
  author={Liu, Sijia and Yao, Yuanshun and Jia, Jinghan and Casper, Stephen and Baracaldo, Nathalie and Hase, Peter and Yao, Yuguang and Liu, Chris Yuhao and Xu, Xiaojun and Li, Hang and others},
  journal={Nature Machine Intelligence},
  pages={1--14},
  year={2025},
  publisher={Nature Publishing Group UK London}
}

@article{jia2024soul,
  title={Soul: Unlocking the power of second-order optimization for llm unlearning},
  author={Jia, Jinghan and Zhang, Yihua and Zhang, Yimeng and Liu, Jiancheng and Runwal, Bharat and Diffenderfer, James and Kailkhura, Bhavya and Liu, Sijia},
  journal={arXiv preprint arXiv:2404.18239},
  year={2024}
}

@article{ji2024reversing,
  title={Reversing the forget-retain objectives: An efficient llm unlearning framework from logit difference},
  author={Ji, Jiabao and Liu, Yujian and Zhang, Yang and Liu, Gaowen and Kompella, Ramana R and Liu, Sijia and Chang, Shiyu},
  journal={Advances in Neural Information Processing Systems},
  volume={37},
  pages={12581--12611},
  year={2024}
}

@article{yao2024large,
  title={Large language model unlearning},
  author={Yao, Yuanshun and Xu, Xiaojun and Liu, Yang},
  journal={Advances in Neural Information Processing Systems},
  volume={37},
  pages={105425--105475},
  year={2024}
}

@article{wang2025rethinking,
  title={Rethinking llm unlearning objectives: A gradient perspective and go beyond},
  author={Wang, Qizhou and Zhou, Jin Peng and Zhou, Zhanke and Shin, Saebyeol and Han, Bo and Weinberger, Kilian Q},
  journal={arXiv preprint arXiv:2502.19301},
  year={2025}
}

@inproceedings{wang2025towards,
  title={Towards effective evaluations and comparisons for LLM unlearning methods},
  author={Wang, Qizhou and Han, Bo and Yang, Puning and Zhu, Jianing and Liu, Tongliang and Sugiyama, Masashi},
  booktitle={The Thirteenth International Conference on Learning Representations},
  year={2025}
}

@article{wang2025uipe,
  title={Uipe: Enhancing llm unlearning by removing knowledge related to forgetting targets},
  author={Wang, Wenyu and Zhang, Mengqi and Ye, Xiaotian and Ren, Zhaochun and Chen, Zhumin and Ren, Pengjie},
  journal={arXiv preprint arXiv:2503.04693},
  year={2025}
}

@article{ren2025general,
  title={A General Framework to Enhance Fine-tuning-based LLM Unlearning},
  author={Ren, Jie and Dai, Zhenwei and Tang, Xianfeng and Liu, Hui and Zeng, Jingying and Li, Zhen and Goutam, Rahul and Wang, Suhang and Xing, Yue and He, Qi},
  journal={arXiv preprint arXiv:2502.17823},
  year={2025}
}

@misc{huutien2025effectssteeringlatentrepresentation,
      title={On Effects of Steering Latent Representation for Large Language Model Unlearning}, 
      author={Dang Huu-Tien and Trung-Tin Pham and Hoang Thanh-Tung and Naoya Inoue},
      year={2025},
      eprint={2408.06223},
      archivePrefix={arXiv},
      primaryClass={cs.CL},
      url={https://arxiv.org/abs/2408.06223}, 
}

@INPROCEEDINGS {hayes2025inexact,
author = { Hayes, Jamie and Shumailov, Ilia and Triantafillou, Eleni and Khalifa, Amr and Papernot, Nicolas },
booktitle = { 2025 IEEE Conference on Secure and Trustworthy Machine Learning (SaTML) },
title = {{ Inexact Unlearning Needs More Careful Evaluations to Avoid a False Sense of Privacy }},
year = {2025},
volume = {},
ISSN = {},
pages = {497-519},
doi = {10.1109/SaTML64287.2025.00034},
url = {https://doi.ieeecomputersociety.org/10.1109/SaTML64287.2025.00034},
publisher = {IEEE Computer Society},
address = {Los Alamitos, CA, USA},
month =apr}

@inproceedings{
schwarzschild2024rethinking,
title={Rethinking {LLM} Memorization through the Lens of Adversarial Compression},
author={Avi Schwarzschild and Zhili Feng and Pratyush Maini and Zachary Chase Lipton and J Zico Kolter},
booktitle={The Thirty-eighth Annual Conference on Neural Information Processing Systems},
year={2024},
url={https://openreview.net/forum?id=KFmRMvzAZy}
}

@inproceedings{
shi2024detecting,
title={Detecting Pretraining Data from Large Language Models},
author={Weijia Shi and Anirudh Ajith and Mengzhou Xia and Yangsibo Huang and Daogao Liu and Terra Blevins and Danqi Chen and Luke Zettlemoyer},
booktitle={The Twelfth International Conference on Learning Representations},
year={2024},
url={https://openreview.net/forum?id=zWqr3MQuNs}
}

@inproceedings{
zhang2025mink,
title={Min-K\%++: Improved Baseline for Pre-Training Data Detection from Large Language Models},
author={Jingyang Zhang and Jingwei Sun and Eric Yeats and Yang Ouyang and Martin Kuo and Jianyi Zhang and Hao Frank Yang and Hai Li},
booktitle={The Thirteenth International Conference on Learning Representations},
year={2025},
url={https://openreview.net/forum?id=ZGkfoufDaU}
}

@misc{patil2023sensitiveinformationdeletedllms,
      title={Can Sensitive Information Be Deleted From LLMs? Objectives for Defending Against Extraction Attacks}, 
      author={Vaidehi Patil and Peter Hase and Mohit Bansal},
      year={2023},
      eprint={2309.17410},
      archivePrefix={arXiv},
      primaryClass={cs.CL},
      url={https://arxiv.org/abs/2309.17410}, 
}

@article{cooper2024machine,
  title={Machine Unlearning Doesn't Do What You Think: Lessons for Generative AI Policy, Research, and Practice},
  author={Cooper, A Feder and Choquette-Choo, Christopher A and Bogen, Miranda and Jagielski, Matthew and Filippova, Katja and Liu, Ken Ziyu and Chouldechova, Alexandra and Hayes, Jamie and Huang, Yangsibo and Mireshghallah, Niloofar and others},
  journal={arXiv preprint arXiv:2412.06966},
  year={2024}
}

@article{liu2024machine,
  title={Machine unlearning in generative ai: A survey},
  author={Liu, Zheyuan and Dou, Guangyao and Tan, Zhaoxuan and Tian, Yijun and Jiang, Meng},
  journal={arXiv preprint arXiv:2407.20516},
  year={2024}
}

@misc{zou2023universal,
      title={Universal and Transferable Adversarial Attacks on Aligned Language Models}, 
      author={Andy Zou and Zifan Wang and J. Zico Kolter and Matt Fredrikson},
      year={2023},
      eprint={2307.15043},
      archivePrefix={arXiv},
      primaryClass={cs.CL}
}

@ARTICLE{phi,
       author = {{Abdin}, Marah and {Aneja}, Jyoti and {Awadalla}, Hany and {Awadallah}, Ahmed and {Awan}, Ammar Ahmad and {Bach}, Nguyen and {Bahree}, Amit and {Bakhtiari}, Arash and {Bao}, Jianmin and {Behl}, Harkirat and {Benhaim}, Alon and {Bilenko}, Misha and {Bjorck}, Johan and {Bubeck}, S{\'e}bastien and {Cai}, Martin and {Cai}, Qin and {Chaudhary}, Vishrav and {Chen}, Dong and {Chen}, Dongdong and {Chen}, Weizhu and {Chen}, Yen-Chun and {Chen}, Yi-Ling and {Cheng}, Hao and {Chopra}, Parul and {Dai}, Xiyang and {Dixon}, Matthew and {Eldan}, Ronen and {Fragoso}, Victor and {Gao}, Jianfeng and {Gao}, Mei and {Gao}, Min and {Garg}, Amit and {Del Giorno}, Allie and {Goswami}, Abhishek and {Gunasekar}, Suriya and {Haider}, Emman and {Hao}, Junheng and {Hewett}, Russell J. and {Hu}, Wenxiang and {Huynh}, Jamie and {Iter}, Dan and {Ade Jacobs}, Sam and {Javaheripi}, Mojan and {Jin}, Xin and {Karampatziakis}, Nikos and {Kauffmann}, Piero and {Khademi}, Mahoud and {Kim}, Dongwoo and {Kim}, Young Jin and {Kurilenko}, Lev and {Lee}, James R. and {Lee}, Yin Tat and {Li}, Yuanzhi and {Li}, Yunsheng and {Liang}, Chen and {Liden}, Lars and {Lin}, Xihui and {Lin}, Zeqi and {Liu}, Ce and {Liu}, Liyuan and {Liu}, Mengchen and {Liu}, Weishung and {Liu}, Xiaodong and {Luo}, Chong and {Madan}, Piyush and {Mahmoudzadeh}, Ali and {Majercak}, David and {Mazzola}, Matt and {C{\'e}sar Teodoro Mendes}, Caio and {Mitra}, Arindam and {Modi}, Hardik and {Nguyen}, Anh and {Norick}, Brandon and {Patra}, Barun and {Perez-Becker}, Daniel and {Portet}, Thomas and {Pryzant}, Reid and {Qin}, Heyang and {Radmilac}, Marko and {Ren}, Liliang and {de Rosa}, Gustavo and {Rosset}, Corby and {Roy}, Sambudha and {Ruwase}, Olatunji and {Saarikivi}, Olli and {Saied}, Amin and {Salim}, Adil and {Santacroce}, Michael and {Shah}, Shital and {Shang}, Ning and {Sharma}, Hiteshi and {Shen}, Yelong and {Shukla}, Swadheen and {Song}, Xia and {Tanaka}, Masahiro and {Tupini}, Andrea and {Vaddamanu}, Praneetha and {Wang}, Chunyu and {Wang}, Guanhua and {Wang}, Lijuan and {Wang}, Shuohang and {Wang}, Xin and {Wang}, Yu and {Ward}, Rachel and {Wen}, Wen and {Witte}, Philipp and {Wu}, Haiping and {Wu}, Xiaoxia and {Wyatt}, Michael and {Xiao}, Bin and {Xu}, Can and {Xu}, Jiahang and {Xu}, Weijian and {Xue}, Jilong and {Yadav}, Sonali and {Yang}, Fan and {Yang}, Jianwei and {Yang}, Yifan and {Yang}, Ziyi and {Yu}, Donghan and {Yuan}, Lu and {Zhang}, Chenruidong and {Zhang}, Cyril and {Zhang}, Jianwen and {Lyna Zhang}, Li and {Zhang}, Yi and {Zhang}, Yue and {Zhang}, Yunan and {Zhou}, Xiren},
        title = "{Phi-3 Technical Report: A Highly Capable Language Model Locally on Your Phone}",
      journal = {arXiv e-prints},
         year = 2024,
        month = apr,
          eid = {arXiv:2404.14219},
        pages = {arXiv:2404.14219},
          doi = {10.48550/arXiv.2404.14219},
archivePrefix = {arXiv},
       eprint = {2404.14219},
 primaryClass = {cs.CL},
       adsurl = {https://ui.adsabs.harvard.edu/abs/2024arXiv240414219A}
}

@inproceedings{
ilharco2023editing,
title={Editing models with task arithmetic},
author={Gabriel Ilharco and Marco Tulio Ribeiro and Mitchell Wortsman and Ludwig Schmidt and Hannaneh Hajishirzi and Ali Farhadi},
booktitle={The Eleventh International Conference on Learning Representations },
year={2023},
url={https://openreview.net/forum?id=6t0Kwf8-jrj}
}

@misc{zlib,
  author = {Gailly, Jean-loup and Adler, Mark},
  title = {{zlib} Compression Library},
  url = {http://zlib.net/}
}

@inproceedings {Carlini2021Extracting,
author = {Nicholas Carlini and Florian Tram{\`e}r and Eric Wallace and Matthew Jagielski and Ariel Herbert-Voss and Katherine Lee and Adam Roberts and Tom Brown and Dawn Song and {\'U}lfar Erlingsson and Alina Oprea and Colin Raffel},
title = {Extracting Training Data from Large Language Models},
booktitle = {30th USENIX Security Symposium (USENIX Security 21)},
year = {2021},
isbn = {978-1-939133-24-3},
pages = {2633--2650},
url = {https://www.usenix.org/conference/usenixsecurity21/presentation/carlini-extracting},
publisher = {USENIX Association},
month = aug
}

@inproceedings{
nasr2025scalable,
title={Scalable Extraction of Training Data from Aligned, Production Language Models},
author={Milad Nasr and Javier Rando and Nicholas Carlini and Jonathan Hayase and Matthew Jagielski and A. Feder Cooper and Daphne Ippolito and Christopher A. Choquette-Choo and Florian Tram{\`e}r and Katherine Lee},
booktitle={The Thirteenth International Conference on Learning Representations},
year={2025},
url={https://openreview.net/forum?id=vjel3nWP2a}
}

@inproceedings{karamolegkou2023copyright,
    title = "Copyright Violations and Large Language Models",
    author = "Karamolegkou, Antonia  and
      Li, Jiaang  and
      Zhou, Li  and
      S{\o}gaard, Anders",
    editor = "Bouamor, Houda  and
      Pino, Juan  and
      Bali, Kalika",
    booktitle = "Proceedings of the 2023 Conference on Empirical Methods in Natural Language Processing",
    month = dec,
    year = "2023",
    address = "Singapore",
    publisher = "Association for Computational Linguistics",
    url = "https://aclanthology.org/2023.emnlp-main.458/",
    doi = "10.18653/v1/2023.emnlp-main.458",
    pages = "7403--7412"
}

@inproceedings{wen2023unveiling,
    title = "Unveiling the Implicit Toxicity in Large Language Models",
    author = "Wen, Jiaxin  and
      Ke, Pei  and
      Sun, Hao  and
      Zhang, Zhexin  and
      Li, Chengfei  and
      Bai, Jinfeng  and
      Huang, Minlie",
    editor = "Bouamor, Houda  and
      Pino, Juan  and
      Bali, Kalika",
    booktitle = "Proceedings of the 2023 Conference on Empirical Methods in Natural Language Processing",
    month = dec,
    year = "2023",
    address = "Singapore",
    publisher = "Association for Computational Linguistics",
    url = "https://aclanthology.org/2023.emnlp-main.84/",
    doi = "10.18653/v1/2023.emnlp-main.84",
    pages = "1322--1338"
}

@inproceedings{thaker2025position,
  author       = {Pratiksha Thaker and
                  Shengyuan Hu and
                  Neil Kale and
                  Yash Maurya and
                  Zhiwei Steven Wu and
                  Virginia Smith},
  title        = {Position: {LLM} Unlearning Benchmarks are Weak Measures of Progress},
  booktitle    = {{IEEE} Conference on Secure and Trustworthy Machine Learning, SaTML
                  2025, Copenhagen, Denmark, April 9-11, 2025},
  pages        = {520--533},
  publisher    = {{IEEE}},
  year         = {2025},
  url          = {https://doi.org/10.1109/SaTML64287.2025.00035},
  doi          = {10.1109/SATML64287.2025.00035},
  bibsource    = {dblp computer science bibliography, https://dblp.org}
}

@inproceedings{
gupta2021adaptive,
title={Adaptive Machine Unlearning},
author={Varun Gupta and Christopher Jung and Seth Neel and Aaron Roth and Saeed Sharifi -Malvajerdi and Christopher Waites},
booktitle={Advances in Neural Information Processing Systems},
editor={A. Beygelzimer and Y. Dauphin and P. Liang and J. Wortman Vaughan},
year={2021},
url={https://openreview.net/forum?id=Goz-qsH1F14}
}

@inproceedings{goodfellow2013an,
  author       = {Ian J. Goodfellow and
                  Mehdi Mirza and
                  Xia Da and
                  Aaron C. Courville and
                  Yoshua Bengio},
  editor       = {Yoshua Bengio and
                  Yann LeCun},
  title        = {An Empirical Investigation of Catastrophic Forgeting in Gradient-Based
                  Neural Networks},
  booktitle    = {2nd International Conference on Learning Representations, {ICLR} 2014,
                  Banff, AB, Canada, April 14-16, 2014, Conference Track Proceedings},
  year         = {2014},
  url          = {http://arxiv.org/abs/1312.6211},
  bibsource    = {dblp computer science bibliography, https://dblp.org}
}

@article{grootendorst2022bertopic,
  title={BERTopic: Neural topic modeling with a class-based TF-IDF procedure},
  author={Grootendorst, Maarten},
  journal={arXiv preprint arXiv:2203.05794},
  year={2022}
}

\newpage
\onecolumn
\appendix
\section*{Appendix}

\section{Experimental Setup}
\label{app:exp-setup}

\subsection{Datasets and Retain/Unlearning Mapping}
\label{app:datasets}

We organize experiments into three tasks: \emph{Novel}, \emph{WMDP}, and \emph{Toxic}. Sizes and dataset identifiers are summarized in Table~\ref{tab:dataset_other}. Below we state, for each task, the precise mapping from the unlearning dataset (also known as forget dataset) to its corresponding retain dataset(s), consistent with the main text and tables.

\paragraph{Novel task.}
We fine-tune on 12 novels and their associated fan fiction. For unlearning, the \textbf{unlearning} dataset is \novel. The \textbf{retain} dataset is the union of the other 11 novels, as listed in Table~\ref{tab:dataset_other}.

\paragraph{WMDP task.}
The \textbf{unlearning} dataset is \emph{WMDP-bio}~\cite{li2024wmdpbenchmarkmeasuringreducing}. The \textbf{retain} dataset is \emph{Salesforce/wikitext} (specifically \texttt{wikitext-2-raw-v1})~\cite{merity2016pointer}, as listed in Table~\ref{tab:dataset_other}.

\paragraph{Toxic task.}
We fine-tune on the union of \texttt{google/civil\_comments} subsets with (i) toxicity$=0$ and (ii) toxicity$\geq$0.5~\cite{DBLP:journals/corr/abs-1903-04561}. For unlearning, the \textbf{unlearning} dataset is the toxicity$\geq$0.5 subset; the \textbf{retain} dataset is the toxicity$=0$ subset, as listed in Table~\ref{tab:dataset_other}.

\begin{table}[h]
\centering
\caption{Dataset sizes for corpora used in the WMDP, Toxic, and Novel tasks.}
\label{tab:dataset_other}
\resizebox{0.45\columnwidth}{!}{%
\begin{tabular}{lcc}
\toprule
\textbf{Dataset} & \textbf{Comment} & \textbf{Size} \\
\midrule
WMDP-bio~\cite{li2024wmdpbenchmarkmeasuringreducing} & bio-forget-corpus & 712.96 MB \\
Salesforce/wikitext~\cite{merity2016pointer} & wikitext-2-raw-v1 & 12.77 MB \\
google/civil\_comments~\cite{DBLP:journals/corr/abs-1903-04561} & toxicity $= 0$ & 348.42 MB \\
google/civil\_comments & toxicity $\ge 0.5$ & 38.38 MB \\
Novel task (fine-tune) & All novels + fan fiction & 81.90 MB \\
Novel task (unlearning) & One novel (\novel) & 0.63 MB \\
Novel task (retain) & Other 11 novels & 33.85 MB \\
\bottomrule
\end{tabular}%
}
\end{table}

\subsection{Models and Training Environment}
\label{app:models-setup}

We conduct experiments with \emph{Meta-Llama-3-8B}~\cite{grattafiori2024llama3herdmodels}, \emph{Qwen2.5-7B}~\cite{yang2024qwen2technicalreport}, and \emph{Zephyr-7B-beta}~\cite{tunstall2023zephyrdirectdistillationlm}. For unlearning, we employ \emph{Negative Preference Optimization (NPO)}~\cite{zhang2024negativepreferenceoptimizationcatastrophic} and \emph{Removal via Model Update (RMU)} as instantiated in the WMDP framework~\cite{li2024wmdpbenchmarkmeasuringreducing}.

\paragraph{Hardware.}
All runs used four NVIDIA A100-SXM4-80GB GPUs (80\,GB each). The host has two AMD EPYC~7643 48-core processors (96 physical cores / 192 threads). The total system RAM was not captured in our environment logs. The OS is Ubuntu~22.04.5~LTS~(64-bit).

\paragraph{Software.}
We used Python~3.11.9 and PyTorch~v2.6.0 with CUDA Toolkit~12.1. The NVIDIA driver is 535.161.08 (supports up to CUDA~12.2).

\section{Is line 7 of \name jailbreaking?}
\label{app:jialbreak}

Although our method adopts a GCG-like optimization, it is fundamentally different from jailbreaking. In jailbreaking, free tokens are optimized to coerce the model into producing a specific harmful or restricted continuation. In contrast, our objective does not prescribe any target output. Instead, the free tokens are optimized solely to maximize separation from $D_u$ in the embedding space. The outcome is not unsafe generations, but natural sentences that remain semantically related to $D_u$ while stress-testing unlearning metrics. The similarity to GCG is thus purely structural, not adversarial in either intent or use.  
In short, \mini borrows the mechanics of GCG but serves a fundamentally different, non-adversarial purpose.
\section{Additional Experiment Results}
\label{app:exp_res}

\begin{figure*}[h!]
    \centering
    \includegraphics[width=1\linewidth]{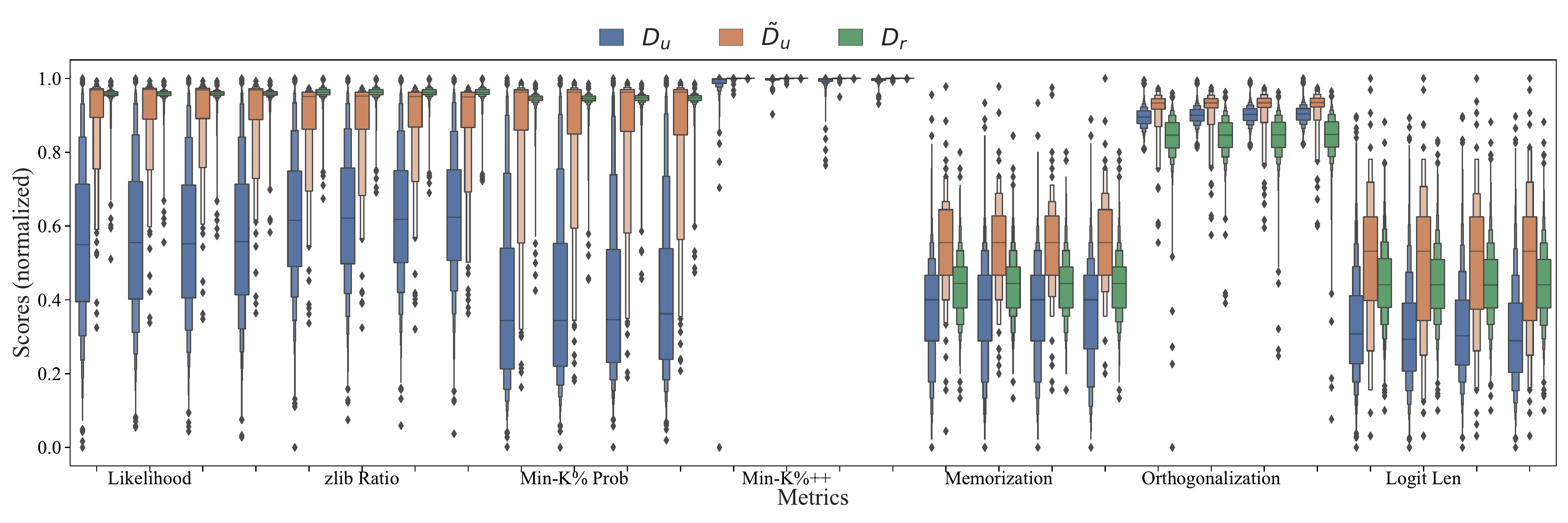}
    \caption{Boxplots for metric scores of $D_u$ (\textit{\novel}), $\tilde{D}_u$, and $D_r$ where the unlearning method is NPO and the model architecture is Llama-3. With the same model before unlearning, we perform unlearning 4 times with identical hyperparameters but different random seeds. Therefore, there are 4 boxes for each of $D_u$, $\tilde{D}_u$, and $D_r$ for every metric. Note that, besides $D_u$ and $D_r$, the $\tilde{D}_u$ is also identical across the 4 runs of unlearning, since \name is not impacted by the unlearned model.
    One can see that \mini performs consistently across different random seeds.}
    \label{fig:repeat}
\end{figure*}
\begin{figure}[t!]
    \centering
    \subfloat[Negative Preference Optimization (NPO)]{
    \includegraphics[width=0.45\linewidth]{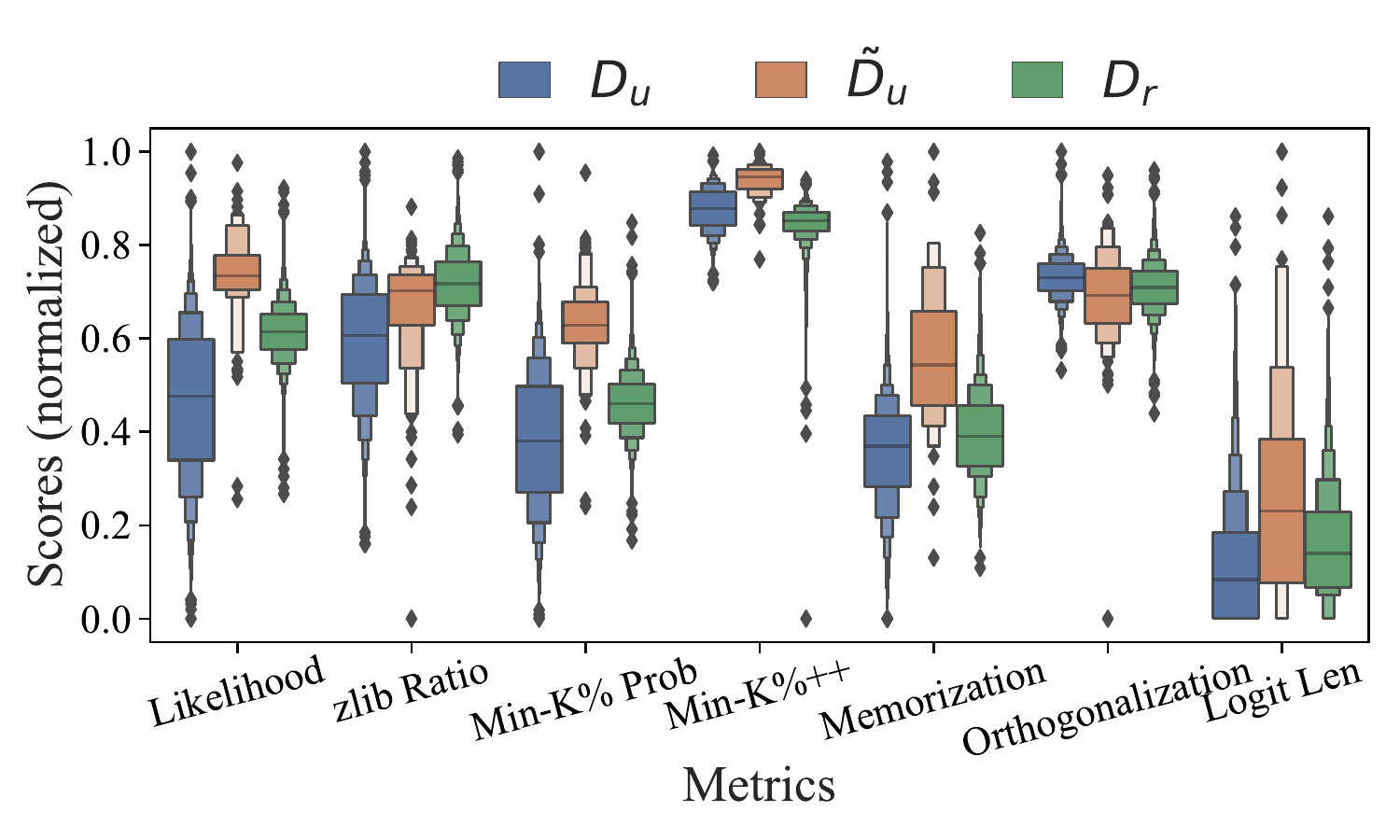}
    }
    \subfloat[Representation Misdirection for Unlearning (RMU)]{
    \includegraphics[width=0.45\linewidth]{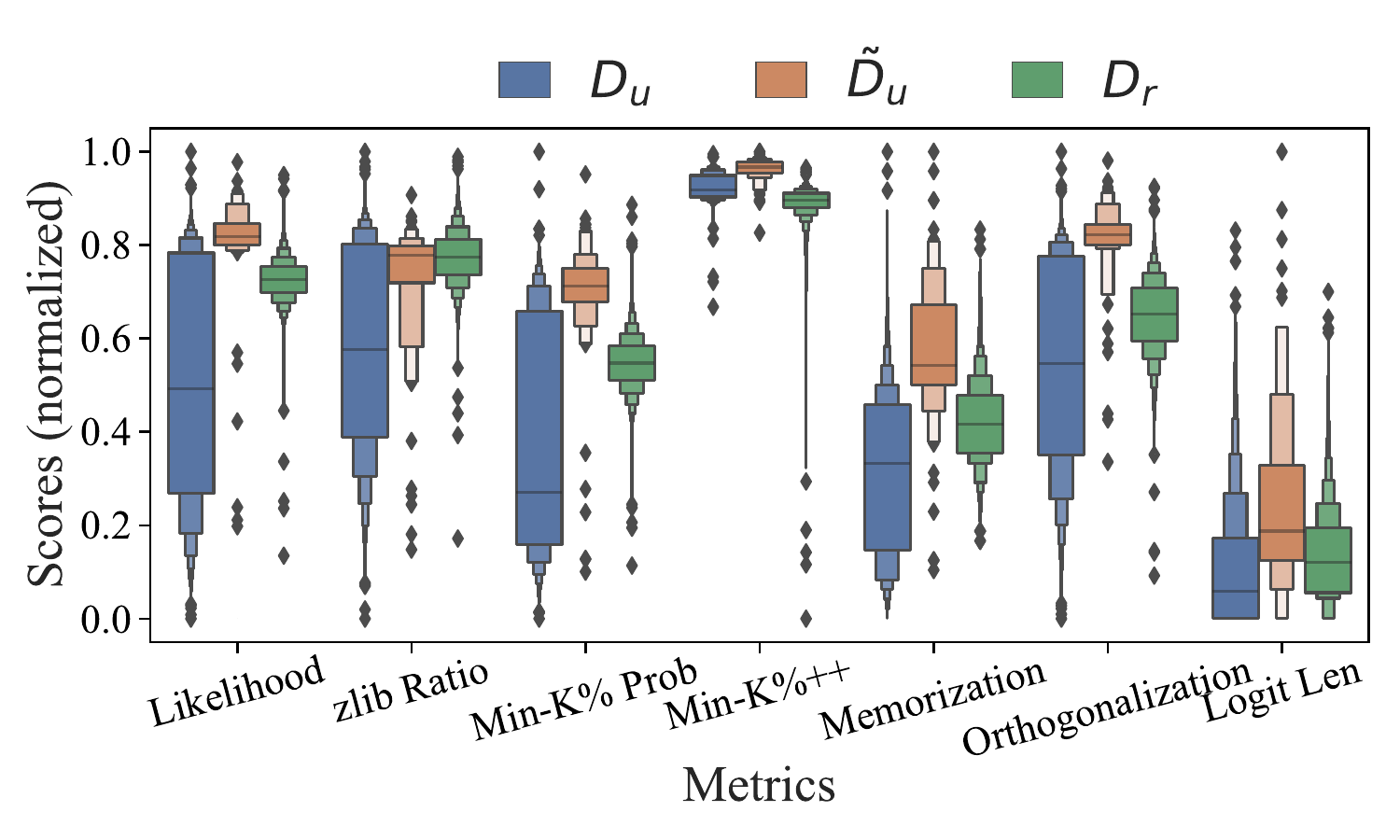}}
    \caption{This is a reproduction of Figure~\ref{fig:hics_llama} with identical settings, except the model architecture is Qwen2.5.}    \label{fig:hics_qwen}
\end{figure}
\begin{figure}[t!]
    \centering
    \subfloat[Negative Preference Optimization (NPO)]{
    \includegraphics[width=0.45\linewidth]{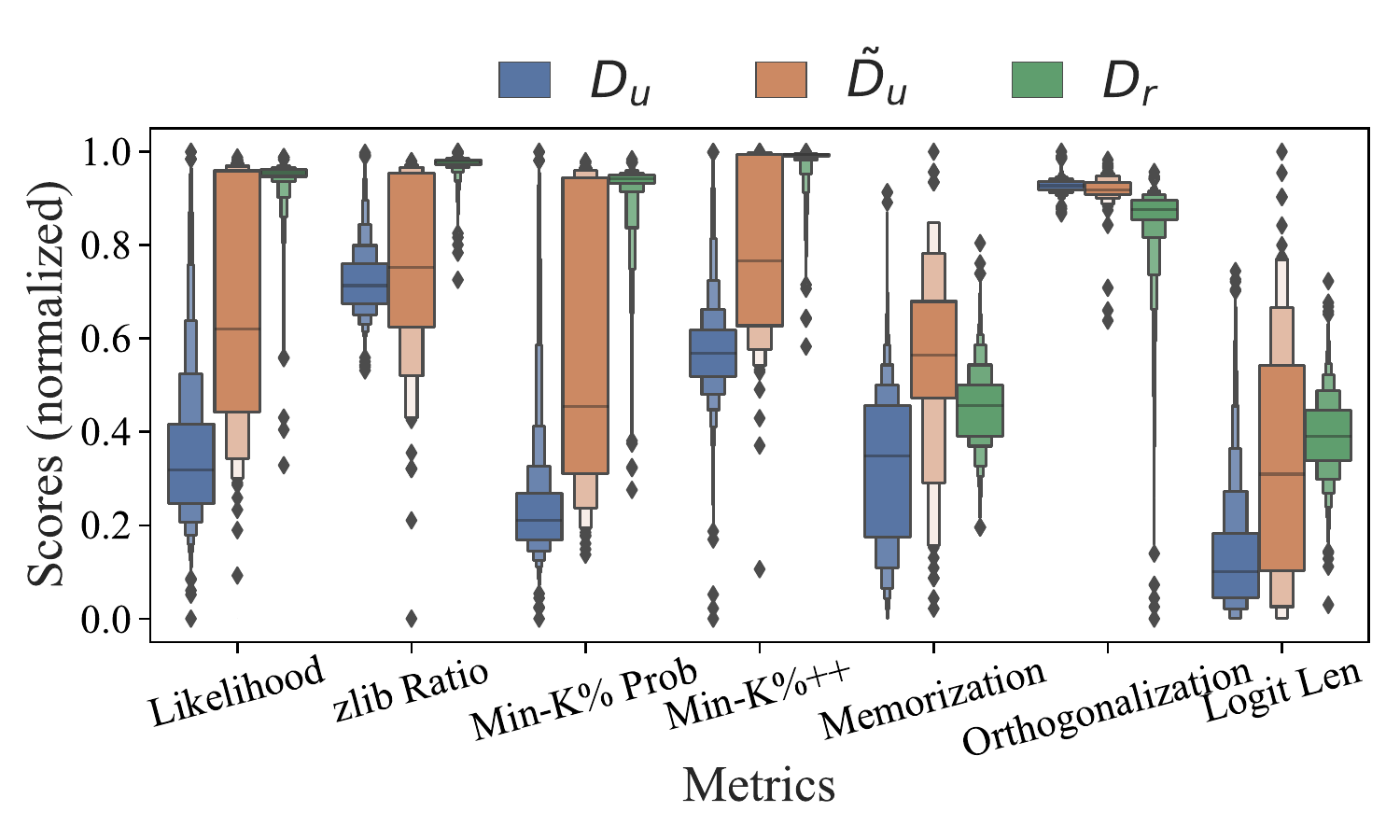}
    }
    \subfloat[Representation Misdirection for Unlearning (RMU)]{
    \includegraphics[width=0.45\linewidth]{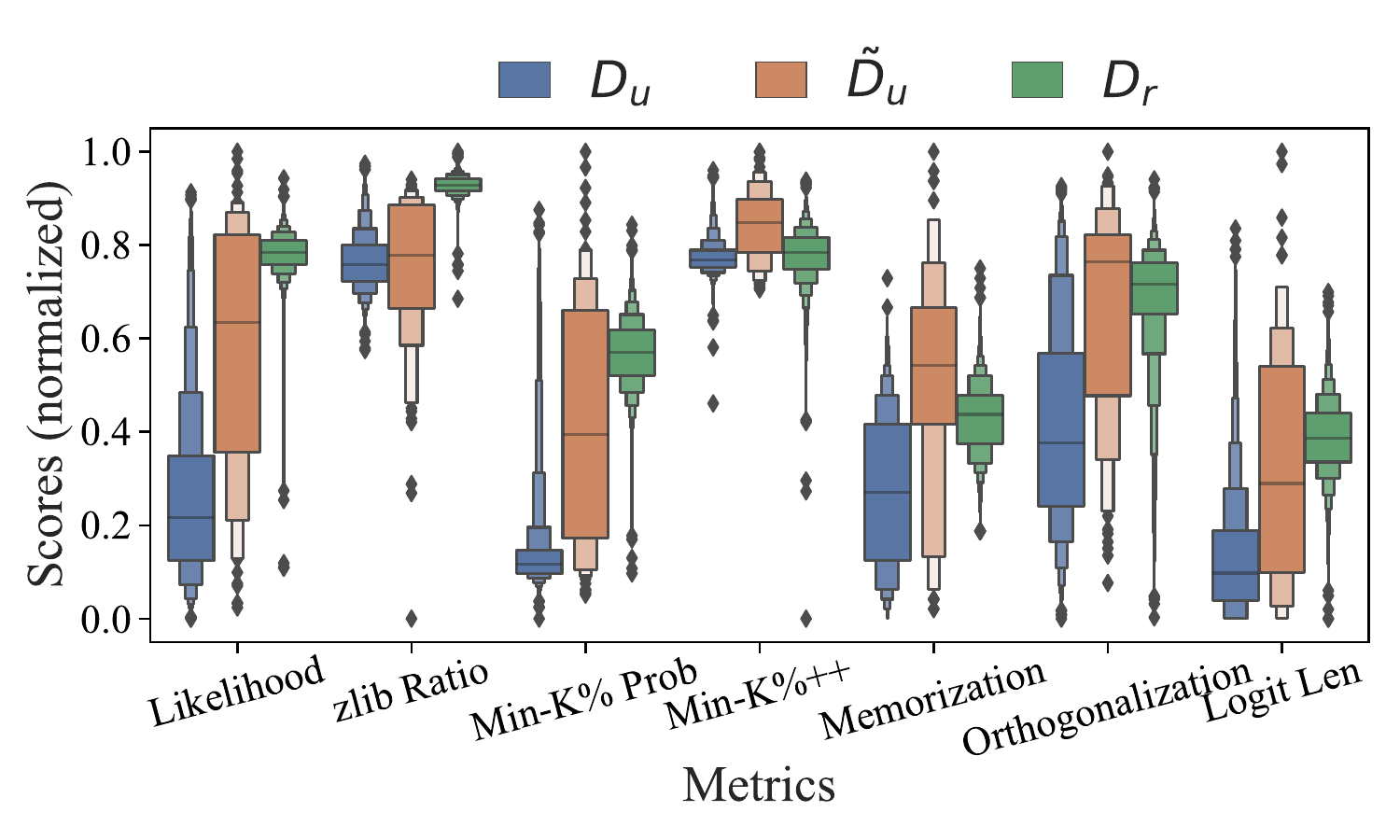}}
    \caption{This is a reproduction of Figure~\ref{fig:hics_llama} with identical settings, except the model architecture is Zephyr.}    \label{fig:hics_zephyr}
\end{figure}
\begin{figure}[t!]
    \centering
    \subfloat[Negative Preference Optimization (NPO)]{
    \includegraphics[width=0.45\linewidth]{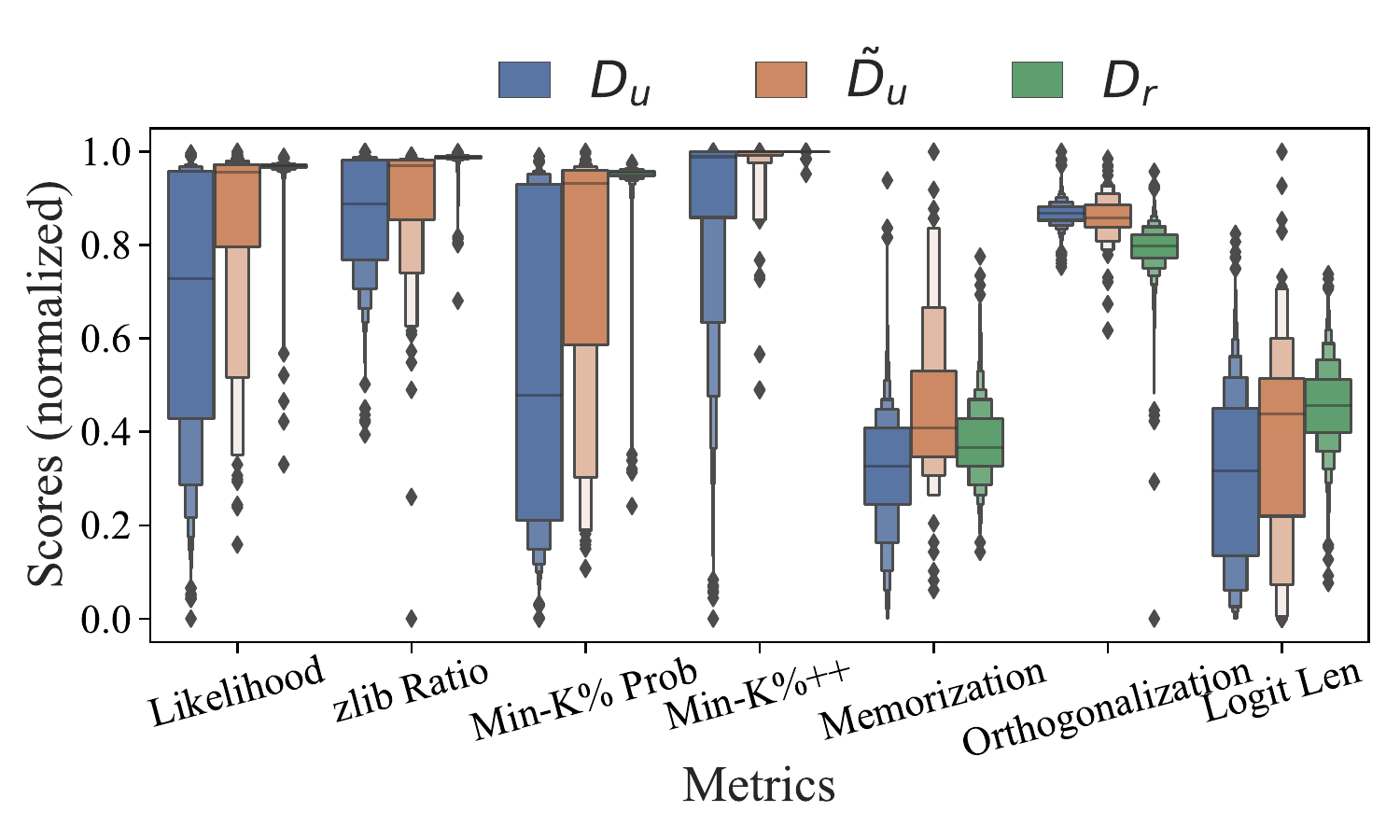}
    }
    \subfloat[Representation Misdirection for Unlearning (RMU)]{
    \includegraphics[width=0.45\linewidth]{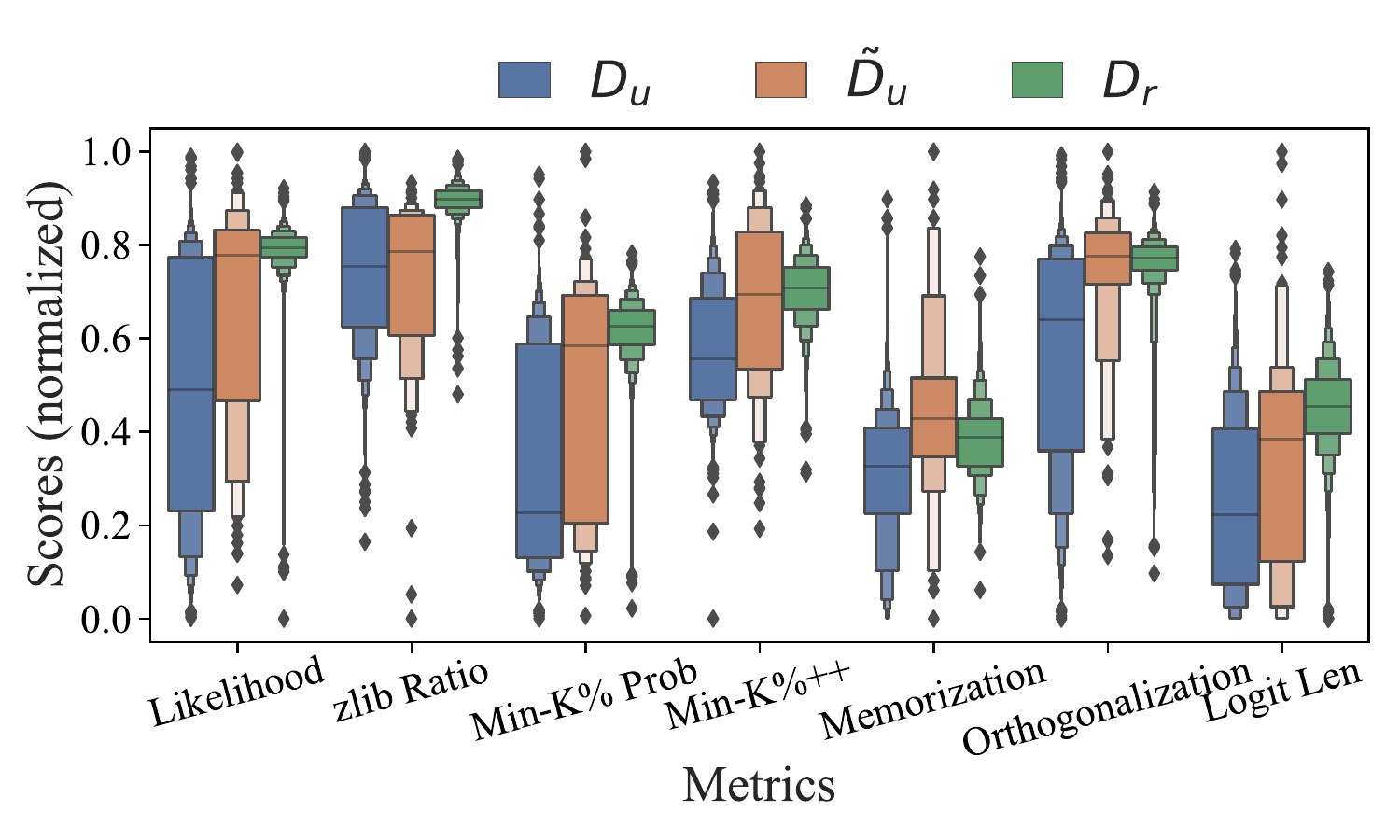}}
    \caption{This is a reproduction of Figure~\ref{fig:hics_llama} with identical settings, except the dataset is Civil Comments. Note that here, for computational efficiency, the metric values are computed for $5\%$ of $D_u$ (randomly sampled), which contain about 13 thousand sentences.}    \label{fig:toxic_llama}
\end{figure}
\begin{figure}[t!]
    \centering
    \subfloat[Negative Preference Optimization (NPO)]{
    \includegraphics[width=0.45\linewidth]{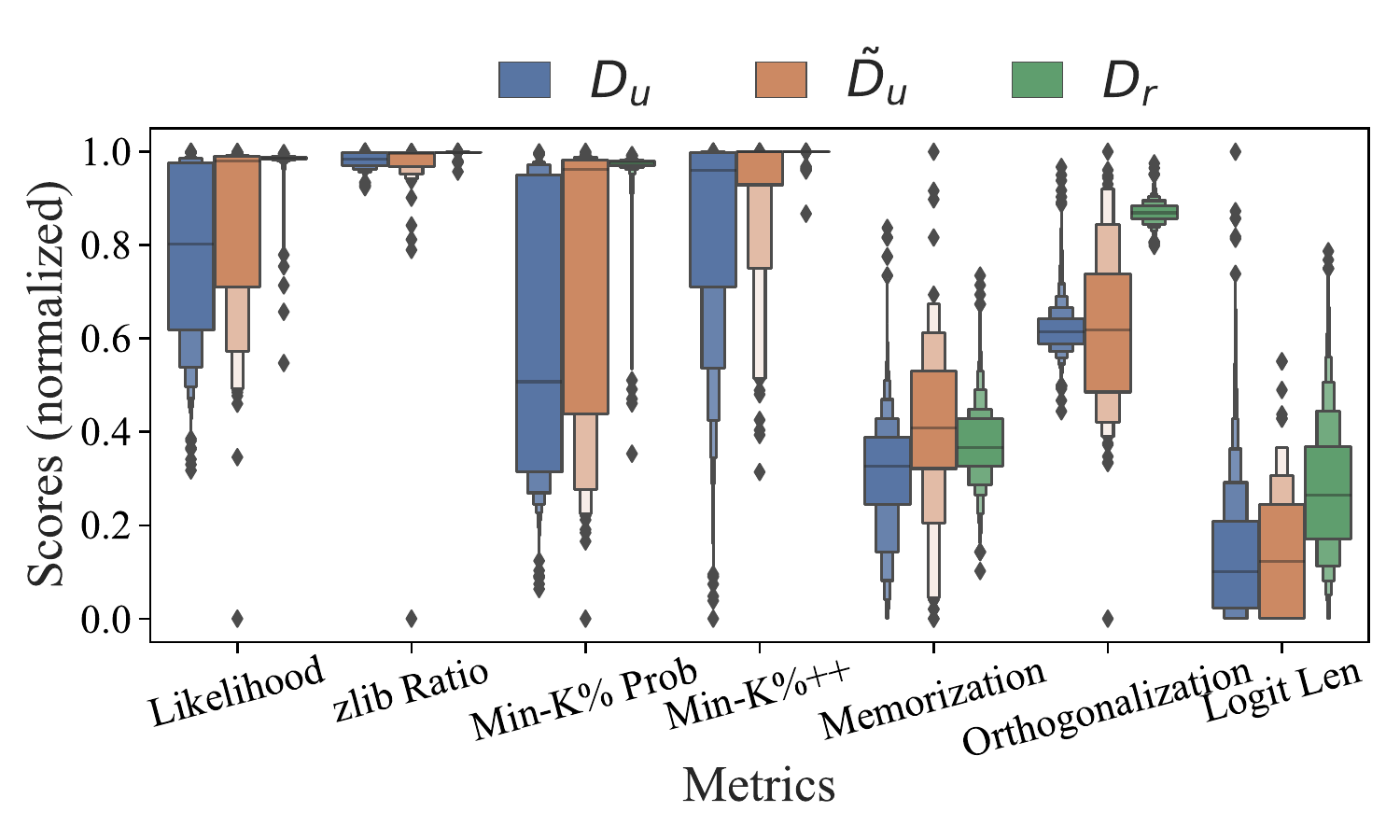}
    }
    \subfloat[Representation Misdirection for Unlearning (RMU)]{
    \includegraphics[width=0.45\linewidth]{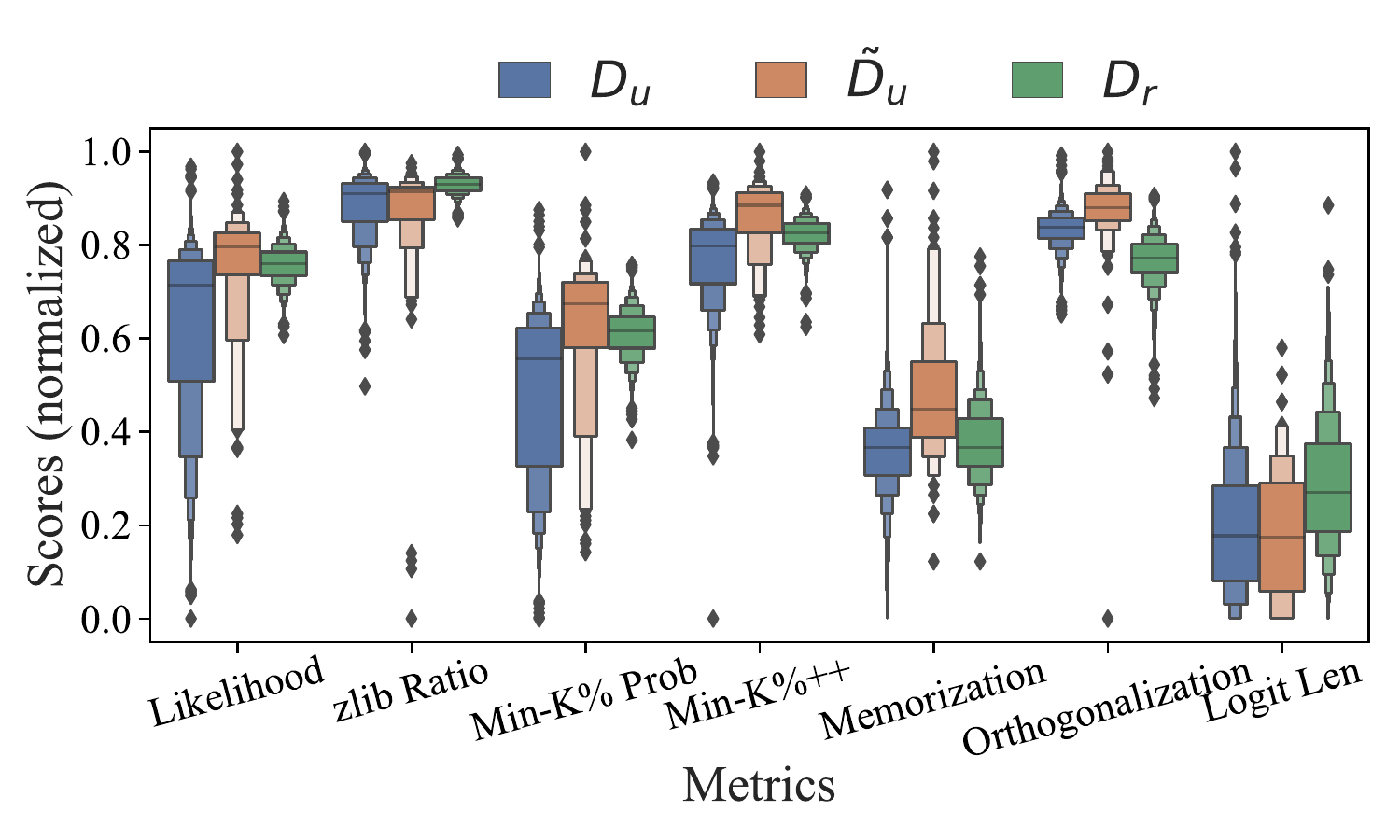}}
    \caption{This is a reproduction of Figure~\ref{fig:hics_llama} with identical settings, except the dataset is Civil Comments, and the model architecture is Qwen2.5.}    \label{fig:toxic_qwen}
\end{figure}
\begin{figure}[t!]
    \centering
    \subfloat[Negative Preference Optimization (NPO)]{
    \includegraphics[width=0.45\linewidth]{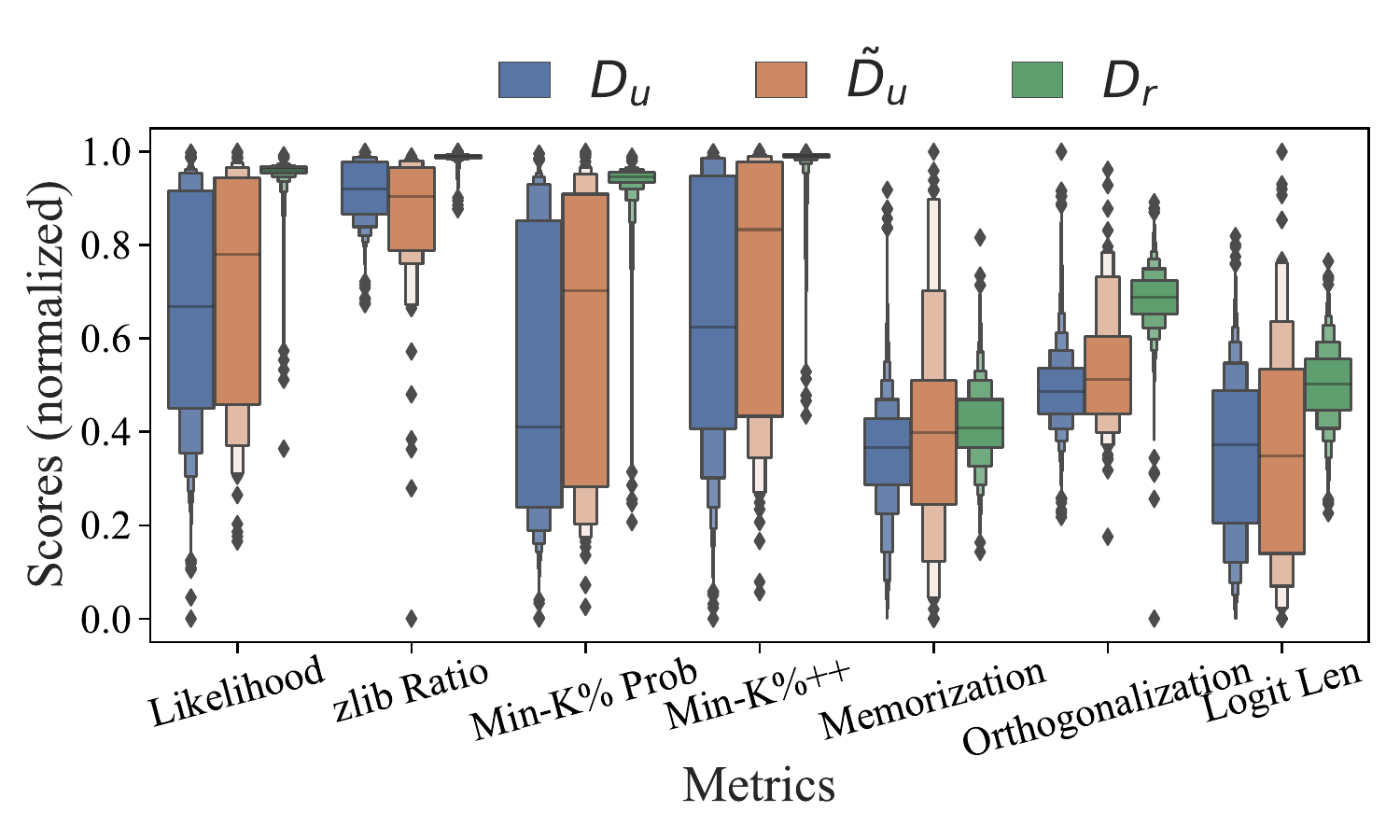}
    }
    \subfloat[Representation Misdirection for Unlearning (RMU)]{
    \includegraphics[width=0.45\linewidth]{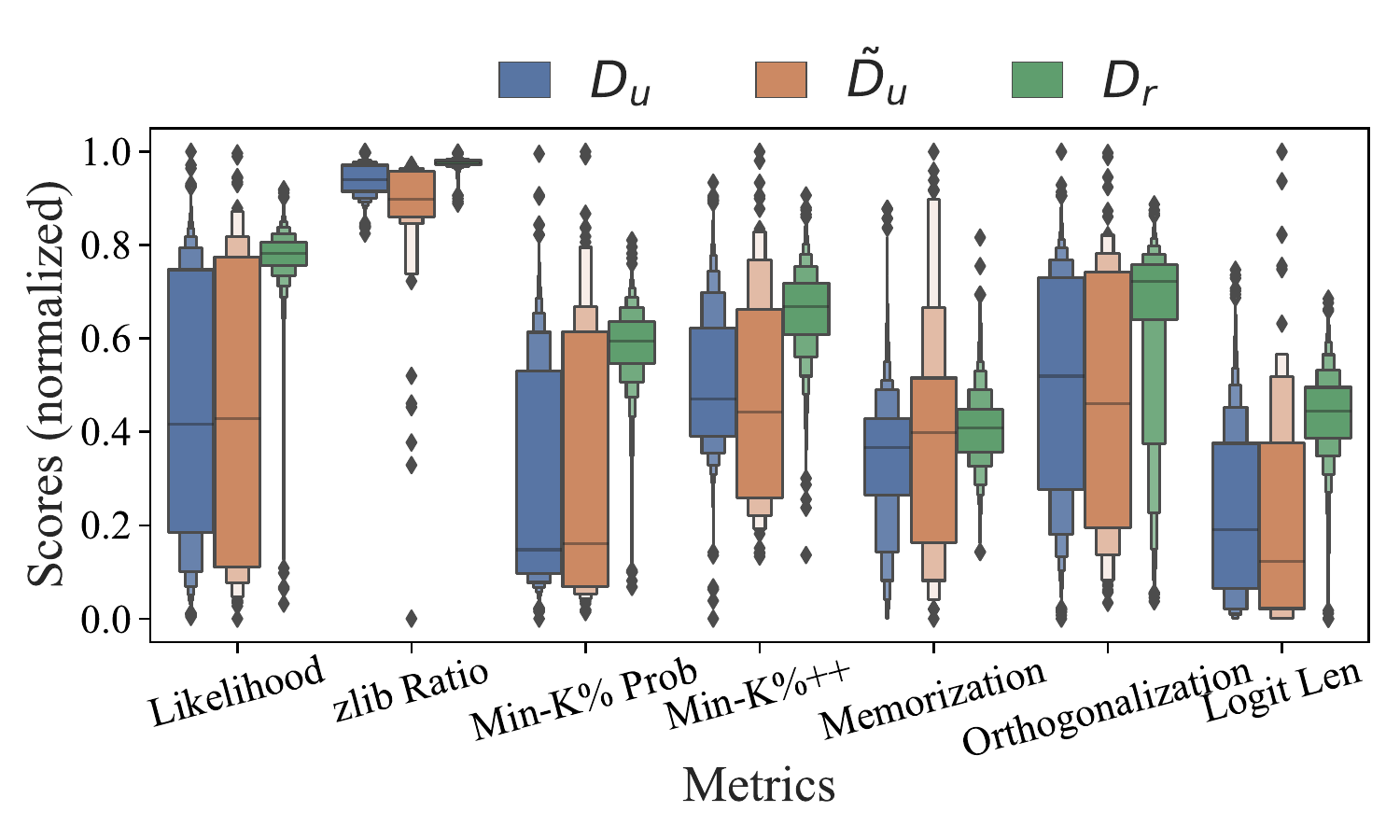}}
    \caption{This is a reproduction of Figure~\ref{fig:hics_llama} with identical settings, except the dataset is Civil Comments, and the model architecture is Zephyr.}    \label{fig:toxic_zephyr}
\end{figure}
\begin{figure}[t!]
    \centering
    \subfloat[Negative Preference Optimization (NPO)]{
    \includegraphics[width=0.45\linewidth]{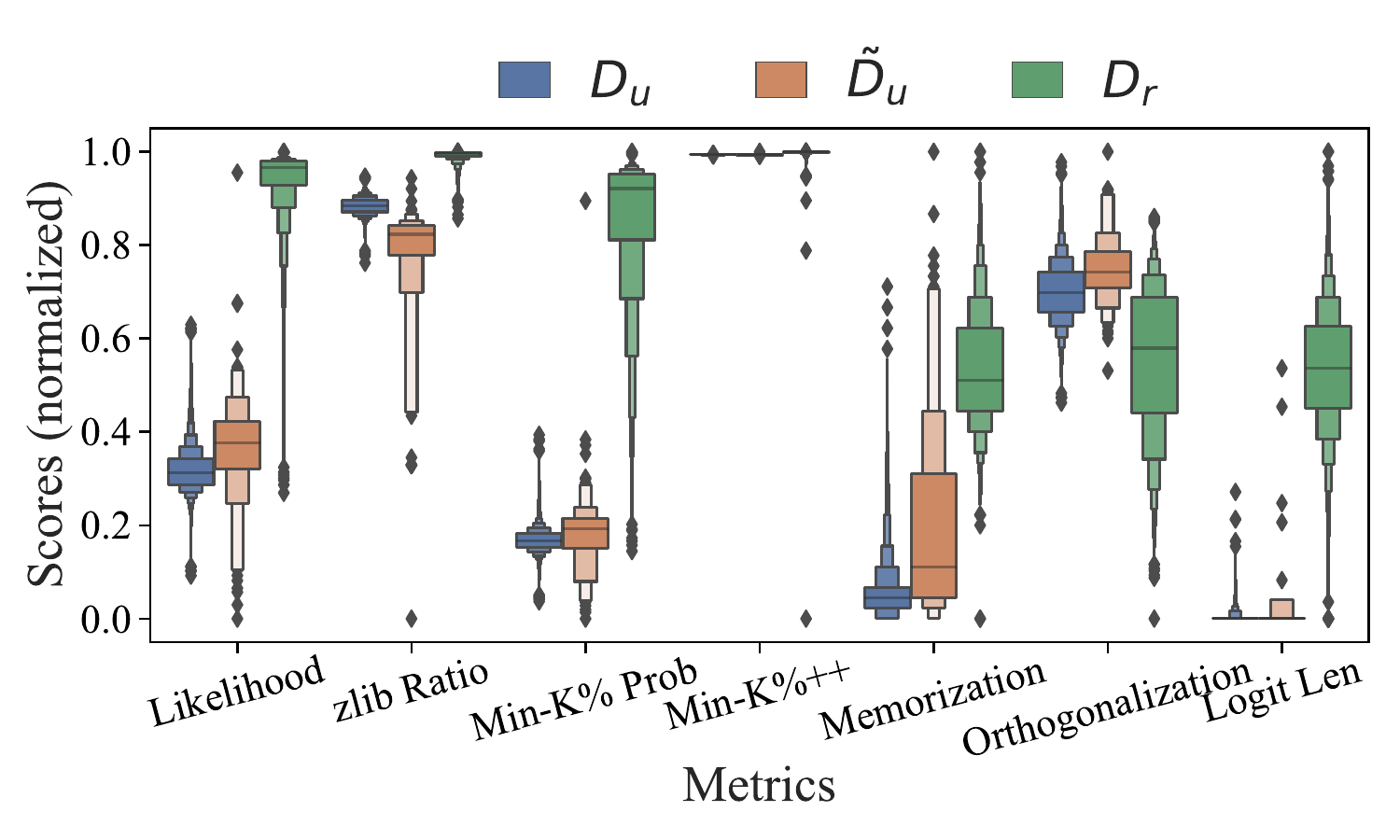}
    }
    \subfloat[Representation Misdirection for Unlearning (RMU)]{
    \includegraphics[width=0.45\linewidth]{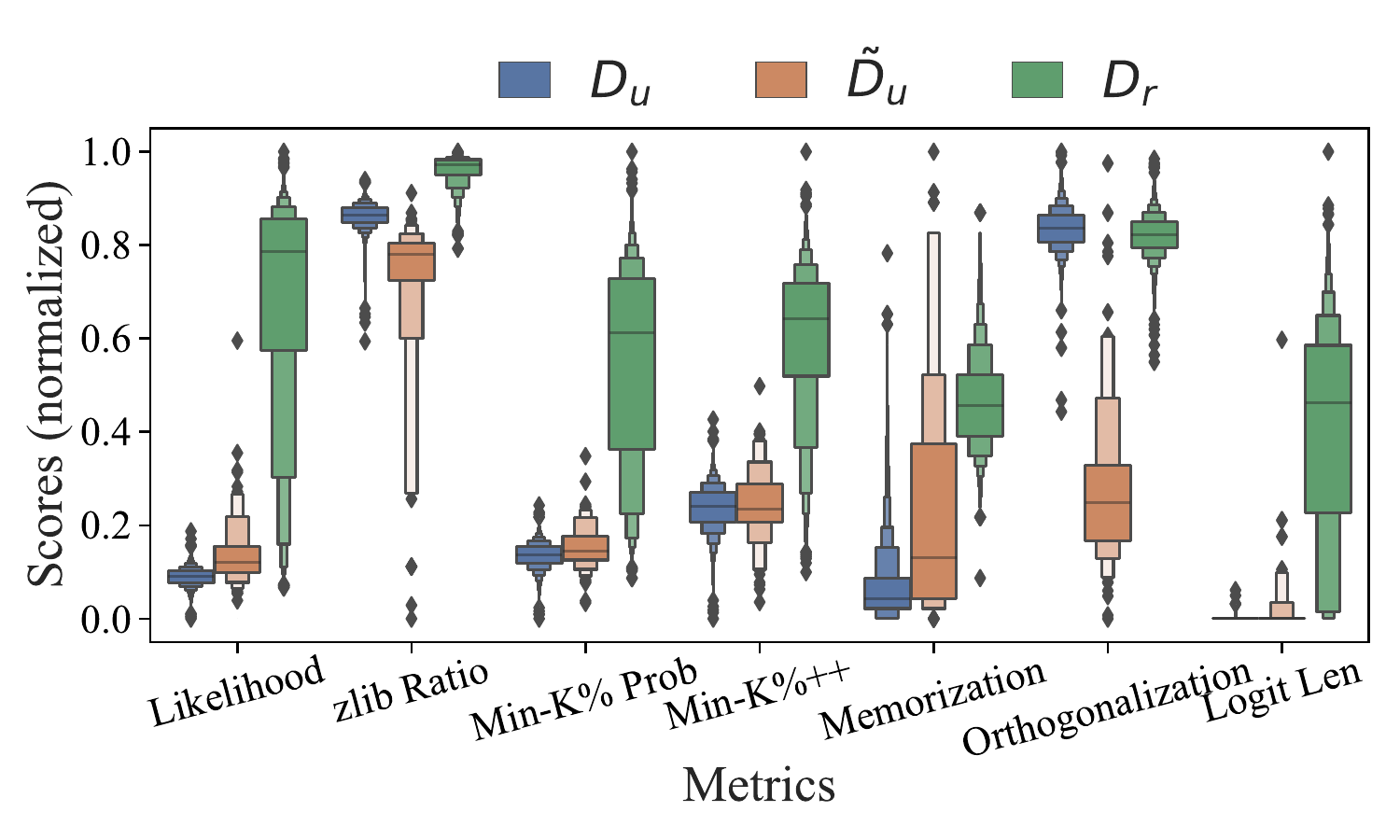}}
    \caption{This is a reproduction of Figure~\ref{fig:hics_llama} with identical settings, except the dataset is WMDP-bio. Note that here, for computational efficiency, the metric values are computed for $1\%$ of $D_u$ (randomly sampled), which contain about 18 thousand sentences.}    \label{fig:wmdp_llama}
\end{figure}
\begin{figure}[t!]
    \centering
    \subfloat[Negative Preference Optimization (NPO)]{
    \includegraphics[width=0.45\linewidth]{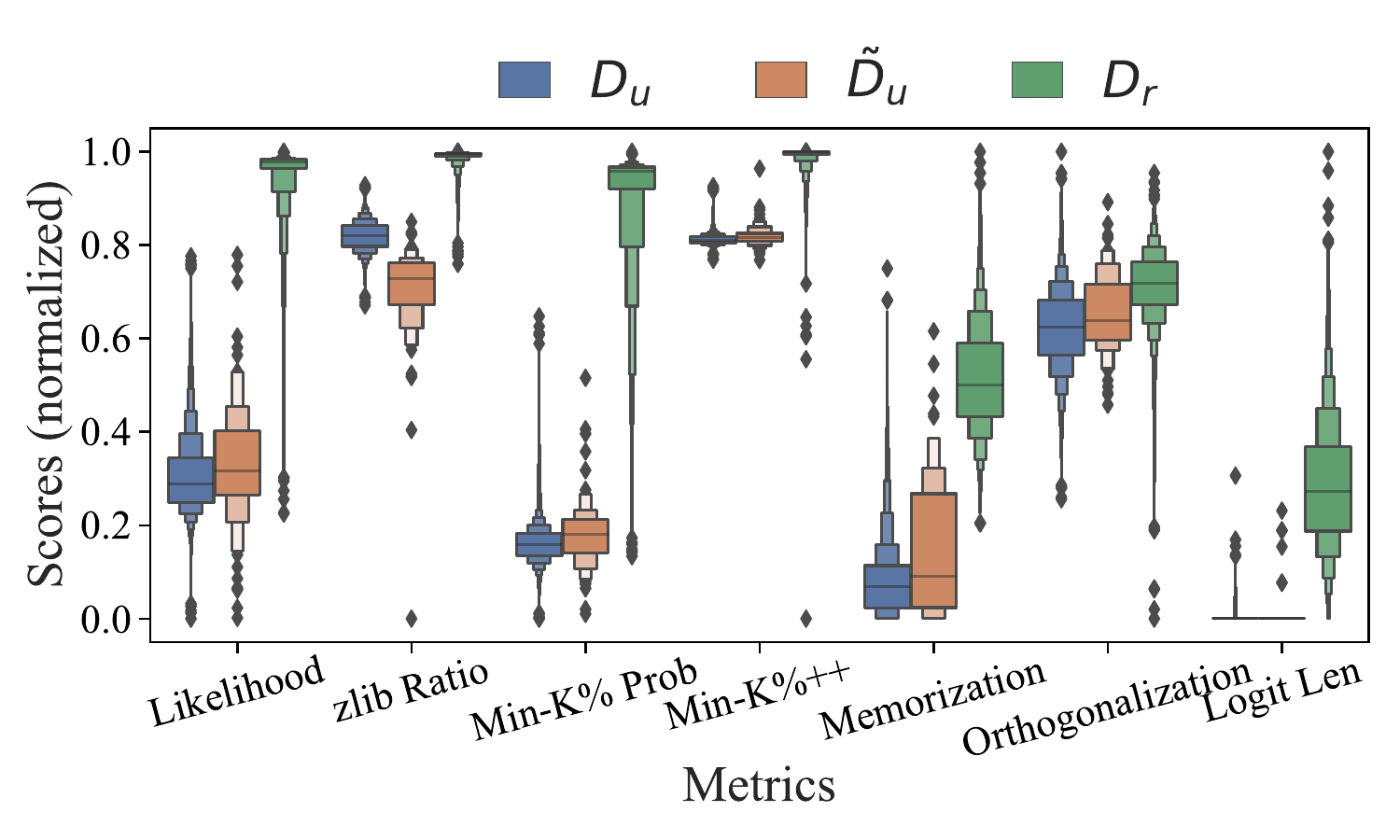}
    }
    \subfloat[Representation Misdirection for Unlearning (RMU)]{
    \includegraphics[width=0.45\linewidth]{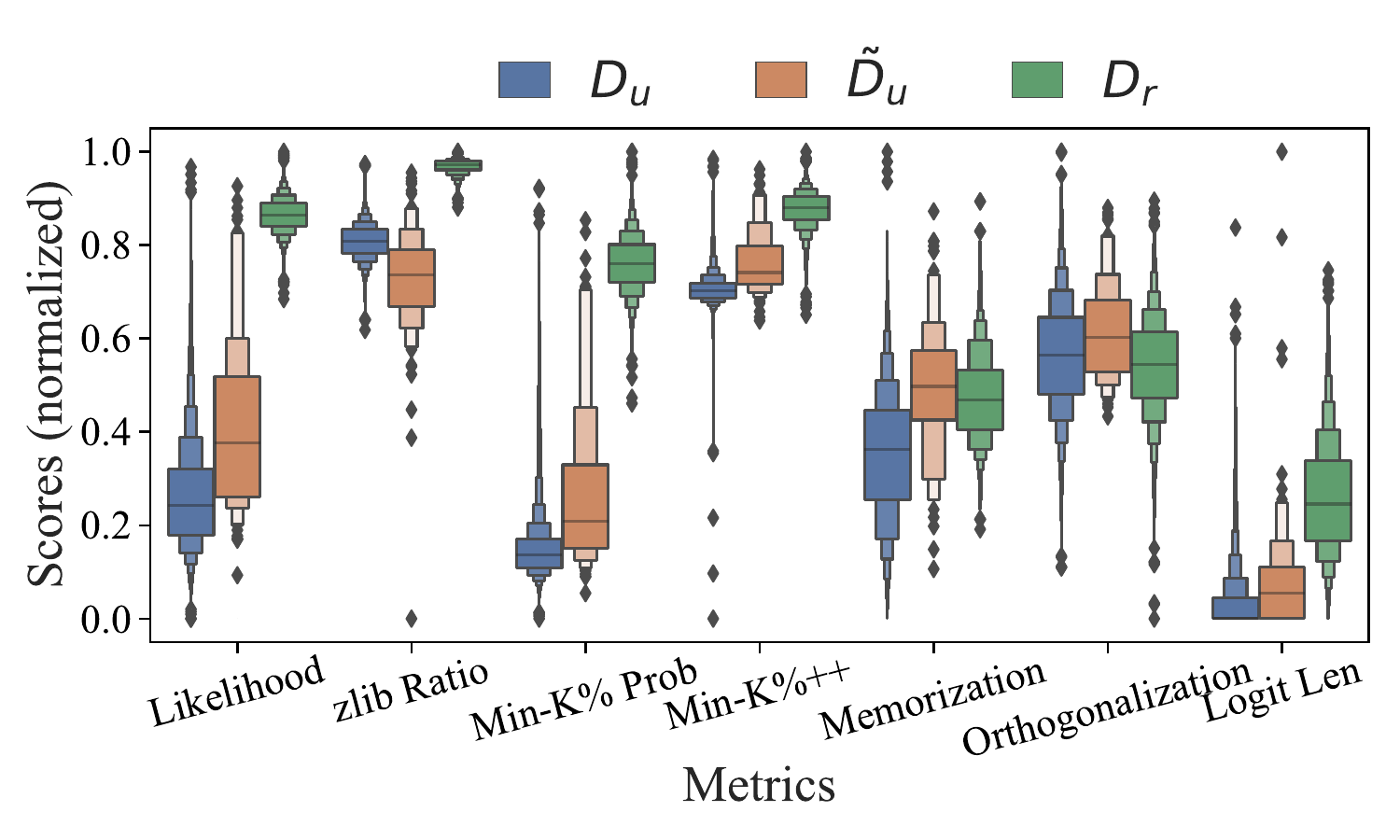}}
    \caption{This is a reproduction of Figure~\ref{fig:hics_llama} with identical settings, except the dataset is WMDP-bio, and the model architecture is Qwen2.5.}    \label{fig:wmdp_qwen}
\end{figure}
\begin{figure}[t!]
    \centering
    \subfloat[Negative Preference Optimization (NPO)]{
    \includegraphics[width=0.45\linewidth]{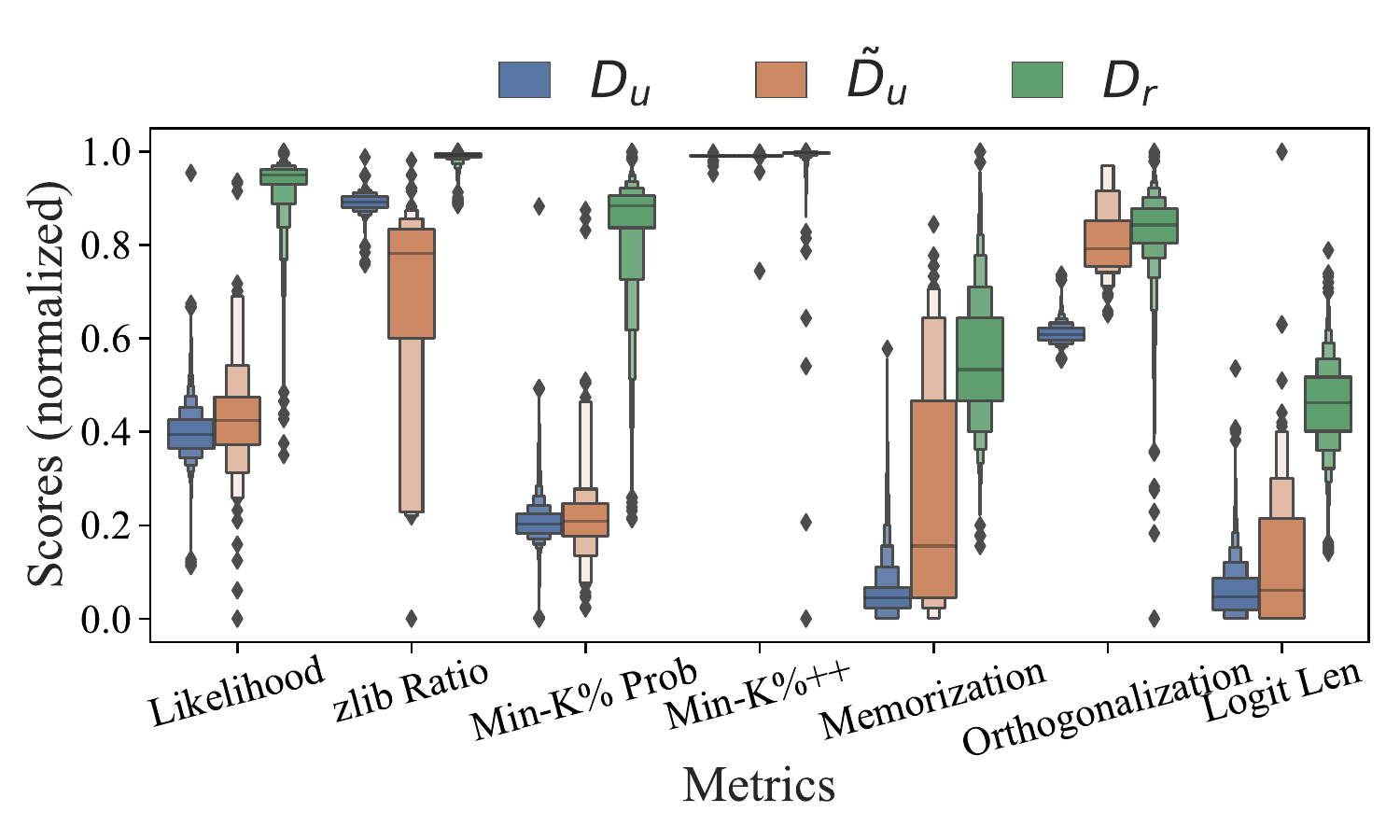}
    }
    \subfloat[Representation Misdirection for Unlearning (RMU)]{
    \includegraphics[width=0.45\linewidth]{plots/WMDP/zephyr/violin_wmdp_bio_percent1_zephyr_npo_l2_continue_1.50_15.00_25.00_467948a8542cc2ff329f9879c49574a0.pdf}}
    \caption{This is a reproduction of Figure~\ref{fig:hics_llama} with identical settings, except the dataset is WMDP-bio, and the model architecture is Zephyr.}    \label{fig:wmdp_zephyr}
\end{figure}
\begin{figure}[h!]
    \centering

    \includegraphics[width=0.6\linewidth]{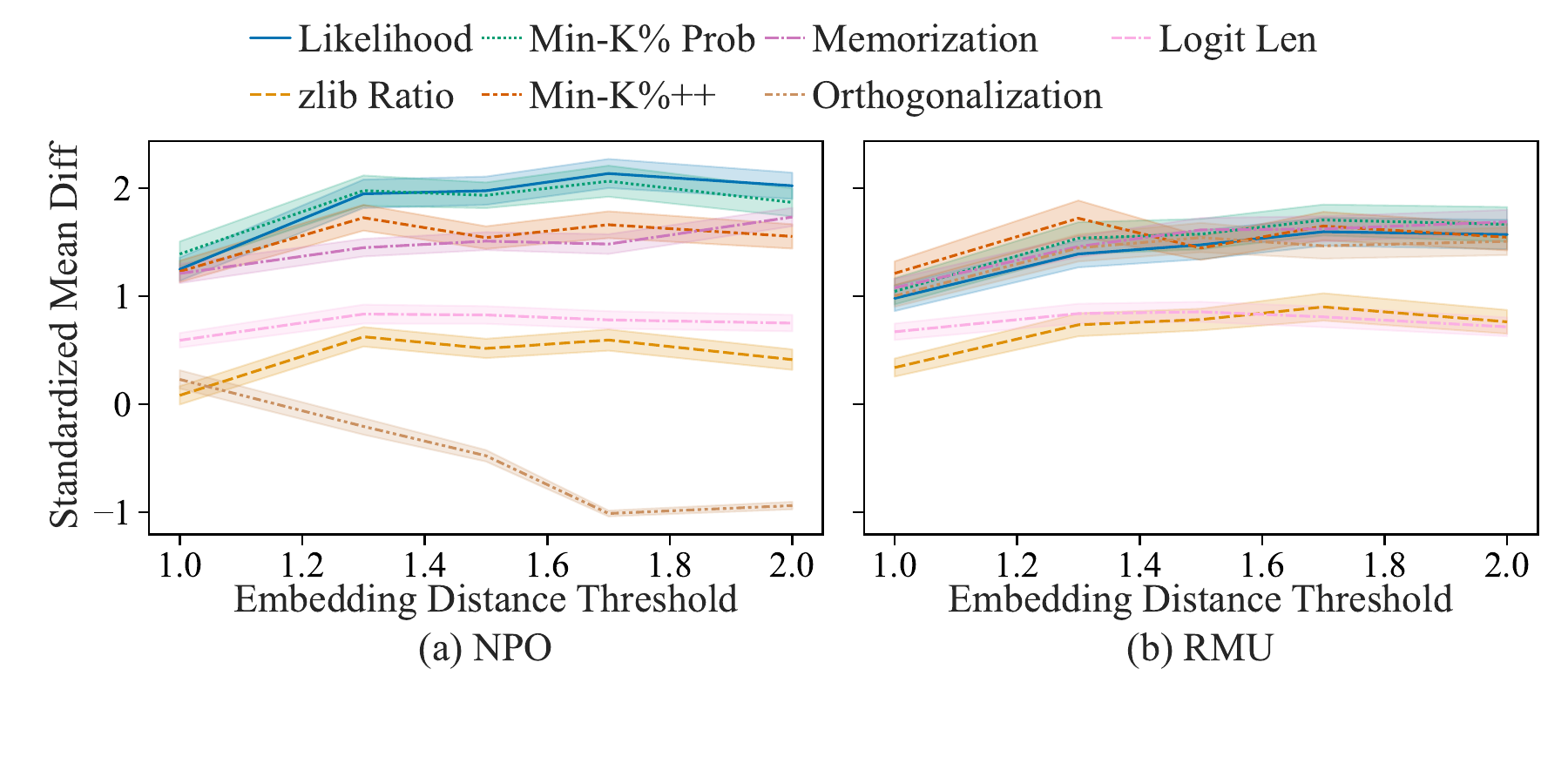}
    
    \caption{This is a reproduction of Figure~\ref{fig:ablation_dist} with identical settings, except the model architecture is Qwen2.5. Similar trends can be observed.}
    \label{fig:hics_qwen_curve}
\end{figure}
\begin{figure}[h!]
    \centering

    \includegraphics[width=0.6\linewidth]{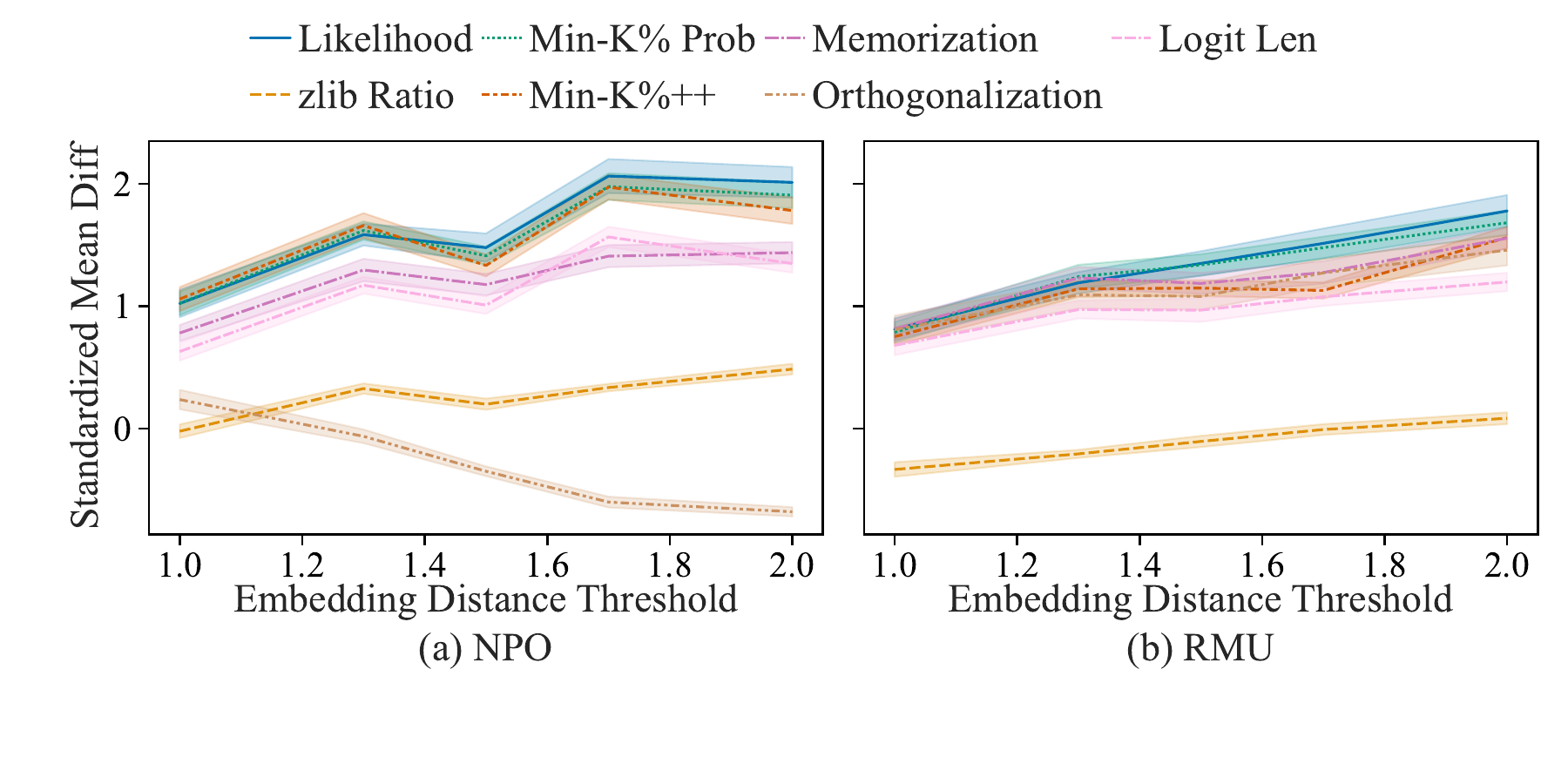}
    
    \caption{This is a reproduction of Figure~\ref{fig:ablation_dist} with identical settings, except the model architecture is Zephyr. Similar trends can be observed.}
    \label{fig:hics_zephyr_curve}
\end{figure}

\end{document}